\documentclass[letterpaper,11pt]{article}

\usepackage{titling}
\usepackage{titlesec}
\usepackage{authblk}

\usepackage[margin=1in]{geometry}
\usepackage[labelfont=bf]{caption}
\usepackage{nowidow}

\usepackage{amsmath}
\usepackage{amssymb}
\usepackage{bm}
\usepackage{bbm}
\usepackage[
  multi-part-units = single,
  range-phrase = --,
  range-units = single,
  separate-uncertainty = true
]{siunitx}
\usepackage{graphicx}
\usepackage{tikz}
\usepackage{natbib}
\usepackage{multirow}
\usepackage[flushleft]{threeparttable}
\usepackage{booktabs}
\usepackage{hyperref}
\usepackage{multibib}

\hypersetup{
  colorlinks=true,
  linkcolor=blue,
  filecolor=blue,
  urlcolor=blue,
  citecolor=blue,
  hypertexnames=false
}

\newcites{supp}{Supplementary references}

\usetikzlibrary{arrows.meta}
\tikzset{
    bigarrow/.style={-{Computer Modern Rightarrow[scale=1]}},
}

\newcolumntype{C}{>{\centering\arraybackslash}p{1.25cm}}

\newcommand{\Phivec}{\boldsymbol{\Phi}}
\newcommand{\Psivec}{\boldsymbol{\Psi}}
\newcommand{\Sigmavec}{\boldsymbol{\Sigma}}

\newcommand{\alphavec}{\bm{\alpha}}

\newcommand{\epsilonvec}{\bm{\epsilon}}
\newcommand{\ellvec}{\bm{\ell}}

\newcommand{\gammavec}{\bm{\gamma}}
\newcommand{\muvec}{\bm{\mu}}

\newcommand{\phivec}{\bm{\phi}}
\newcommand{\pivec}{\bm{\pi}}
\newcommand{\psivec}{\bm{\psi}}
\newcommand{\rhovec}{\bm{\rho}}
\newcommand{\thetavec}{\bm{\theta}}
\newcommand{\xivec}{\bm{\xi}}

\newcommand{\zerovec}{\mathbf{0}}
\newcommand{\onevec}{\mathbf{1}}

\newcommand{\bvec}{\mathbf{b}}
\newcommand{\dvec}{\mathbf{d}}

\newcommand{\svec}{\mathbf{s}}

\newcommand{\wvec}{\mathbf{w}}
\newcommand{\xvec}{\mathbf{x}}

\newcommand{\Pvec}{\mathbf{P}}

\newcommand{\Zvec}{\mathbf{Z}}

\newcommand{\corr}{\textrm{corr}}
\newcommand{\cov}{\textrm{cov}}
\newcommand{\var}{\textrm{var}}

\newcommand{\titlevar}{WOMBAT \lowercase{v}2.S: A Bayesian inversion framework for attributing global CO\textsubscript{2} flux components from multiprocess data}

\makeatletter
\def\@maketitle{
  \newpage
  \null
  \vspace{-1em}
  \begin{center}
  \let \footnote \thanks
    {\LARGE \@title \par}
    \vskip 2em
    {\large \@author \par}
    \vskip 1em
    {\small {}*Corresponding author: Josh Jacobson (\href{mailto:joshj@uow.edu.au}{joshj@uow.edu.au}) \par}
  \end{center}
  \vspace{1ex}
}
\makeatother

\title{\titlevar}

\author[1,\textasteriskcentered]{Josh Jacobson}
\author[2]{Michael Bertolacci}
\author[1]{Andrew Zammit-Mangion}
\author[3]{\authorcr Andrew Schuh}
\author[1,4]{Noel Cressie}

\affil[1]{School of Mathematics and Applied Statistics, University of Wollongong, Australia}
\affil[2]{School of Physics, Mathematics, and Computing, University of Western Australia, Australia}
\affil[3]{Cooperative Institute for Research in the Atmosphere (CIRA), Colorado State University, USA}
\affil[4]{Jet Propulsion Laboratory, California Institute of Technology, USA}

\begin{document}

\maketitle

\begin{abstract}
  Contributions from photosynthesis and other natural components of the carbon cycle present the largest uncertainties in our understanding of carbon dioxide (CO\textsubscript{2}) sources and sinks.
  While the global spatiotemporal distribution of the net flux (the sum of all contributions) can be inferred from atmospheric CO\textsubscript{2} concentrations through flux inversion, attributing the net flux to its individual components remains challenging.
  The advent of solar-induced fluorescence (SIF) satellite observations provides an opportunity to isolate natural components by anchoring gross primary productivity (GPP), the photosynthetic component of the net flux.
  Here, we introduce a novel statistical flux-inversion framework that simultaneously assimilates observations of SIF and CO\textsubscript{2} concentration, extending WOMBAT~v2.0 (WOllongong Methodology for Bayesian Assimilation of Trace-gases, version 2.0) with a hierarchical model of spatiotemporal dependence between GPP and SIF processes.
  We call the new framework WOMBAT~v2.S, and we apply it to SIF and CO\textsubscript{2} data from NASA's Orbiting Carbon Observatory-2 (OCO-2) satellite and other instruments to estimate natural fluxes over the globe during a recent six-year period.
  In a simulation experiment that matches OCO-2's retrieval characteristics, the inclusion of SIF improves accuracy and uncertainty quantification of component flux estimates.
  Comparing estimates from WOMBAT~v2.S, v2.0, and the independent FLUXCOM initiative, we observe that linking GPP to SIF has little effect on net flux, as expected, but leads to spatial redistribution and more realistic seasonal structure in natural flux components.
\end{abstract}

\vspace{2ex}


\vspace{2ex}

\section{Introduction}

Carbon dioxide (CO\textsubscript{2}) is continuously exchanged between Earth's surface and its atmosphere.
The rate of exchange at a specific location and time is called net surface flux of CO\textsubscript{2}, and it is determined by both natural phenomena and human activities.
The largest natural flux components are gross primary productivity (GPP) and ecosystem respiration \citep[e.g.,][]{BeerEtAl2010}.
GPP is the absorption of CO\textsubscript{2} by plants during photosynthesis (negative flux), and ecosystem respiration releases CO\textsubscript{2} back into the atmosphere (positive flux), primarily through plant respiration and microbial decomposition.
Over land, the sum of these two components is termed net ecosystem exchange (NEE), which currently acts as a global CO\textsubscript{2} sink, absorbing roughly one-quarter of human CO\textsubscript{2} emissions; another quarter is absorbed by the oceans \citep{FriedlingsteinEtAl2023}.
Despite their significance, the spatial and temporal distributions of natural fluxes are poorly characterized \citep{AnavEtAl2015}, as their direct observation over large areas and long timeframes is not possible \citep{SchimelEtAl2015}.
Thus, natural fluxes remain a major source of uncertainty in our understanding of Earth's carbon cycle \citep{SitchEtAl2015}.
Quantifying the spatiotemporal distributions of GPP and other natural fluxes is critical for advancing carbon cycle research and developing adaptation strategies.

Flux inversion is a widely used method for estimating surface fluxes of trace gases like CO\textsubscript{2} \citep[e.g.,][]{GurneyEtAl2002, CiaisEtAl2010, LiuEtAl2017}.
The method uses an atmospheric chemical transport model to link atmospheric mole-fraction observations to net surface fluxes from upwind of the observation locations.
Due to the rapid mixing of the atmosphere, even abundant atmospheric observations provide imperfect information about surface fluxes, and consequently the high-dimensional inversion problem is often ill-posed \citep{Tarantola2005a}.
To address this challenge, most inversion frameworks adopt a Bayesian approach \citep[see][for several examples]{ByrneEtAl2023}, incorporating strong prior information derived from ``bottom-up'' estimates of individual flux components.
Bottom-up approaches apply mechanistic principles and diverse data sources, such as flux-tower measurements and land-cover classifications, to estimate and scale up flux components from local to regional scales \citep[e.g.,][]{HaynesEtAl2019, JungEtAl2020}.
Hence, their accuracy varies by component \citep{ByrneEtAl2023}, with emissions due to human activities considered to be relatively reliable, whereas estimates of natural CO\textsubscript{2} flux components are substantially more uncertain \citep[e.g.,][]{BasuEtAl2013}.

Even with informative priors on fossil-fuel and other human emissions, CO\textsubscript{2} flux inversions struggle to attribute flux to natural components such as GPP and respiration.
This is because CO\textsubscript{2} mole-fraction observations are a function of the net flux, which is small compared with the quantity of CO\textsubscript{2} absorbed through GPP and emitted through respiration.
Promisingly, satellite observations of solar-induced fluorescence (SIF), introduced over the past decade, now provide a proxy for photosynthetic activity and, by extension, GPP \citep{FrankenbergEtAl2011, JoinerEtAl2014, SunEtAl2017}.
SIF is a pathway by which plants dissipate excess light energy during photosynthesis, and so SIF observations are a dynamic indicator of the physiological processes that drive GPP \citep{Porcar-CastellEtAl2014, FrankenbergBerry2018}.

SIF observations are increasingly being used to inform bottom-up estimates of GPP \citep[e.g.,][]{ParazooEtAl2014, BaiEtAl2021, KiraEtAl2021, DoughtyEtAl2023}, including in mechanistic models of the terrestrial biosphere \citep{MacBeanEtAl2018, BacourEtAl2019, NortonEtAl2019}, but their integration into flux inversions is largely untapped.
In an early effort, \citet{LiuEtAl2017} used SIF observations to characterize GPP after inferring the net flux via inversion.
Since this post-inversion approach does not assimilate SIF directly, it implicitly treats SIF and CO\textsubscript{2} mole-fraction observations as independent sources of information about GPP.
More recently, gap-filled SIF data products \citep[e.g.,][]{ZhangEtAl2018} have been assimilated by \citet{ShigaEtAl2018} and \citet{ZhangEtAl2023a} to constrain variability in the net flux, but without a direct link to GPP.
Consequently, these inversion methods cannot distinguish individual flux components.

In this paper, we introduce a novel flux-inversion framework that, for the first time, jointly assimilates satellite observations of SIF alongside satellite and in~situ observations of CO\textsubscript{2} mole fraction.
We do this by extending the WOllongong Methodology for Bayesian Assimilation of Trace-gases (WOMBAT) statistical flux-inversion framework.
The initial version, WOMBAT~v1.0 \citep{Zammit-MangionEtAl2022a}, was designed to estimate net fluxes while tracking uncertainties stemming from discretization errors \citep[e.g.,][]{KaminskiEtAl2001}, inaccuracies in transport models \citep[e.g.,][]{BasuEtAl2018, SchuhEtAl2019}, and biases and dependencies in measurements \citep[e.g.,][]{Chevallier2007, ODellEtAl2018}.
Subsequently, WOMBAT~v2.0 \citep{BertolacciEtAl2024} introduced multiple advances, including the ability to decompose NEE into GPP and respiration components by imposing physical constraints.
Despite this effort, v2.0 struggled to disentangle GPP and respiration due to its reliance on CO\textsubscript{2} mole-fraction data alone.

We call our new framework WOMBAT~v2.S, where ``S'' stands for ``SIF,'' since it extends v2.0 to include a spatially and temporally varying link between GPP and SIF.
This link takes the form of a linear relationship motivated through a Taylor approximation, in which the expansion point and coefficients derive from formal analyses of mechanistic bottom-up estimates of GPP and SIF.
Through this relationship, SIF observations provide a direct constraint on GPP, thereby complementing CO\textsubscript{2} mole-fraction observations.
We show that including SIF observations in the inversion improves both the accuracy of GPP and respiration flux estimates and the quantification of their uncertainties.

The paper is organized as follows.
In Section~\ref{sec:model}, we present the Bayesian hierarchical model that underpins WOMBAT~v2.S, including the novel spatiotemporal relationship we establish between the GPP flux process, the SIF process, and the SIF observations.
Section~\ref{sec:application} describes the configuration and validation of WOMBAT~v2.S, and presents results from applying the framework in a global inversion over 2015--2020.
WOMBAT~v2.S flux estimates and their uncertainties are compared to those of WOMBAT~v2.0 and to bottom-up estimates from the FLUXCOM initiative \citep{NelsonEtAl2024}.
We find that including SIF has little effect on NEE (which is well constrained by CO\textsubscript{2} mole-fraction observations), but leads to a spatial redistribution and more realistic seasonal structure in GPP and respiration.
Section~\ref{sec:conclusion} summarizes our main contributions, discusses the implications of our findings, and suggests avenues for future research.
Supplementary Material provides further details of the model and inversion setup, and presents additional figures and tables that support the results.
Software is openly available at \url{https://github.com/joshhjacobson/wombat-v2s}.

\section{Multivariate Bayesian hierarchical model}
\label{sec:model}

Our new multivariate inversion framework evolves from the Bayesian hierarchical model established in WOMBAT~v2.0 \citep{BertolacciEtAl2024}.
There are four levels in the WOMBAT~v2.0 hierarchy: a process model for CO\textsubscript{2} fluxes, a process model that links the CO\textsubscript{2} mole-fraction process to the CO\textsubscript{2} flux process, a data model for CO\textsubscript{2} mole-fraction observations, and a parameter model for unknown parameters in the process and data models.
For WOMBAT~v2.S, we build on this with both a process model that links the SIF process to the CO\textsubscript{2} flux process, and a data model for SIF satellite observations.
Section~\ref{sec:flux-process-model} presents the flux process model, which decomposes the net CO\textsubscript{2} surface flux into individual component fluxes such as GPP and respiration.
Section~\ref{sec:mole-frac-process-model} gives a brief overview of the CO\textsubscript{2} mole-fraction process model.
Section~\ref{sec:sif-process-model} introduces the new SIF process model, which is linked to the flux process through its GPP component.
Section~\ref{sec:data-model} presents the SIF data model and details how SIF and CO\textsubscript{2} mole-fraction observations are combined into a single, grouped data model.
Section~\ref{sec:parameter-model} discusses the updated parameter model, and Section~\ref{sec:inference} describes inference for all unknowns in the hierarchical model.

\subsection{Flux process model}
\label{sec:flux-process-model}

Let $X(\svec, t)$ be the net CO\textsubscript{2} surface flux at location $\svec \in \mathbb{S}^2$ and time $t \in \mathcal{T} \subset \mathbb{R}$, where $\mathbb{S}^2$ is Earth's surface and $\mathcal{T} \equiv [t_0, t_1]$ is the time period of interest.
We model $X(\svec, t)$ as the sum of natural and other fluxes:
\begin{equation}
  \label{eq:flux-components}
  X(\svec, t) = X_{\textup{gpp}}(\svec, t) + X_{\textup{resp}}(\svec, t) + X_{\textup{ocean}}(\svec, t) + X_{\textup{other}}^0(\svec, t), \quad \svec \in \mathbb{S}^2,\ t \in \mathcal{T},
\end{equation}
where $X_c(\cdot\,, \cdot)$, $c \in \mathcal{C} \equiv \{\textup{gpp}, \textup{resp}, \textup{ocean}\}$, are GPP and respiration of CO\textsubscript{2} from the terrestrial biosphere, and ocean-air exchange due to CO\textsubscript{2} pressure differences between the ocean and the atmosphere, respectively. 
The term $X_{\textup{other}}^0(\cdot\,, \cdot)$ contains the sum of all other component fluxes, including human activities like the burning of fossil fuels, biofuels, and other carbon-based materials, as well as biomass burning due to wildfires.
The zero superscript signifies that the ``other'' fluxes, which have relatively small uncertainties \citep{BasuEtAl2013}, are fixed to a bottom-up estimate (see Section~\ref{supp:bottom-up} in the Supplementary Material).
We thereby focus inference on the primary natural fluxes.

The first three terms in \eqref{eq:flux-components} are unknown and subject to the following constraints: $X_{\textup{gpp}}(\cdot\,, \cdot)$ is always negative or zero, $X_{\textup{resp}}(\cdot\,, \cdot)$ is always positive or zero, and $X_{\textup{ocean}}(\cdot\,, \cdot)$ can be positive or negative.
Further, GPP and respiration fluxes are zero at ocean locations, and ocean fluxes are zero at land locations.
In practice, we discretize the flux process into grid cells (see Section~\ref{supp:transport-model} in the Supplementary Material), some of which contain both land and ocean.
For these cells, any component flux is allowed to be nonzero.

Because GPP and respiration are closely related biological processes, they exhibit similar spatial and temporal patterns.
This makes them difficult to disentangle using only CO\textsubscript{2} mole-fraction observations, which are representative of the net flux.
For this reason, most inversion frameworks estimate NEE, which we define as
\begin{equation}
  \label{eq:nee-definition}
  X_{\textup{nee}}(\svec, t) = X_{\textup{gpp}}(\svec, t) + X_{\textup{resp}}(\svec, t), \quad \svec \in \mathbb{S}^2,\ t \in \mathcal{T}.
\end{equation}
In principle, WOMBAT~v2.0 can separately attribute GPP and respiration using strong prior information and by incorporating numerical constraints on the sign of each component.
In practice, the attribution problem is ill-posed and further simplifying assumptions are required (see Section~\ref{sec:parameter-model}).

WOMBAT~v2.0 represents component fluxes using a spatially varying time-series decomposition.
For each component $c \in \mathcal{C}$, the decomposition is given by
\begin{equation}
\label{eq:time-decomposition}
  \begin{aligned}
    X_c(\svec, t) = \beta_{c,0}(\svec) + \beta_{c,1}(\svec) t
    + \sum_{k=1}^{K_c} &(\beta_{c,2,k}(\svec) + \beta_{c,3,k}(\svec)t) \cos(2\pi k t / 365.25) \\
    &+ \sum_{k=1}^{K_c} (\beta_{c,4,k}(\svec) + \beta_{c,5,k}(\svec)t) \sin(2\pi k t / 365.25)
    + \epsilon_c(\svec, t),
  \end{aligned}
\end{equation}
for $\svec \in \mathbb{S}^2$ and  $t \in \mathcal{T}$ (with $t$ measured in days), where $\beta_{c,\cdot}(\svec)$, $\beta_{c,\cdot, \cdot}(\svec)$, and $\epsilon_{c}(\svec, t)$ are unknown.
This decomposition comprises three broad terms: (1) a \emph{linear term} represented by the intercept and the temporal trend; (2) a \emph{seasonal term} represented by the $2K_c$ harmonics of the solar cycle, with a period of 365.25 days; and (3) a \emph{residual term}, $\epsilon_{c}(\cdot\,, \cdot)$.
The linear term captures any long-term change in the flux \citep[e.g., increasing GPP due to the CO\textsubscript{2} fertilization effect;][]{ZhuEtAl2016}, while the seasonal term captures the intra-annual cycle.
The residual term captures any remaining variability, including both climate-driven anomalies \citep[e.g., fluxes driven by El~Ni\~{n}o;][]{LiuEtAl2017} and high-frequency cycles.
For instance, some component fluxes exhibit a diurnal cycle with a 24-hour period that cannot be represented by annual harmonics and thus appears in the residual term.

The spatial coefficients $\beta_{c,\cdot}(\svec)$ and $\beta_{c,\cdot, \cdot}(\svec)$, and spatiotemporal residual term $\epsilon_{c}(\svec, t)$ are high-dimensional and can be highly variable.
To incorporate prior information, WOMBAT~v2.0 centers the distribution for the coefficients and the residual terms on estimates $\beta_{c,\cdot}^0(\svec)$, $\beta_{c,\cdot, \cdot}^0(\svec)$, and $\epsilon_c^0(\svec, t)$, which derive from bottom-up flux estimates as described in Section~\ref{supp:bottom-up} in the Supplementary Material.
To keep computations tractable, space and time are partitioned into $R$ regions and $Q$ time periods, respectively.
The bottom-up estimates and the space-time partitions are used together to define a basis-function model for component flux $X_c(\cdot\,, \cdot)$ as
\begin{equation}
\label{eq:basis-decomposition}
  X_c(\svec, t)
  = \phivec_c(\svec, t)'(\onevec + \alphavec_c)
  \equiv X_c^0(\svec, t) + \phivec_c(\svec, t)'\alphavec_c,
  \quad c \in \mathcal{C},\ \svec \in \mathbb{S}^2,\ t \in \mathcal{T},
\end{equation}
where $\phivec_c(\cdot\,, \cdot)$ is a vector of basis functions of dimension $(2R + 4K_cR + QR)$ that we obtain from bottom-up estimates.
Specifically, each element of $\phivec_c(\svec, t)$ corresponds to the bottom-up estimate for one term of \eqref{eq:time-decomposition}, for example $\beta_{c,0}^0(\svec)$, $\beta_{c,1}^0(\svec)t$, $\beta_{c,2,1}^0(\svec)\cos(2\pi t / 365.25)$, or $\epsilon_c(\svec, t)$.
Each element of $\phivec_c(\svec, t)$ is evaluated over a spatial or spatiotemporal partition, and is equal to zero outside that partition.
We define $X_c^0(\cdot\,, \cdot) \equiv \phivec_c(\cdot\,, \cdot)'\onevec$ as the sum of bottom-up estimates of a component's linear, seasonal, and residual terms (i.e., the bottom-up estimate of the component flux).
The random and unknown vector $\alphavec_c \equiv (\alphavec_{c,0}', \alphavec_{c,1}', \alphavec_{c,2,1}', \ldots, \alphavec_{c,5,K_c}', \alphavec_{c,6}')'$ has the same dimension as $\phivec_c(\cdot\,, \cdot)$.
Consequently, $\alphavec_c$ is a vector of basis-function coefficients that spatially adjusts the known bottom-up estimates of the linear and seasonal coefficient fields, $\beta_{c,\cdot}^0(\cdot)$ and $\beta_{c,\cdot,\cdot}^0(\cdot)$, respectively, and spatiotemporally adjusts the bottom-up estimate of the residual term, $\epsilon_c^0(\cdot\,, \cdot)$.
We discuss the model for the $\alphavec_c$-vectors in Section~\ref{sec:parameter-model} and give full mathematical details of the basis-function decomposition in Section~\ref{supp:basis-function-model} in the Supplementary Material.
See also \citet{BertolacciEtAl2024} for a detailed discussion of the choices underlying \eqref{eq:time-decomposition} and \eqref{eq:basis-decomposition}.

Populating the sum in \eqref{eq:flux-components} with the basis-function model for each component flux from \eqref{eq:basis-decomposition} yields a process model for the net flux:
\begin{equation}
\label{eq:net-flux}
  X(\svec, t) = X^0(\svec, t) + \phivec(\svec, t)'\alphavec, \quad \svec \in \mathbb{S}^2,\ t \in \mathcal{T},
\end{equation}
where $X^0(\cdot\,, \cdot) \equiv X_{\textup{other}}^0(\cdot\,, \cdot) + \sum_{c \in \mathcal{C}} X^0_{c}(\cdot\,, \cdot)$, which is the combined flux from all bottom-up estimates;
$\phivec(\cdot\,, \cdot) \equiv (\phivec_{\textup{gpp}}(\cdot\,, \cdot)', \phivec_{\textup{resp}}( \cdot\,, \cdot)', \phivec_{\textup{ocean}}( \cdot\,, \cdot)')'$;
and $\alphavec \equiv (\alphavec_{\textup{gpp}}', \alphavec_{\textup{resp}}', \alphavec_{\textup{ocean}}')'$.
The flux process model in \eqref{eq:net-flux} is flexible and scientifically interpretable: it contains the individual component fluxes, and it is tailored to the study of the natural cycles of each component or of aggregated components like NEE \citep{BertolacciEtAl2024}.
While the flux process is of primary interest in CO\textsubscript{2} flux inversion, inference on flux is typically made using observations relating to the CO\textsubscript{2} mole-fraction process.

\subsection{Mole-fraction process model}
\label{sec:mole-frac-process-model}

At a given location $\svec \in \mathbb{S}^2$, geopotential height $h \geq 0$, and time $t \in \mathcal{T}$, the WOMBAT~v2.0 mole-fraction process model links surface flux, $X(\svec, t)$, to atmospheric mole fraction, $Y_{\textup{co}2}(\svec, h, t)$, through a transport model, $\hat{\mathcal{H}}_{\textup{co}2}$, that approximates the movement of CO\textsubscript{2} throughout the atmosphere.
Details of the mole-fraction process model are supplied in Section~\ref{supp:mole-frac-process-model} in the Supplementary Material.
The transport operator, $\hat{\mathcal{H}}_{\textup{co}2}$, describes the relationship between CO\textsubscript{2} mole fraction in the atmosphere and \emph{net} CO\textsubscript{2} fluxes at Earth's surface.
In the next section, we show how linking GPP to SIF helps to isolate both GPP and respiration components of the net flux.

\subsection{SIF process model}
\label{sec:sif-process-model}

We now extend WOMBAT~v2.0 to v2.S through the addition of a SIF process model.
During photosynthesis, a fraction of the light absorbed by chlorophyll is re-emitted as fluorescence (i.e., SIF) at longer wavelengths \citep{Baker2008}.
While SIF is a physical process that occurs during light reactions within individual leaves, it is the aggregate signal from an entire canopy that is observed by satellites.
We denote the latent, canopy-scale SIF process at location $\svec \in \mathbb{S}^2$ and time $t \in \mathcal{T}$ by $Y_{\textup{sif}}(\svec, t)$.
Since SIF is a by-product of photosynthesis, there is a physical relationship between the SIF process, $Y_{\textup{sif}}(\cdot\,, \cdot)$, and the GPP component flux process, $X_{\textup{gpp}}(\cdot\,, \cdot)$.
The relationship is local and instantaneous, and hence the SIF process at a given location and time depends on the GPP process at the same location and time.
We express this through a model for SIF that conditions on GPP as follows:
\begin{equation}
  \label{eq:sif-gpp-relationship}
  Y_{\textup{sif}}(\svec, t) = \mathcal{H}_{\textup{sif}}(X_{\textup{gpp}}(\svec, t);\; \svec, t) + v_{\textup{sif}}(\svec, t), \quad \svec \in \mathbb{S}^2,\ t \in \mathcal{T},
\end{equation}
where the function $\mathcal{H}_{\textup{sif}}$ represents the biochemical and physiological processes that yield SIF during GPP \citep[e.g.,][]{HanEtAl2022, SunEtAl2023a}.
The term $v_{\textup{sif}}(\cdot\,, \cdot)$ is a spatiotemporal random process that captures all aspects of canopy SIF emission that do not depend on GPP.
These include stochastic factors such as the cumulative probability that emitted fluorescence escapes the canopy without reabsorption, as well as responses to stressors like temperature and water availability that can impact SIF and GPP differently \citep{Porcar-CastellEtAl2014}.

The relationship between GPP and SIF, $\mathcal{H}_{\textup{sif}}$, varies across biomes and seasons due to differences in plant physiology and canopy structure \citep{JoinerEtAl2014, DammEtAl2015, ZhangEtAl2018a, PierratEtAl2022a}, but for a given location and season it is close to linear at the coarse spatial and temporal scales relevant to flux inversion \citep{SunEtAl2017, ByrneEtAl2018, LiEtAl2018, MagneyEtAl2020}.
To adopt a linear relationship, we approximate $\mathcal{H}_{\textup{sif}}$ using a spatiotemporally varying first-order Taylor-series expansion around a bottom-up estimate of GPP, $X^0_{\textup{gpp}}(\svec, t)$.
That is,
\begin{equation}
\label{eq:taylor-expansion}
  Y_{\textup{sif}}(\svec, t) = 
  \mathcal{H}_{\textup{sif}}\left(X_{\textup{gpp}}^0(\svec, t);\; \svec, t\right) + 
  \mathcal{H}^{(1)}_{\textup{sif}}\left(X_{\textup{gpp}}^0(\svec, t);\; \svec, t\right)\left(X_{\textup{gpp}}(\svec, t) - X_{\textup{gpp}}^0(\svec, t)\right) + 
  v_{\textup{sif}}^{*}(\svec, t),
\end{equation}
for $\svec \in \mathbb{S}^2$ and $t \in \mathcal{T}$, where $\mathcal{H}^{(1)}_{\textup{sif}}\left(X_{\textup{gpp}}^0(\svec, t);\; \svec, t\right)$ is the derivative of $\mathcal{H}_{\textup{sif}}$ with respect to $X_{\textup{gpp}}(\svec, t)$, evaluated at $X_{\textup{gpp}}^0(\svec, t)$, and $v_{\textup{sif}}^{*}(\cdot\,, \cdot)$ contains $v_{\textup{sif}}(\cdot\,, \cdot)$ plus any error introduced by the truncation of higher-order (nonlinear) terms in the Taylor-series expansion.
In what follows, we simplify notation and write the Taylor-series coefficients as $\eta_0(\svec, t) \equiv \mathcal{H}_{\textup{sif}}\left(X_{\textup{gpp}}^0(\svec, t);\; \svec, t\right)$ and $\eta_1(\svec, t) \equiv \mathcal{H}^{(1)}_{\textup{sif}}\left(X_{\textup{gpp}}^0(\svec, t);\; \svec, t\right)$.

Bottom-up estimates of GPP, $X_{\textup{gpp}}^0(\cdot\,, \cdot)$, and SIF, $Y_{\textup{sif}}^0(\cdot\,, \cdot)$, are both provided by the SiB4 terrestrial biosphere model \citep[see][and Section~\ref{supp:bottom-up} in the Supplementary Material]{HaynesEtAl2019, HaynesEtAl2019a}.
Let $\hat{\mathcal{H}}_{\textup{sif}}$ denote the mechanistic SiB4 relationship between GPP and SIF, such that $Y_{\textup{sif}}^0(\svec, t) = \hat{\mathcal{H}}_{\textup{sif}}\left(X_{\textup{gpp}}^0(\svec, t);\; \svec, t\right) + v_{\textup{sif}}^0(\svec, t)$, where the residual $v_{\textup{sif}}^0(\cdot\,, \cdot)$ is a SiB4-based estimate of $v_{\textup{sif}}(\cdot\,, \cdot)$ in \eqref{eq:sif-gpp-relationship}.
Using values simulated from this relationship, we obtain $\hat{\eta}_0(\svec, t) = Y_{\textup{sif}}^0(\svec, t) - v_{\textup{sif}}^0(\svec, t)$ as an approximation of $\eta_0(\cdot\,, \cdot)$ and derive $\hat{\eta}_1(\cdot\,, \cdot)$ as the slope in a spatiotemporally varying regression; see Section~\ref{supp:sif-gpp-sensitivity} in the Supplementary Material for details.
Substituting these expressions into \eqref{eq:taylor-expansion} gives
\begin{equation}
  \label{eq:linear-model}
  \begin{aligned}
    Y_{\textup{sif}}(\svec, t)
    &= \hat{\eta}_0(\svec, t) +
    \hat{\eta}_1(\svec, t)\left(X_{\textup{gpp}}(\svec, t) - X_{\textup{gpp}}^0(\svec, t)\right) +
    v_{\textup{sif}}^{**}(\svec, t) \\
    &= Y_{\textup{sif}}^0(\svec, t) +
    \hat{\eta}_1(\svec, t)\left(X_{\textup{gpp}}(\svec, t) - X_{\textup{gpp}}^0(\svec, t)\right) +
    v_{\textup{sif}}^{**}(\svec, t) - v_{\textup{sif}}^0(\svec, t),
  \end{aligned}
\end{equation}
for $\svec \in \mathbb{S}^2$ and $t \in \mathcal{T}$, where $v_{\textup{sif}}^{**}(\cdot\,, \cdot)$ equals $v_{\textup{sif}}^{*}(\cdot\,, \cdot)$ plus any error incurred in approximation of the Taylor-series coefficients.

Now, using \eqref{eq:basis-decomposition} we can write $X_{\textup{gpp}}(\svec, t) - X_{\textup{gpp}}^0(\svec, t) = \phivec_{\textup{gpp}}(\svec, t)'\alphavec_{\textup{gpp}}$, where $\phivec_{\textup{gpp}}(\svec, t)$ is a vector of basis functions for GPP and $\alphavec_{\textup{gpp}}$ is a corresponding vector of unknown coefficients.
Substituting this relationship into \eqref{eq:linear-model}, we obtain a basis-function representation of the SIF process:
\begin{equation}
\label{eq:sif-process-model}
  Y_{\textup{sif}}(\svec, t) =
  Y_{\textup{sif}}^0(\svec, t) + \hat{\psivec}_{\textup{sif}}(\svec, t)'\alphavec_{\textup{gpp}} + \xi_{Y_\textup{sif}}(\svec, t), \quad \svec \in \mathbb{S}^2,\ t \in \mathcal{T},
\end{equation}
where $\hat{\psivec}_{\textup{sif}}(\svec, t) \equiv \hat{\eta}_1(\svec, t)\phivec_{\textup{gpp}}(\svec, t)$, which is called a sensitivity vector because $\hat{\eta}_1(\cdot\,, \cdot)$ represents the sensitivity of SIF to GPP.
The term $\xi_{Y_\textup{sif}}(\svec, t) = v_{\textup{sif}}^{**}(\svec, t) - v_{\textup{sif}}^0(\svec, t)$ is a mean-zero spatiotemporal process that accommodates potentially correlated error in the SIF--GPP relationship.
Through \eqref{eq:sif-process-model}, we link the unknown random vector $\alphavec_{\textup{gpp}}$ to the SIF process.
Next, we connect the SIF process to SIF observations.

\subsection{Grouped data model}
\label{sec:data-model}

In this section, we introduce the data model for SIF observations and combine it with the WOMBAT~v2.0 data model for CO\textsubscript{2} mole-fraction observations.
Let $Z_{\textup{sif}, i}$ denote the $i$th SIF observation, where $i = 1, \ldots, N_{\textup{sif}}$, and $N_{\textup{sif}}$ is the total number of SIF observations available.
A satellite SIF observation is made at a specific location $\svec_i$ and time $t_i$, and can be affected by both bias and correlated measurement error.
Therefore, we use the following SIF data model:
\begin{equation}
\label{eq:sif-data-model}
  Z_{\textup{sif}, i} = Y_{\textup{sif}}(\svec_i, t_i) + b_{\textup{sif}, i} + \xi_{Z_{\textup{sif}, i}} + \epsilon_{\textup{sif}, i}, \quad i = 1, \ldots, N_{\textup{sif}},
\end{equation}
where $b_{\textup{sif}, i}$ is a bias term, $\xi_{Z_{\textup{sif}, i}}$ is a spatiotemporally correlated random error term, and $\epsilon_{\textup{sif}, i}$ is a mean-zero uncorrelated random error term independent of $\xi_{Z_{\textup{sif}, i}}$.
The bias term, $b_{\textup{sif}, i}$, accounts for any systematic error in the retrieval process; the correlated error term, $\xi_{Z_{\textup{sif}, i}}$, accounts for physical processes like aerosols and clouds that can affect the atmospheric column; and the uncorrelated error term, $\epsilon_{\textup{sif}, i}$, captures any remaining variability in the observations.
Combining \eqref{eq:sif-process-model} with \eqref{eq:sif-data-model} yields a relationship between the SIF observations and $\alphavec_{\textup{gpp}}$, the GPP basis-function coefficients:
\begin{equation}
\label{eq:sif-basis-data-model}
  Z_{\textup{sif}, i} = Z_{\textup{sif}, i}^0 + \hat{\psivec}_{\textup{sif}, i}'\alphavec_{\textup{gpp}} + b_{\textup{sif}, i} + \xi_{\textup{sif}, i} + \epsilon_{\textup{sif}, i}, \quad i = 1, \ldots, N_{\textup{sif}},
\end{equation}
where $Z_{\textup{sif}, i}^0 \equiv Y_{\textup{sif}}^0(\svec_i, t_i)$; $\hat{\psivec}_{\textup{sif}, i} \equiv \hat{\psivec}_{\textup{sif}}(\svec_i, t_i)$; and $\xi_{\textup{sif}, i} \equiv \xi_{Y_\textup{sif}}(\svec_i, t_i) + \xi_{Z_{\textup{sif}, i}}$, which combines all possible sources of correlated random error into a general error term whose constituents cannot be separately inferred.

Mole-fraction observations of CO\textsubscript{2} derive from both column-averaged satellite retrievals and point-referenced in~situ and flask measurements.
WOMBAT further splits these two observation types into a set of groups, denoted by $\mathcal{G}_{\textup{co}2}$, according to bias and error properties \citep{Zammit-MangionEtAl2022a}.
For instance, point-referenced in~situ observations are grouped by instrument.
As explained in Section~\ref{supp:mole-frac-data-model} in the Supplementary Material, each observation group is accommodated using a mole-fraction data model that has a similar structure to the SIF data model in \eqref{eq:sif-basis-data-model}.
In WOMBAT~v2.S, we treat SIF observations as a new group and combine the SIF and mole-fraction data models into a unified data model, where the combined set of groups is given by $\mathcal{G} = \mathcal{G}_{\textup{co}2} \cup \{\textup{sif}\}$.
The resulting data model allows for different bias and error terms across the different groups, but it maintains a consistent structure of the form
\begin{equation}
\label{eq:overall-data-vector}
  \Zvec_g = \Zvec^0_g + \hat{\Psivec}_g'\alphavec + \bvec_g + \xivec_g + \epsilonvec_g, \quad g \in \mathcal{G},
\end{equation}
where the vector $\Zvec_g \equiv (Z_{g,1}, \ldots, Z_{g,N_g})'$ contains the data in group $g$; $\hat{\Psivec}_g$ is a matrix with $N_g$ rows in which the $i$th row is the sensitivity vector $\hat{\psivec}_{g, i}'$; $\bvec_g \equiv (b_{g,1}, \ldots, b_{g,N_g})'$ is a vector of bias terms; $\xivec_g \equiv (\xi_{g,1}, \ldots, \xi_{g,N_g})'$ is a vector of correlated random error terms; and $\epsilonvec_g \equiv (\epsilon_{g,1}, \ldots, \epsilon_{g,N_g})'$ is a vector of uncorrelated random error terms that are independent of $\xivec_g$.
Details of how the data model is tailored to each observation group are given in Section~\ref{supp:error-properties} in the Supplementary Material.

\subsection{Parameter model}
\label{sec:parameter-model}

Distributional properties of the random and unknown terms that occur at different levels of the hierarchy are captured by the parameter model.
The multivariate distribution of the unknown basis-function coefficients, $\alphavec$, determines the fluxes and drives our analysis.
To enable correlation between component fluxes and to constrain the estimated fluxes to satisfy physical properties, we model the basis-function coefficients as $\alphavec \sim \textup{ConstrGau}(\zerovec, \Sigmavec_{\alpha}, F_{\alpha})$, where $\textup{ConstrGau}(\muvec, \Sigmavec, F)$ is a multivariate Gaussian distribution with renormalized density on the set $F$ and zero outside $F$.
The constraint set $F_{\alpha}$ is a linear inequality of the form $\dvec + \Phivec \alphavec \geq \zerovec$ (applied elementwise), which ensures that physical properties, such as GPP nonpositivity and respiration nonnegativity, are respected in our inferences.
Due to the constraint set $F_{\alpha}$, $\alphavec = \zerovec$ will typically be the mode (not the mean) of the multivariate distribution over $\alphavec$.
This is because $\alphavec = \zerovec$ in \eqref{eq:basis-decomposition} corresponds to $X_c(\cdot\,, \cdot) = X_c^0(\cdot\,, \cdot)$ for $c \in \mathcal{C}$, which will generally satisfy the physical constraints encoded in $F_{\alpha}$.
The covariance matrix $\Sigmavec_{\alpha}$ is governed by several unknown parameters, denoted by $\thetavec_{\alpha}$, which are all given hyperprior distributions.
The structure of $\Sigmavec_{\alpha}$ induces correlations between basis-function coefficients, both across different component fluxes and across time periods within and between components.
Section~\ref{supp:alpha-covariance} in the Supplementary Material gives the specification of $\Sigmavec_{\alpha}$.

When attributing flux to GPP and respiration, the linear term of the decomposition in \eqref{eq:time-decomposition} is weakly identifiable.
This is because the basis functions for the GPP and respiration linear terms are very similar in their spatial and temporal structure.
In WOMBAT~v2.0, the identifiability issue was addressed by fixing the respiration linear term (RLT) to be known and equal to its bottom-up estimate.
This amounts to fixing the corresponding elements of $\alphavec_{\textup{resp}}$ to zero for all regions, which also results in a strong constraint on the GPP linear term.
In WOMBAT~v2.S, we can relax this fixed RLT assumption since, from \eqref{eq:nee-definition}, $X_{\textup{resp}}(\cdot\,, \cdot) = X_{\textup{nee}}(\cdot\,, \cdot) - X_{\textup{gpp}}(\cdot\,, \cdot)$, which is now informed by the combination of SIF and CO\textsubscript{2} mole-fraction data.

With this RLT relaxation, we found that the prior means of the linear terms for GPP and respiration were distant from their bottom-up estimates.
This occurs because the sign constraints on $\alphavec$ (induced through $F_{\alpha}$) are more readily satisfied by larger flux magnitudes.
To remedy this, we update the WOMBAT~v2.0 parameterization of $\alphavec$ in two ways.
First, we enforce the constraints to hold at a coarser spatial resolution than in WOMBAT~v2.0, which helps draw the prior mean of $\alphavec$ closer to zero.
Second, we calibrate $\Sigmavec_{\alpha}$ through a reparameterization of the elements of $\alphavec$ that correspond to the GPP and respiration linear terms.
These modifications are described in Sections~\ref{supp:sign-constraints} and \ref{supp:reparameterization} in the Supplementary Material, respectively.

The remaining unknowns are the bias and error terms, $\bvec_g$, $\xivec_g$, and $\epsilonvec_g$, for each observation group $g \in \mathcal{G}$.
Observations used in flux inversions are commonly accompanied by error-variance estimates $\{V_{g,i} : i = 1, \ldots, N_g\}$.
Each $V_{g,i}$ represents an estimate of the total variance of the errors $\xi_{g,i} + \epsilon_{g,i}$, so we call $V_{g,i}$ an ``error budget.''
Since error budgets are approximations of varying quality, we adjust them as $\var(\xi_{g,i} + \epsilon_{g,i}) = (\gamma^Z_g)^{-1}V_{g,i}$, where $\gamma^Z_g > 0$ is an unknown parameter that scales the error budgets in group $g$.
We then partition the scaled error-variance of each observation into temporally correlated and uncorrelated components using a group-specific, unknown parameter $\rho^Z_g \in [0, 1]$; that is, $\var(\xi_{g,i}) = \rho^Z_g(\gamma^Z_g)^{-1}V_{g,i}$ and $\var(\epsilon_{g,i}) = (1 - \rho^Z_g)(\gamma^Z_g)^{-1}V_{g,i}$.
Temporal correlation within group $g$ is modeled using an exponential covariance function with $e$-folding length given by the unknown parameter $\ell^Z_g \geq 0$.
For the group-specific bias term, we construct a linear model with known geophysical covariates (e.g., aerosol content) and unknown coefficients $\pivec_g$.
All remaining details about these unknown parameters, including their prior distributions, are given in Section~\ref{supp:parameter-model} in the Supplementary Material.

\subsection{Inference}
\label{sec:inference}

\begin{figure}[t!]
  \centering
  \begin{tikzpicture}[scale=0.9,line width=0.8pt,every node/.style={transform shape}]
    \input{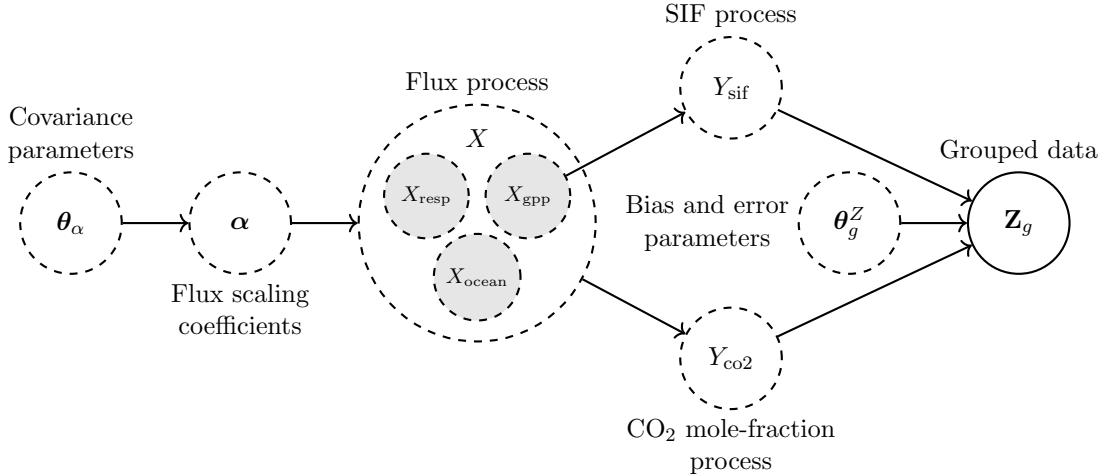}
  \end{tikzpicture}
  \caption{
    Graphical model summarizing the relationships between the parameters, processes, and grouped data $\{\Zvec_g : g \in \mathcal{G}\}$ in WOMBAT~v2.S.
    Dashed lines indicate unknown quantities that are inferred from the data.
    Shaded nodes are the primary subject of interest in the inversion.
  }
  \label{fig:graphical-model}
\end{figure}

Figure~\ref{fig:graphical-model} shows a graphical model summarizing the hierarchical structure of WOMBAT~v2.S.
The unknowns to be inferred include: the random vector of basis-function coefficients $\alphavec$ and the parameters that specify its distribution, notated $\thetavec_{\alpha}$;
the bias coefficients $\pivec \equiv (\pivec_1', \ldots, \pivec_G')'$;
and the parameters that model the error properties, namely $\gammavec^Z \equiv (\gamma^Z_1, \ldots, \gamma^Z_G)'$, $\rhovec^Z \equiv (\rho^Z_1, \ldots, \rho^Z_G)'$, and $\ellvec^Z \equiv (\ell^Z_1, \ldots, \ell^Z_G)'$, where $G$ is the number of groups in $\mathcal{G}$.
To simplify exposition, in Figure~\ref{fig:graphical-model} we denote each set of bias and error parameters as $\thetavec^Z_g \equiv (\gamma^Z_g, \rho^Z_g, \ell^Z_g, \pivec_g')'$ for observation group $g \in \mathcal{G}$.

Since the bias and error parameters are general, the SIF observation group does not require special consideration in the inference procedure.
As in WOMBAT~v2.0, inference proceeds in two stages.
The first stage is a preliminary inversion from which we obtain estimates of $\rhovec^Z$ and $\ellvec^Z$.
The second stage is a full inversion where $\rhovec^Z$ and $\ellvec^Z$ are fixed to their first-stage estimates $\hat{\rhovec}^Z$ and $\hat{\ellvec}^Z$, respectively, and the remaining parameters are inferred using a Markov chain Monte Carlo (MCMC) algorithm.
The Gibbs scheme uses standard slice sampling \citep{Neal2003}, except when sampling the basis-function coefficients $\alphavec$, which are constrained to the set $F_{\alpha}$.
To draw samples that satisfy $F_{\alpha}$ in this step, we use the exact Hamiltonian Monte Carlo (HMC) scheme proposed by \citet{PakmanPaninski2014} for constrained multivariate Gaussian distributions.
The full MCMC algorithm can be found in Section~S.1.4 in the Supplementary Material of \citet{BertolacciEtAl2024}.

By substituting posterior samples of $\alphavec$ into \eqref{eq:basis-decomposition}, we obtain posterior samples of the component fluxes, $\{X_c(\cdot\,, \cdot)\}$.
In the next section, we discuss how these posterior samples are impacted by the inclusion of SIF observations in the inversion.

\section{\texorpdfstring{Flux inversion with SIF and CO\textsubscript{2} data}{Flux inversion with SIF and CO2 data}}
\label{sec:application}

We apply the hierarchical model of Section~\ref{sec:model} to make top-down, global inference on GPP, respiration, and other natural components of CO\textsubscript{2} surface flux over a six-year period.
Following WOMBAT~v2.0 \citep{BertolacciEtAl2024}, we align our inversion with the OCO-2 model intercomparison project (MIP), an organized effort to evaluate flux-inversion frameworks under a common protocol.
The latest round is called the v10 MIP \citep{ByrneEtAl2023}, since it prescribes version 10 OCO-2 mole-fraction data for use in inversions; see \citet{CrowellEtAl2019} and \citet{PeiroEtAl2022} for past rounds.
The v10 MIP protocol also prescribes bottom-up fossil-fuel flux estimates that are assumed fixed and known, and a period from January 2015 to December 2020 for reporting inferred fluxes.
Here, we augment the v10 MIP protocol by including version 10 OCO-2 SIF data in our inversion.
Section~\ref{sec:data} describes all data sources used in WOMBAT~v2.S.
In Section~\ref{sec:osse}, we validate the model using an observing system simulation experiment (OSSE), which also assesses the utility of including SIF observations.
In Section~\ref{sec:results}, we present results from applying WOMBAT~v2.S to the real data, and we compare the inferred fluxes with those from WOMBAT~v2.0 and with estimates from the independent FLUXCOM initiative.

\subsection{Data}
\label{sec:data}

We use both SIF and CO\textsubscript{2} mole-fraction data in our inversion.
SIF observations are from NASA's OCO-2 satellite, and mole-fraction observations are from a variety of instruments, including OCO-2 retrievals, as well as in~situ and flask measurements.

\emph{SIF observations.}
The OCO-2 satellite, launched in July 2014, makes global retrievals of SIF and atmospheric CO\textsubscript{2} mole fraction \citep{CrispEtAl2017, ElderingEtAl2017a, SunEtAl2018}.
The onboard instrument measures reflected sunlight in three spectral bands, one for oxygen absorption (\qtyrange{757}{773}{nm}) and two for CO\textsubscript{2} absorption (\qtyrange{1.59}{1.62}{\micro\meter} and \qtyrange{2.04}{2.08}{\micro\meter}).
The SIF emission spectrum (\qtyrange{660}{850}{nm}) has peaks at \qtylist{680;740}{nm}, the latter of which is near the OCO-2 oxygen band.
Using spectral absorption lines within the oxygen band, the OCO-2 SIF retrieval algorithm separates the fluorescence signal from reflected sunlight in narrow bands centered at \qtylist{757;771}{nm} \citep{FrankenbergEtAl2011a}.
SIF retrievals at these two wavelengths are then combined in a normalizing weighted average to produce SIF at \qty{740}{nm}, allowing comparability with other satellite missions \citep{ParazooEtAl2019}.

The OCO-2 satellite follows a near-polar orbit, returning to the same location every 16 days.
High-latitude orbit tracks are spaced closer than those at mid-latitudes and, at the equator, adjacent orbit tracks are separated by about \qty{160}{km}.
Orbit tracks have a swath width of about \qty{10}{km} and, with eight measurements across track, each retrieval has a spatial footprint of a few square kilometers.
Accordingly, SIF retrievals from OCO-2 have the finest spatial footprint of all SIF satellite retrievals to date.
OCO-2 retrievals are made in three modes: nadir mode, with the instrument pointed at Earth's surface directly below the satellite; glint mode, with the instrument pointed at the reflection of the sun on the surface; and target mode, with the instrument fixed on a specific location, typically a validation site.

We obtain bias-corrected SIF retrievals from the version 10r SIF Lite files \citep{OCO-2ScienceTeamEtAl2020}, which implement the debiasing method described in \citet{DoughtyEtAl2022}.
As recommended in the data user's guide, we screen these retrievals for quality flags 0 and 1 and remove cases where the retrieved value plus three times the retrieval standard deviation is negative.
For consistency with the OCO-2 CO\textsubscript{2} observations prescribed by the MIP, we use SIF retrievals made in nadir and glint modes, and we average these by mode into 10-second bands along orbit tracks.
We compute 10-second averages of both the retrievals and their error variances using the approach of the original MIP \citep{CrowellEtAl2019}, and we retain averages based on at least five retrievals.
Hereafter, we refer to these averaged values as SIF observations and observation-error variances, respectively.

Each retained SIF observation is then matched by location and time to an estimate of SIF--GPP model-error variance, which is described in Section~\ref{supp:sif-gpp-sensitivity} in the Supplementary Material.
Similar to the convention used for CO\textsubscript{2} observations in the v10 MIP, we add the model-error variance to the observation-error variance to construct an ``error budget'' for each SIF observation.
We discard SIF observations when the assumption of a linear SIF--GPP relationship is not supported by bottom-up SiB4 estimates or when SIF observations are outliers relative to bottom-up SIF estimates (see Section~\ref{supp:sif-gpp-sensitivity} in the Supplementary Material).
As a result, we retain about \qty{10}{\percent} of the SIF observations available for assimilation.

\begin{figure}[t!]
  \centering
  \includegraphics{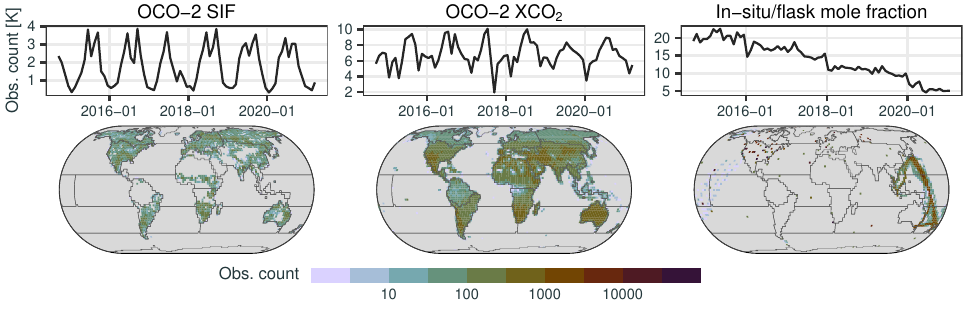}
  \caption{
    Processed and retained observations for OCO-2 SIF, OCO-2 XCO$_2$, and in-situ/flask mole-fraction observation types.
    (Top row) The total number of observations (in thousands) in each month of the study period, by observation type.
    (Bottom row) The total number of observations over the study period, counted in \qtyproduct{2 x 2.5}{\degree} grid cells by observation type.
    In the bottom panels, the color scale is log-transformed, gray cells have zero observations, and gray lines delineate the 23 regions used to partition the spatial domain.
  }
  \label{fig:observation-count}
\end{figure}

The final 138,590 processed SIF observations span the period from September 6, 2014 to March 31, 2021, the same period as the OCO-2 CO\textsubscript{2} observations described below.
Figure~\ref{fig:observation-count} shows the spatial and temporal observation density for SIF and both CO\textsubscript{2} observation types.
Retained SIF observations are generally less abundant during winter months when SIF is negligible and cannot inform GPP.
They are also sparse in desert and tropical regions (particularly in the Amazon), where the bottom-up SIF--GPP relationship tends to be absent or nonlinear, which prevents assimilation of SIF observations.
Hence, SIF observations offer limited information on GPP for some regions and months.

\emph{CO\textsubscript{2} mole-fraction observations.}
Our inversion uses CO\textsubscript{2} mole-fraction observations that derive from two sources: column-averaged CO\textsubscript{2} mole fraction, or XCO\textsubscript{2}, retrieved by the OCO-2 satellite, and in~situ and flask measurements collected from various instruments around the globe.
Both types of observations are prescribed by the v10 MIP protocol \citep{ByrneEtAl2023} and are the same as those used in WOMBAT~v2.0.
In total, there are 530,201 XCO\textsubscript{2} satellite observations, which span the period from September 6, 2014 to March 31, 2021, and 1,053,457 in~situ and flask observations, which span the period from September 1, 2014 to January 31, 2021.
As shown in Figure~\ref{fig:observation-count}, spatial coverage is much broader for XCO\textsubscript{2} observations than for in~situ and flask observations, but the latter are more abundant, particularly in the Northern Hemisphere.
Temporal coverage of XCO\textsubscript{2} observations is seasonal throughout the study period, while the number of in~situ and flask observations declines steadily due to longer processing times needed for these data sets.
Additional strengths and weaknesses of both observation types are discussed in Section~\ref{supp:mole-frac-observations} in the Supplementary Material.

The resulting collection of SIF and CO\textsubscript{2} mole-fraction data covers the period from September 2014 to March 2021, and while we estimate fluxes across this entire period, we report fluxes on the prescribed period from January 2015 to December 2020.
The data buffer on either side of the reporting period helps improve reliability of flux estimates at the start and end of the period.
In addition to observational data, the WOMBAT~v2.S framework depends on a transport model, bottom-up flux estimates, and prior distributions on the parameters in the process and data models.
These details are provided in Section~\ref{supp:application-details} in the Supplementary Material, along with a description of the computing resources used to run the inversion.
\nowidow

\subsection{Observing system simulation experiment (OSSE)}
\label{sec:osse}

To quantify the benefit of assimilating SIF observations in WOMBAT~v2.S, we conduct an OSSE in which the true flux is known and the same hierarchical model is used for both simulation and inversion.
We construct four true-flux cases according to the model in \eqref{eq:net-flux} by specifying the flux basis-function coefficients, $\alphavec$, in four different ways.
In the first case, we set $\alphavec = \zerovec$, which means the true flux is the bottom-up estimate.
We refer to this case as ``Bottom-up.''
The second case sets $\alphavec$ equal to the posterior mean obtained from the WOMBAT~v2.0 inversion in \citet{BertolacciEtAl2024}.
We label this case ``v2.0 mean.''
In the first two cases, the respiration linear term (RLT; see \eqref{eq:time-decomposition}) is equal to its bottom-up estimate.
The final two cases differ from this pattern.
In these cases, we maintain NEE at its ``v2.0 mean'' value while adjusting both the GPP intercept and trend by either positive \qty{10}{\percent} (``Positive shift'') or negative \qty{10}{\percent} (``Negative shift'').
To preserve the NEE value, we make compensating adjustments to the RLT, causing the RLT to deviate from its bottom-up estimate.
Full details are provided in Section~\ref{supp:alpha-setup} in the Supplementary Material.
Section~\ref{supp:alpha-setup} also describes how we simulate noisy observations of SIF and CO\textsubscript{2} mole fraction using the four true-flux cases and error-parameter estimates from the first stage of the real-data inversion (see Section~\ref{sec:inference}), thereby inducing the error properties of the real data.
Figure~\ref{fig:osse-true-flux} in the Supplementary Material plots the true GPP, respiration, NEE, and ocean component fluxes for the four cases, aggregated to monthly global totals.

We make inference on the true fluxes by running the second stage of the inversion under four setups, determined by the interaction of two factors: whether SIF observations are included or not, and whether the RLT is \emph{fixed} to its bottom-up estimate in all regions or \emph{inferred} for land regions in the inversion.
As an exception, the RLT is always fixed in the basis-function region for tropical South America because nonlinearity in the bottom-up SIF--GPP relationship prevents us from assimilating SIF observations across most of this region (see Figure~\ref{fig:observation-count} and Section~\ref{supp:sif-gpp-sensitivity} in the Supplementary Material for details).
Across the four true-flux cases and the four inversion setups, we consider a total of 16 experiments in the OSSE design.

We evaluate the ability of each inversion setup to recover the true component fluxes using the root-mean-squared error (RMSE) and the continuous ranked probability score \citep[CRPS;][]{GneitingRaftery2007} on a monthly and regionally aggregated basis.
The RMSE quantifies the accuracy of point estimates, while the CRPS also assesses the calibration of posterior prediction intervals.
These statistics are based on the same regions used to construct the flux basis functions, which are detailed in Section~\ref{supp:basis-function-setup} in the Supplementary Material.
To compare the 16 experiments, we compute the RMSE and mean CRPS over all combinations of these 23 regions and the 72 months in the evaluation period (January 2015 to December 2020).

Table~\ref{tab:osse-metrics-rmse} reports the RMSE for each component flux in all 16 OSSE experiments.
Reassuringly, estimation of NEE and ocean component fluxes is not affected by whether the RLT is fixed or inferred, nor whether SIF observations are included or not.
This is expected because NEE is well identified from terrestrial CO\textsubscript{2} mole-fraction observations, and ocean flux is not directly related to GPP, which is where SIF adds information.

\begin{table}[t!]
  \renewcommand{\arraystretch}{0.9}
  \renewrobustcmd{\bfseries}{\fontseries{b}\selectfont}
  \centering
  \begin{threeparttable}
    \begin{tabular}{lllCCCC}
\toprule
True Flux & \multicolumn{2}{c}{Inversion Setup} & \multicolumn{4}{c}{RMSE [PgC/year]} \\
\cmidrule(lr){2-3} \cmidrule(l{10pt}){4-7}
& RLT\textsuperscript{a} & Includes SIF & GPP & Resp. & NEE & Ocean \\
\midrule
Bottom-up\textsuperscript{b} & Fixed & Yes & \textbf{0.05} & \textbf{0.05} & 0.03 & 0.06 \\
 &  & No & 0.08 & 0.08 & 0.03 & 0.06 \\
 & Inferred & Yes & 0.17 & 0.16 & 0.03 & 0.06 \\
 &  & No & 1.62 & 1.62 & 0.03 & 0.06 \\
\midrule
v2.0 mean\textsuperscript{b} & Fixed & Yes & \textbf{0.28} & \textbf{0.29} & 0.11 & 0.06 \\
 &  & No & 0.41 & 0.41 & 0.12 & 0.06 \\
 & Inferred & Yes & 0.49 & 0.49 & 0.11 & 0.06 \\
 &  & No & 1.93 & 1.93 & 0.12 & 0.06 \\
\midrule
Positive shift & Fixed & Yes & 0.71 & 0.70 & 0.10 & 0.07 \\
 &  & No & 0.73 & 0.73 & 0.11 & 0.06 \\
 & Inferred & Yes & \textbf{0.45} & \textbf{0.44} & 0.10 & 0.07 \\
 &  & No & 2.36 & 2.37 & 0.11 & 0.06 \\
\midrule
Negative shift & Fixed & Yes & 0.69 & 0.71 & 0.11 & 0.06 \\
 &  & No & 0.71 & 0.72 & 0.12 & 0.06 \\
 & Inferred & Yes & \textbf{0.42} & \textbf{0.42} & 0.11 & 0.06 \\
 &  & No & 2.32 & 2.33 & 0.12 & 0.06 \\
\bottomrule
\end{tabular}

    \begin{tablenotes}
      \item [a] {\scriptsize The respiration linear term (RLT) is either fixed to its bottom-up estimate or inferred for land regions.}
      \item [b] {\scriptsize The bottom-up RLT estimate is the true RLT in these cases.}
    \end{tablenotes}
  \end{threeparttable}
  \caption{
    Root-mean-squared error (RMSE) when estimating monthly regional flux components in the OSSE of Section~\ref{sec:osse}.
    Smaller RMSE indicates better performance, and the smallest RMSE for each true-flux case is given in bold unless there is a tie.
    We obtain each value over the same regions and time periods as those used for constructing flux basis functions (see Section~\ref{supp:basis-function-setup} in the Supplementary Material).
  }
  \label{tab:osse-metrics-rmse}
\end{table}

\begin{figure}[t!]
  \centering
  \includegraphics{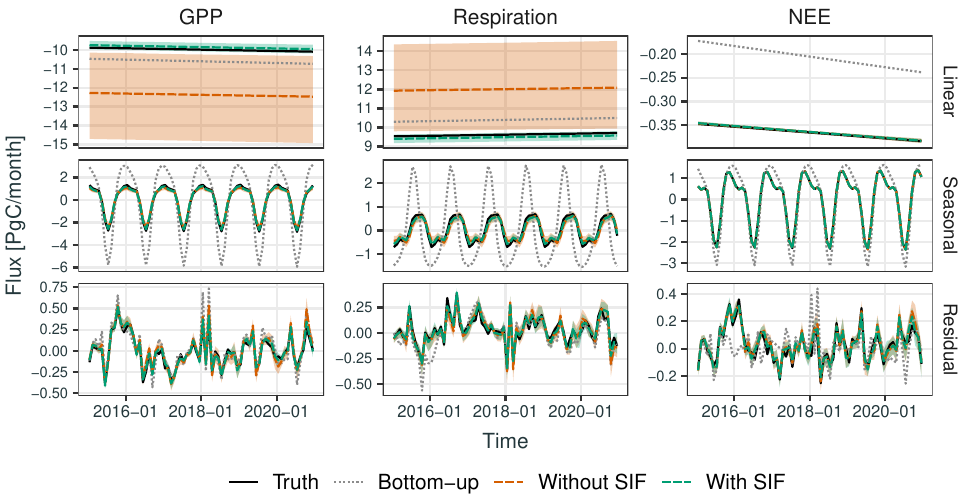}
  \caption{
    Decomposition of monthly global total OSSE posterior flux estimates into linear, seasonal, and residual terms when the true flux (solid line) is the ``Negative shift'' case.
    Posterior estimates with and without SIF are shown, and the respiration linear term (RLT) is inferred for land regions in both setups.
    Shaded areas show the region between the 2.5th and 97.5th posterior percentiles (i.e., \qty{95}{\percent} prediction intervals).
    Bottom-up estimates are included for reference.
  }
  \label{fig:osse-flux-decomposition}
\end{figure}

The inversion setup mainly affects estimation of GPP and respiration components.
Including SIF observations consistently improves RMSE, but treatment of the RLT plays a key role.
When the true-flux case is ``Bottom-up'' or ``v2.0 mean,'' setups with a fixed RLT achieve the smallest RMSE for GPP and respiration components, reflecting the advantage of the RLT being fixed to its true value in these cases.
However, in the ``Positive shift'' and ``Negative shift'' cases, the true RLT differs from its bottom-up estimate by \qty{10}{\percent}, so fixing the RLT to its bottom-up estimate becomes a disadvantage.
In these cases, the setup of inferred RLT with SIF outperforms the setup of fixed RLT with SIF by nearly \qty{40}{\percent} in RMSE.
The true RLT will typically differ from its bottom-up estimate in practice; therefore, inferring the RLT is beneficial, provided that SIF observations are used to anchor GPP.

The setup of inferred RLT without SIF performs poorly, with RMSE an order of magnitude larger than other setups across all true-flux cases.
This is primarily due to an inability to separately identify linear terms for GPP and respiration when SIF is not available.
Figure~\ref{fig:osse-flux-decomposition} illustrates this for the ``Negative shift'' case: although seasonal and residual terms are estimated similarly with or without SIF, the true linear terms for GPP and respiration are only recovered when SIF is included.
Without SIF, these terms fall just outside the much wider posterior prediction intervals, underscoring the challenge of separately estimating GPP and respiration linear terms.
As expected, the NEE linear term remains minimally affected, confirming that SIF observations primarily improve estimates of GPP and respiration.

The CRPS results are reported in Table~\ref{tab:osse-metrics-crps} in the Supplementary Material.
They are consistent with the RMSE results and with the posterior prediction intervals shown in Figure~\ref{fig:osse-flux-decomposition}.
In summary, the OSSE shows that inferring the RLT with the aid of SIF observations, as we do in WOMBAT~v2.S, produces more accurate and better-calibrated estimates of GPP and respiration, particularly regarding the linear terms in their time series decomposition.
We therefore expect WOMBAT~v2.S to provide better estimates of GPP and respiration fluxes than WOMBAT~v2.0 when applied to real data.

\subsection{Carbon cycle findings using satellite and in~situ observations}
\label{sec:results}

We now present results from applying the WOMBAT~v2.S framework in the context of the v10 MIP using the real observations of SIF and CO\textsubscript{2} mole fraction described in Section~\ref{sec:data}.
The OSSE in Section~\ref{sec:osse} showed that ocean flux estimates are not affected by the inclusion of SIF observations, so here we focus on GPP, respiration, and NEE fluxes.
We compare the posterior flux estimates from WOMBAT~v2.S with those from WOMBAT~v2.0, and with the bottom-up SiB4 estimates that form the GPP and respiration prior in both inversions.
For additional context, we include estimates from the FLUXCOM X-BASE product \citep{NelsonEtAl2024}, part of the FLUXCOM initiative \citep{TramontanaEtAl2016, JungEtAl2020}.

The FLUXCOM initiative uses machine learning to scale in~situ NEE flux measurements from 294 flux-tower sites to global flux products using satellite and meteorological data.
The FLUXCOM X-BASE product (hereafter, FLUXCOM) generates NEE estimates at \qty{0.05}{\degree}, hourly resolution, with GPP derived using the \citet{ReichsteinEtAl2005} partitioning method.
Respiration estimates are calculated by subtracting GPP from NEE.
For comparison with WOMBAT, we spatially aggregate FLUXCOM estimates to the \qtyproduct{2 x 2.5}{\degree} resolution of the transport model.
While products from the FLUXCOM initiative are widely used as a benchmark \citep[e.g.,][]{AnavEtAl2015}, we treat these estimates as a plausible representation of carbon fluxes rather than validation data.

In what follows, we compare flux estimates from WOMBAT~v2.S, v2.0, and FLUXCOM.
The bottom-up SiB4 estimates we use are also included for reference.
We proceed from coarse to fine spatial scales, starting with global totals and seasonal cycles, then moving to zonal characteristics, and finally to examining spatial patterns.

\begin{figure}[t!]
  \centering
  \includegraphics{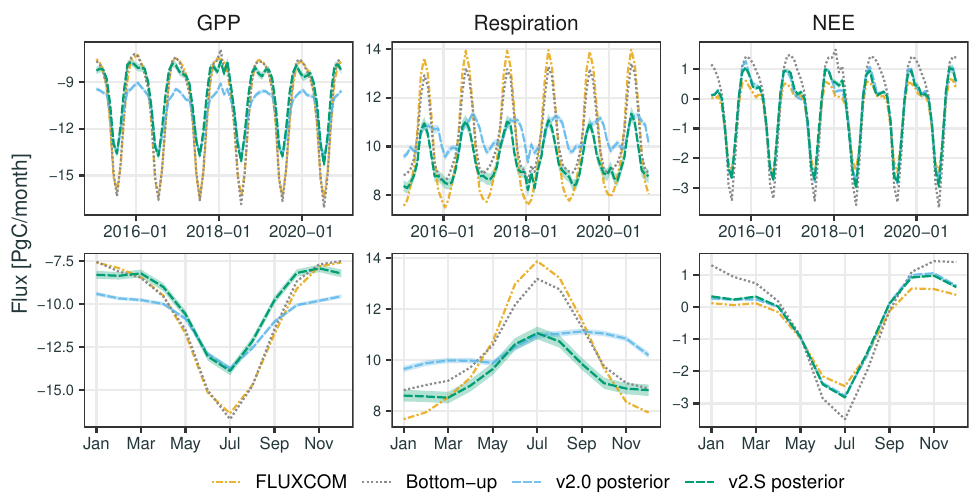}
  \caption{
    (Top row) Global monthly total GPP, respiration, and NEE component fluxes during the six-year period from January 2015 to December 2020.
    (Bottom row) Estimated seasonal profile of each component flux, where each month shows the average of the six corresponding estimates during the six-year period.
    Lines show the FLUXCOM and bottom-up (SiB4) estimates, and the WOMBAT~v2.0 and v2.S posterior means.
    Shaded areas around the WOMBAT estimates depict the region between the 2.5th and 97.5th posterior percentiles (i.e., \qty{95}{\percent} prediction intervals).
  }
  \label{fig:flux-global}
\end{figure}

\emph{Global totals and seasonal cycles.}
Averaged over the six years in the study period, WOMBAT~v2.S estimates a global annual GPP flux of \qty{-117.48 \pm 1.15}{PgC.yr^{-1}}, which is smaller than the v2.0 estimate of \qty{-129.15 \pm 0.06}{PgC.yr^{-1}}.
Both estimates are within the range of \qtyrange{110}{165}{PgC.yr^{-1}} typically seen in the literature \citep{AnavEtAl2015}.
The FLUXCOM and bottom-up estimates are both around \qty{-127}{PgC.yr^{-1}} in annual GPP flux.
Annual average NEE is between \qtyrange[range-phrase = { and }]{-5.5}{-4}{PgC.yr^{-1}} for all estimates considered, except for the bottom-up estimate, which is \qty{-2.5}{PgC.yr^{-1}}.

Figure~\ref{fig:flux-global} shows monthly global total GPP, respiration, and NEE component fluxes, as well as the average seasonal cycle for each component.
The largest GPP sink is consistently in July, driven by the boreal growing season (see supplementary Figure~\ref{fig:seasonal-cycle-zonal} for the average seasonal cycle of different latitude bands).
The primary difference between GPP estimates is the amplitude of the seasonal cycle, which is larger in WOMBAT~v2.S than in v2.0.
The FLUXCOM and bottom-up GPP estimates are nearly identical, with even larger seasonal amplitude than either of the WOMBAT estimates.
Seasonal differences are more pronounced in estimates of the respiration component.
Compared with WOMBAT~v2.0, the inclusion of SIF in v2.S creates an amplified respiration cycle that is more consistent with FLUXCOM and bottom-up estimates, showing the largest annual source occurring in July.
Regarding NEE, WOMBAT~v2.S estimates are essentially unchanged from v2.0 estimates, which is consistent with our findings in the OSSE.
Both are similar to FLUXCOM NEE estimates.
In contrast, bottom-up estimates show a larger seasonal amplitude for NEE, but recall that they do not incorporate observational data.

Although subtle, a key difference between WOMBAT estimates is the width of the posterior prediction intervals, which are wider for WOMBAT~v2.S than for v2.0.
The RLT is inferred in WOMBAT~v2.S, and the wider intervals reflect the additional uncertainty involved when separately estimating GPP and respiration linear terms.
This effect aligns with the OSSE results and is more evident in Figure~\ref{fig:flux-decomposition} in the Supplementary Material.

\emph{Zonal fluxes.}
To investigate temporal patterns in different regions, Figure~\ref{fig:flux-net-zonal} in the Supplementary Material shows monthly total GPP, respiration, and NEE fluxes for zonal bands covering the northern boreal (\ang{50}N--\ang{90}N), the northern temperate (\ang{23}N--\ang{50}N), the tropics (\ang{23}S--\ang{23}N), the northern tropics (\ang{0}--\ang{23}N), the southern tropics (\ang{23}S--\ang{0}), and the southern extratropics (\ang{90}S--\ang{23}S).
Similarly, Figure~\ref{fig:seasonal-cycle-zonal} shows the average seasonal cycle for GPP, respiration, and NEE in each zonal band.
The global cycles in Figure~\ref{fig:flux-global} are also reflected in the northern boreal and northern temperate bands, but there are important deviations in tropical and southern bands.
For example, the largest GPP sink in the southern extratropics occurs in January and is relatively small (as expected), while in the tropics its timing varies by hemisphere.
As is well established, this hemispheric offset produces relatively flat seasonal cycles for GPP, respiration, and NEE across the tropics as a whole \citep[e.g.,][]{AnavEtAl2015}, although the WOMBAT~v2.S estimates exhibit slightly more pronounced seasonal cycles than other estimates.

Additionally, Figure~\ref{fig:seasonal-cycle-zonal} shows that assimilating SIF in WOMBAT~v2.S has minimal impact on the seasonal cycle of NEE across all zonal bands.
This is to be expected, but it contrasts with \citet{ZhangEtAl2023a}, who report a reduced-magnitude NEE seasonal cycle in the tropics after assimilating gap-filled SIF data products.
However, their approach is not hierarchical and treats SIF as a covariate for NEE, omitting the confounding relationship between SIF and GPP.

Figure~\ref{fig:seasonal-latitude-profile} in the Supplementary Material shows the average latitudinal profile of GPP, respiration, and NEE fluxes for each season.
Profiles for the different estimates of GPP and respiration are generally similar, but compared with FLUXCOM and bottom-up estimates in northern temperate latitudes, both WOMBAT estimates are larger in magnitude during boreal winter and smaller during summer.
Compared with WOMBAT~v2.0, v2.S shows a slight zonal redistribution of GPP during the boreal growing season (June--August), leading to less uptake in the tropics and more uptake in northern temperate latitudes.
\citet{ParazooEtAl2014} reported the opposite effect in a similar study using only SIF observations to optimize GPP estimates in isolation.
The different result in WOMBAT~v2.S can be attributed to several factors: the use of a hierarchical model, different SIF observations, the impact of CO\textsubscript{2} mole-fraction observations, and joint estimation of other fluxes in the inversion.
The zonal redistribution of GPP in WOMBAT~v2.S also brings the estimates closer to FLUXCOM in northern temperate latitudes, where more extensive flux-tower data coverage makes FLUXCOM relatively well-informed.

\begin{figure}[t!]
  \centering
  \includegraphics{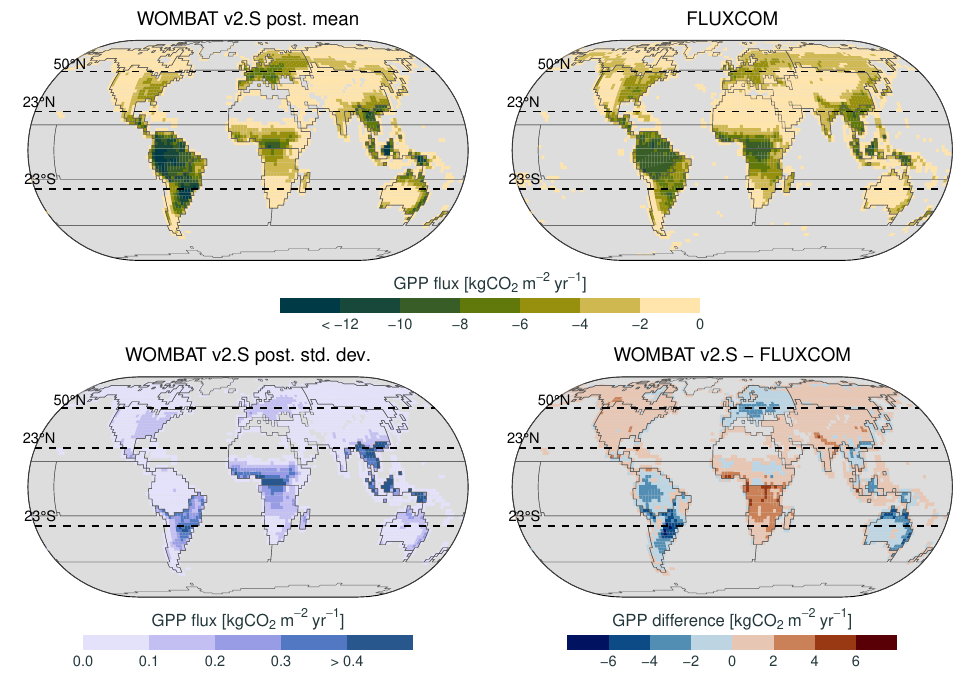}
  \caption{
    (Left column) WOMBAT~v2.S posterior mean and posterior standard deviation of the average GPP flux over January 2015--December 2020.
    (Right column) FLUXCOM estimate of the average GPP flux, and average difference between the WOMBAT~v2.S posterior mean and the FLUXCOM estimate.
    Grid cells are \qtyproduct{2 x 2.5}{\degree} and those shown in gray have zero GPP flux.
    Solid gray lines delineate the 23 regions used to partition the spatial domain.
  }
  \label{fig:average-maps-gpp}
\end{figure}

\emph{Spatial patterns.}
To examine spatial variability in flux estimates, Figure~\ref{fig:average-maps-gpp} displays grid-scale posterior means and standard deviations of the average GPP flux over the period from January 2015 to December 2020.
Also included are FLUXCOM grid-scale GPP flux averages and differences between WOMBAT~v2.S and FLUXCOM averages.
In agreement with \citet{AnavEtAl2015}, the largest GPP uptake occurs in the tropics, particularly in the Amazon and tropical Asia.
Notable uptake also occurs in the eastern United States, in Europe, the Congo Basin, along the eastern coast of Australia, and in New Zealand.
These regions also exhibit larger uncertainties in flux estimates (except for tropical South America, where the RLT is fixed), as indicated by the WOMBAT~v2.S posterior standard deviations.
Fixing the RLT to its bottom-up estimate in tropical South America limits the variability of GPP estimates in this region (see Section~\ref{supp:basis-function-setup} in the Supplementary Material).

The map of differences in Figure~\ref{fig:average-maps-gpp} shows that WOMBAT~v2.S estimates larger (more negative) GPP fluxes than FLUXCOM across most of South America, tropical Asia, and coastal Australia, and it estimates smaller (less negative) GPP fluxes in North America, Central Africa, and temperate to boreal Asia.
Figure~\ref{fig:average-maps-gpp-v2} in the Supplementary Material shows a comparison to WOMBAT~v2.0.
In both comparisons, differences are relatively smooth and occur over large areas, generally with boundaries along the spatial partitions of the basis functions used in the inversion.
This is because the inferred basis-function coefficients adjust the basis functions at the regional scale.

In the Supplementary Material, Figures \ref{fig:average-maps-resp} and \ref{fig:average-maps-nee} present grid-scale estimates of the average respiration and NEE fluxes, respectively.
As expected, the spatial pattern of respiration generally mirrors that of GPP in magnitude, which is why separating the two from NEE is challenging.
WOMBAT~v2.S and v2.0 NEE estimates are in relative agreement, and uncertainty in tropical regions is lower for NEE than for corresponding estimates of GPP and respiration.
These findings corroborate the OSSE result that NEE can be identified from terrestrial CO\textsubscript{2} mole-fraction observations more easily than its GPP and respiration components.

\section{Conclusions}
\label{sec:conclusion}

The WOMBAT~v2.S framework marks a significant step forward in attributing individual components of natural CO\textsubscript{2} flux across the globe.
The framework extends WOMBAT~v2.0 with a hierarchical statistical model that relates SIF satellite observations to the GPP component flux.
Using a first-order Taylor-series expansion about a bottom-up estimate of GPP, the model captures the linear nature of the SIF--GPP relationship while allowing for spatial and temporal variation within biomes and across seasons.
As the first flux-inversion framework to simultaneously assimilate SIF and CO\textsubscript{2} mole-fraction observations, WOMBAT~v2.S places a direct observational lens on the statistical distribution of GPP and other components of natural CO\textsubscript{2} flux.

In Section~\ref{sec:model}, we demonstrated how SIF observations are incorporated into WOMBAT~v2.S, and we explained that doing so allows us to relax strict modeling assumptions that were required in WOMBAT~v2.0.
Through the OSSE in Section~\ref{sec:osse}, we showed that SIF consistently improves the accuracy and uncertainty characterization of GPP and respiration estimates.
The inclusion of SIF does not impact the net flux, which is already well constrained by CO\textsubscript{2} mole-fraction observations.
When applied to SIF and CO\textsubscript{2} observations from the OCO-2 satellite and other sources in Section~\ref{sec:results}, WOMBAT~v2.S yields GPP and respiration estimates that better align with the current understanding of carbon-cycle seasonality at boreal and midlatitudes than those from WOMBAT~v2.0.
Incorporating SIF also leads to spatial redistribution of GPP and respiration estimates, with WOMBAT~v2.S showing a shift in GPP from tropical to northern temperate latitudes during June to August.

To take full advantage of satellite SIF observations in CO\textsubscript{2} flux inversions, there are remaining challenges to address.
A limitation of most approaches, including WOMBAT~v2.S, is the assumption of a linear relationship between GPP and SIF.
While this assumption is often a valid one, deviations from linearity can occur \citep{MagneyEtAl2020}.
For example, bottom-up SiB4 estimates of GPP and SIF tend to exhibit nonlinear relationships in tropical regions.
In these cases, WOMBAT~v2.S discards the corresponding OCO-2 SIF observations, which forfeits information in regions like the Amazon where observational insight is especially valuable.
In a future version, this could be mitigated by including higher-order terms in the Taylor-series expansion, or perhaps by introducing a mechanistic Gaussian process model.
Accounting for plant type in the SIF process model could also help isolate any piecewise linear relationships that may arise due to differences in canopy structure at fine scales.
The ability to assimilate SIF observations in the presence of nonlinearities will be crucial to improving GPP and respiration estimates in tropical regions, where dense canopies, humidity, and other factors can complicate the SIF--GPP relationship.

A major assumption in WOMBAT~v2.S is that of a known relationship between GPP and SIF.
While we derive this relationship using bottom-up estimates from the SiB4 biosphere model (see Section~\ref{supp:sif-gpp-sensitivity} in the Supplementary Material), alternative models and data sources could yield different results.
For example, an empirical relationship could be obtained by relating gap-filled SIF data products \citep[e.g.,][]{ZhangEtAl2018} with FLUXCOM GPP estimates \citep{NelsonEtAl2024}.
Uncertainties in the SIF--GPP relationship should also be incorporated into the SIF process model, especially in tropical regions where the relationship is not well understood.
Implementing greater flexibility in the SIF--GPP relationship would enable assimilation of SIF observations in regions and seasons where the relationship is currently treated as missing.

Other limitations of WOMBAT~v2.S are shared by most inversion frameworks, including the reliance on a single transport model and the use of a relatively coarse spatiotemporal parameterization.
The transport model we use is widely validated, but differences between transport models should increase variability of flux estimates \citep{BasuEtAl2018, SchuhEtAl2019}.
Although a coarse parameterization reduces the computational burden, it limits the scale at which reliable inferences can be made.
\citet{Zammit-MangionEtAl2022a} and \citet{BertolacciEtAl2024} discuss these challenges in the context of the WOMBAT framework, and they remain key areas for further research.

We conclude by noting that the WOMBAT~v2.S framework could also produce posterior estimates of the spatiotemporal SIF process, which have promising applications across multiple domains.
For instance, SIF has proven to be a reliable indicator of environmental stresses such as heatwaves and drought \citep[e.g.,][]{SunEtAl2023}, making its global estimation a valuable tool for informing climate-impact mitigation and adaptation strategies.

\section*{Software and data availability}
The software implementation of WOMBAT~v2.S is openly available at \url{https://github.com/joshhjacobson/wombat-v2s}.
The software includes guidance on acquiring the input data and reproducing the results in the manuscript.
Outputs from the WOMBAT~v2.S framework are also available.
These include MCMC samples from the WOMBAT~v2.S posterior distribution of the model parameters for the real-data inversion and all OSSE inversions; bottom-up estimates and WOMBAT~v2.S samples from the posterior distribution of the fluxes; and bottom-up estimates and WOMBAT~v2.S samples from the posterior distribution of the coefficients in the linear and seasonal terms of time-series decomposition.

\section*{Acknowledgments}
This research was made possible through resources provided by the National Computational Infrastructure (NCI), supported by the Australian Government, and additional resources funded by a 2021 University of Wollongong Major Equipment Grant.
The OCO-2 SIF and XCO\textsubscript{2} data used in this study were produced by the OCO-2 mission at the Jet Propulsion Laboratory, California Institute of Technology, with SIF data accessed through NASA's Goddard Earth Sciences Data and Information Services Center (GES DISC) at \url{https://disc.gsfc.nasa.gov/datasets/OCO2_L2_Lite_SIF_10r/summary}.
The processed XCO\textsubscript{2} data are available, alongside in~situ and flask data, from the OCO-2 v10 MIP portal (\url{https://gml.noaa.gov/ccgg/OCO2_v10mip/download.php}).
Our thanks go to the members of the OCO-2 Flux and SIF groups for their long-standing efforts and engagement in this line of research.
We also incorporated MERRA-2 data provided by NASA Goddard Space Flight Center's Global Modeling and Assimilation Office (GMAO).
The FLUXCOM X-BASE data product may be accessed through the Integrated Carbon Observation System (ICOS) Carbon Portal (\url{https://doi.org/10.18160/5NZG-JMJE}).
We extend our appreciation to Markus Reichstein, Martin Jung, and Ulrich Weber for their valuable insights regarding the FLUXCOM data.

\section*{Funding}
This work was supported by the Australian Research Council (ARC) Discovery Project (DP) DP190100180 and by NASA ROSES grant no.\ 23-OCOST23-0001.
J.J.\ acknowledges a University Postgraduate Award from the University of Wollongong, Australia; A.Z.-M.\ was also supported by ARC Discovery Early Career Research Award (DECRA) DE180100203; and A.S.\ was supported by NASA grant no.\ 80NSSC24K0741.

\bibliographystyle{apalike-maxbibnames20}
\bibliography{references}

\begin{thebibliography}{}

\bibitem[Anav et~al., 2015]{AnavEtAl2015}
Anav, A., Friedlingstein, P., Beer, C., Ciais, P., Harper, A., Jones, C.,
  Murray-Tortarolo, G., Papale, D., Parazoo, N.~C., Peylin, P., Piao, S.,
  Sitch, S., Viovy, N., Wiltshire, A., and Zhao, M. (2015).
\newblock Spatiotemporal patterns of terrestrial gross primary production:
  {{A}} review.
\newblock {\em Reviews of Geophysics}, 53(3):785--818.

\bibitem[Bacour et~al., 2019]{BacourEtAl2019}
Bacour, C., Maignan, F., MacBean, N., Porcar-Castell, A., Flexas, J.,
  Frankenberg, C., Peylin, P., Chevallier, F., Vuichard, N., and Bastrikov, V.
  (2019).
\newblock Improving estimates of gross primary productivity by assimilating
  solar-induced fluorescence satellite retrievals in a terrestrial biosphere
  model using a process-based {{SIF}} model.
\newblock {\em Journal of Geophysical Research: Biogeosciences},
  124(11):3281--3306.

\bibitem[Bai et~al., 2021]{BaiEtAl2021}
Bai, Y., Liang, S., and Yuan, W. (2021).
\newblock Estimating global gross primary production from sun-induced
  chlorophyll fluorescence data and auxiliary information using machine
  learning methods.
\newblock {\em Remote Sensing}, 13(5):963.

\bibitem[Baker, 2008]{Baker2008}
Baker, N.~R. (2008).
\newblock Chlorophyll fluorescence: A probe of photosynthesis in vivo.
\newblock {\em Annual Review of Plant Biology}, 59(1):89--113.

\bibitem[Basu et~al., 2018]{BasuEtAl2018}
Basu, S., Baker, D.~F., Chevallier, F., Patra, P.~K., Liu, J., and Miller,
  J.~B. (2018).
\newblock The impact of transport model differences on
  {{CO}}{\textsubscript{2}} surface flux estimates from {{OCO-2}} retrievals of
  column average {{CO}}{\textsubscript{2}}.
\newblock {\em Atmospheric Chemistry and Physics}, 18(10):7189--7215.

\bibitem[Basu et~al., 2013]{BasuEtAl2013}
Basu, S., Guerlet, S., Butz, A., Houweling, S., Hasekamp, O., Aben, I.,
  Krummel, P., Steele, P., Langenfelds, R., Torn, M., Biraud, S., Stephens, B.,
  Andrews, A., and Worthy, D. (2013).
\newblock Global {{CO}}{\textsubscript{2}} fluxes estimated from {{GOSAT}}
  retrievals of total column {{CO}}{\textsubscript{2}}.
\newblock {\em Atmospheric Chemistry and Physics}, 13(17):8695--8717.

\bibitem[Beer et~al., 2010]{BeerEtAl2010}
Beer, C., Reichstein, M., Tomelleri, E., Ciais, P., Jung, M., Carvalhais, N.,
  R{\"o}denbeck, C., Arain, M.~A., Baldocchi, D., Bonan, G.~B., Bondeau, A.,
  Cescatti, A., Lasslop, G., Lindroth, A., Lomas, M., Luyssaert, S., Margolis,
  H., Oleson, K.~W., Roupsard, O., Veenendaal, E., \ldots\ Papale, D. (2010).
\newblock Terrestrial gross carbon dioxide uptake: Global distribution and
  covariation with climate.
\newblock {\em Science}, 329(5993):834--838.

\bibitem[Bertolacci et~al., 2024]{BertolacciEtAl2024}
Bertolacci, M., {Zammit-Mangion}, A., Schuh, A., Bukosa, B., Fisher, J.~A.,
  Cao, Y., Kaushik, A., and Cressie, N. (2024).
\newblock Inferring changes to the global carbon cycle with {{WOMBAT}} v2.0, a
  hierarchical flux-inversion framework.
\newblock {\em The Annals of Applied Statistics}, 18(1):303--327.

\bibitem[Byrne et~al., 2023]{ByrneEtAl2023}
Byrne, B., Baker, D.~F., Basu, S., Bertolacci, M., Bowman, K.~W., Carroll, D.,
  Chatterjee, A., Chevallier, F., Ciais, P., Cressie, N., Crisp, D., Crowell,
  S., Deng, F., Deng, Z., Deutscher, N.~M., Dubey, M.~K., Feng, S., Garc{\'i}a,
  O.~E., Griffith, D. W.~T., Herkommer, B., \ldots\ Zeng, N. (2023).
\newblock National {{CO}}{\textsubscript{2}} budgets (2015--2020) inferred from
  atmospheric {{CO}}{\textsubscript{2}} observations in support of the global
  stocktake.
\newblock {\em Earth System Science Data}, 15(2):963--1004.

\bibitem[Byrne et~al., 2018]{ByrneEtAl2018}
Byrne, B., Wunch, D., Jones, D. B.~A., Strong, K., Deng, F., Baker, I.,
  K{\"o}hler, P., Frankenberg, C., Joiner, J., Arora, V.~K., Badawy, B.,
  Harper, A.~B., Warneke, T., Petri, C., Kivi, R., and Roehl, C.~M. (2018).
\newblock Evaluating {{GPP}} and respiration estimates over northern
  midlatitude ecosystems using solar-induced fluorescence and atmospheric
  {{CO}}{\textsubscript{2}} measurements.
\newblock {\em Journal of Geophysical Research: Biogeosciences},
  123(9):2976--2997.

\bibitem[Chevallier, 2007]{Chevallier2007}
Chevallier, F. (2007).
\newblock Impact of correlated observation errors on inverted
  {{CO}}{\textsubscript{2}} surface fluxes from {{OCO}} measurements.
\newblock {\em Geophysical Research Letters}, 34(24):2007GL030463.

\bibitem[Ciais et~al., 2010]{CiaisEtAl2010}
Ciais, P., Rayner, P., Chevallier, F., Bousquet, P., Logan, M., Peylin, P., and
  Ramonet, M. (2010).
\newblock Atmospheric inversions for estimating {{CO}}{\textsubscript{2}}
  fluxes: Methods and perspectives.
\newblock {\em Climatic Change}, 103(1-2):69--92.

\bibitem[Crisp et~al., 2017]{CrispEtAl2017}
Crisp, D., Pollock, H.~R., Rosenberg, R., Chapsky, L., Lee, R. A.~M., Oyafuso,
  F.~A., Frankenberg, C., O'Dell, C.~W., Bruegge, C.~J., Doran, G.~B.,
  Eldering, A., Fisher, B.~M., Fu, D., Gunson, M.~R., Mandrake, L., Osterman,
  G.~B., Schwandner, F.~M., Sun, K., Taylor, T.~E., Wennberg, P.~O., \ldots\
  Wunch, D. (2017).
\newblock The on-orbit performance of the {{Orbiting Carbon Observatory-2}}
  ({{OCO-2}}) instrument and its radiometrically calibrated products.
\newblock {\em Atmospheric Measurement Techniques}, 10(1):59--81.

\bibitem[Crowell et~al., 2019]{CrowellEtAl2019}
Crowell, S., Baker, D., Schuh, A., Basu, S., Jacobson, A.~R., Chevallier, F.,
  Liu, J., Deng, F., Feng, L., McKain, K., Chatterjee, A., Miller, J.~B.,
  Stephens, B.~B., Eldering, A., Crisp, D., Schimel, D., Nassar, R., O'Dell,
  C.~W., Oda, T., Sweeney, C., \ldots\ Jones, D. B.~A. (2019).
\newblock The 2015--2016 carbon cycle as seen from {{OCO-2}} and the global in
  situ network.
\newblock {\em Atmospheric Chemistry and Physics}, 19(15):9797--9831.

\bibitem[Damm et~al., 2015]{DammEtAl2015}
Damm, A., Guanter, L., {Paul-Limoges}, E., van~der Tol, C., Hueni, A.,
  Buchmann, N., Eugster, W., Ammann, C., and Schaepman, M. (2015).
\newblock Far-red sun-induced chlorophyll fluorescence shows ecosystem-specific
  relationships to gross primary production: {{An}} assessment based on
  observational and modeling approaches.
\newblock {\em Remote Sensing of Environment}, 166:91--105.

\bibitem[Doughty et~al., 2022]{DoughtyEtAl2022}
Doughty, R., Kurosu, T.~P., Parazoo, N., K{\"o}hler, P., Wang, Y., Sun, Y., and
  Frankenberg, C. (2022).
\newblock Global {{GOSAT}}, {{OCO-2}}, and {{OCO-3}} solar-induced chlorophyll
  fluorescence datasets.
\newblock {\em Earth System Science Data}, 14(4):1513--1529.

\bibitem[Doughty et~al., 2023]{DoughtyEtAl2023}
Doughty, R., Wang, Y., Johnson, J., Parazoo, N., Magney, T., Pierrat, Z., Xiao,
  X., Guanter, L., K{\"o}hler, P., Frankenberg, C., Somkuti, P., Ma, S., Qin,
  Y., Crowell, S., and Moore, B. (2023).
\newblock A novel data-driven global model of photosynthesis using
  solar-induced chlorophyll fluorescence.
\newblock ESS Open Archive.
\newblock \url{https://doi.org/10.22541/essoar.168167172.20799710/v1}.

\bibitem[Eldering et~al., 2017]{ElderingEtAl2017a}
Eldering, A., O'Dell, C.~W., Wennberg, P.~O., Crisp, D., Gunson, M.~R., Viatte,
  C., Avis, C., Braverman, A., Castano, R., Chang, A., Chapsky, L., Cheng, C.,
  Connor, B., Dang, L., Doran, G., Fisher, B., Frankenberg, C., Fu, D., Granat,
  R., Hobbs, J., \ldots\ Yoshimizu, J. (2017).
\newblock The {{Orbiting Carbon Observatory-2}}: First 18 months of science
  data products.
\newblock {\em Atmospheric Measurement Techniques}, 10(2):549--563.

\bibitem[Frankenberg and Berry, 2018]{FrankenbergBerry2018}
Frankenberg, C., and Berry, J. (2018).
\newblock Solar induced chlorophyll fluorescence: Origins, relation to
  photosynthesis and retrieval.
\newblock In Liang, S., editor, {\em Comprehensive {{Remote Sensing}}}, pages
  143--162. Elsevier, Oxford, UK.

\bibitem[Frankenberg et~al., 2011a]{FrankenbergEtAl2011a}
Frankenberg, C., Butz, A., and Toon, G.~C. (2011a).
\newblock Disentangling chlorophyll fluorescence from atmospheric scattering
  effects in {{O2 A-band}} spectra of reflected sun-light.
\newblock {\em Geophysical Research Letters}, 38(3):L03801.

\bibitem[Frankenberg et~al., 2011b]{FrankenbergEtAl2011}
Frankenberg, C., Fisher, J.~B., Worden, J., Badgley, G., Saatchi, S.~S., Lee,
  J.-E., Toon, G.~C., Butz, A., Jung, M., Kuze, A., and Yokota, T. (2011b).
\newblock New global observations of the terrestrial carbon cycle from
  {{GOSAT}}: {{Patterns}} of plant fluorescence with gross primary
  productivity.
\newblock {\em Geophysical Research Letters}, 38(17):L17706.

\bibitem[Friedlingstein et~al., 2023]{FriedlingsteinEtAl2023}
Friedlingstein, P., O'Sullivan, M., Jones, M.~W., Andrew, R.~M., Bakker, D.
  C.~E., Hauck, J., Landsch{\"u}tzer, P., Le~Qu{\'e}r{\'e}, C., Luijkx, I.~T.,
  Peters, G.~P., Peters, W., Pongratz, J., Schwingshackl, C., Sitch, S.,
  Canadell, J.~G., Ciais, P., Jackson, R.~B., Alin, S.~R., Anthoni, P.,
  Barbero, L., \ldots\ Zheng, B. (2023).
\newblock Global carbon budget 2023.
\newblock {\em Earth System Science Data}, 15(12):5301--5369.

\bibitem[Gneiting and Raftery, 2007]{GneitingRaftery2007}
Gneiting, T., and Raftery, A.~E. (2007).
\newblock Strictly proper scoring rules, prediction, and estimation.
\newblock {\em Journal of the American Statistical Association},
  102(477):359--378.

\bibitem[Gurney et~al., 2002]{GurneyEtAl2002}
Gurney, K.~R., Law, R.~M., Denning, A.~S., Rayner, P.~J., Baker, D., Bousquet,
  P., Bruhwiler, L., Chen, Y.-H., Ciais, P., Fan, S., Fung, I.~Y., Gloor, M.,
  Heimann, M., Higuchi, K., John, J., Maki, T., Maksyutov, S., Masarie, K.,
  Peylin, P., Prather, M., \ldots\ Yuen, C.-W. (2002).
\newblock Towards robust regional estimates of {{CO}}{\textsubscript{2}}
  sources and sinks using atmospheric transport models.
\newblock {\em Nature}, 415(6872):626--630.

\bibitem[Han et~al., 2022]{HanEtAl2022}
Han, J., Chang, C. Y.-Y., Gu, L., Zhang, Y., Meeker, E.~W., Magney, T.~S.,
  Walker, A.~P., Wen, J., Kira, O., McNaull, S., and Sun, Y. (2022).
\newblock The physiological basis for estimating photosynthesis from
  {{Chl}}{\emph{a}} fluorescence.
\newblock {\em New Phytologist}, 234(4):1206--1219.

\bibitem[Haynes et~al., 2019a]{HaynesEtAl2019}
Haynes, K.~D., Baker, I.~T., Denning, A.~S., St{\"o}ckli, R., Schaefer, K.,
  Lokupitiya, E.~Y., and Haynes, J.~M. (2019a).
\newblock Representing grasslands using dynamic prognostic phenology based on
  biological growth stages: 1. {{Implementation}} in the simple biosphere model
  ({{SiB4}}).
\newblock {\em Journal of Advances in Modeling Earth Systems},
  11(12):4423--4439.

\bibitem[Haynes et~al., 2019b]{HaynesEtAl2019a}
Haynes, K.~D., Baker, I.~T., Denning, A.~S., Wolf, S., Wohlfahrt, G., Kiely,
  G., Minaya, R.~C., and Haynes, J.~M. (2019b).
\newblock Representing grasslands using dynamic prognostic phenology based on
  biological growth stages: {{Part}} 2. {{Carbon}} cycling.
\newblock {\em Journal of Advances in Modeling Earth Systems},
  11(12):4440--4465.

\bibitem[Joiner et~al., 2014]{JoinerEtAl2014}
Joiner, J., Yoshida, Y., Vasilkov, A., Schaefer, K., Jung, M., Guanter, L.,
  Zhang, Y., Garrity, S., Middleton, E., Huemmrich, K., Gu, L., and
  Belelli~Marchesini, L. (2014).
\newblock The seasonal cycle of satellite chlorophyll fluorescence observations
  and its relationship to vegetation phenology and ecosystem atmosphere carbon
  exchange.
\newblock {\em Remote Sensing of Environment}, 152:375--391.

\bibitem[Jung et~al., 2020]{JungEtAl2020}
Jung, M., Schwalm, C., Migliavacca, M., Walther, S., {Camps-Valls}, G.,
  Koirala, S., Anthoni, P., Besnard, S., Bodesheim, P., Carvalhais, N.,
  Chevallier, F., Gans, F., Goll, D.~S., Haverd, V., K{\"o}hler, P., Ichii, K.,
  Jain, A.~K., Liu, J., Lombardozzi, D., Nabel, J. E. M.~S., \ldots\
  Reichstein, M. (2020).
\newblock Scaling carbon fluxes from eddy covariance sites to globe: Synthesis
  and evaluation of the {{FLUXCOM}} approach.
\newblock {\em Biogeosciences}, 17(5):1343--1365.

\bibitem[Kaminski et~al., 2001]{KaminskiEtAl2001}
Kaminski, T., Rayner, P.~J., Heimann, M., and Enting, I.~G. (2001).
\newblock On aggregation errors in atmospheric transport inversions.
\newblock {\em Journal of Geophysical Research: Atmospheres},
  106(D5):4703--4715.

\bibitem[Kira et~al., 2021]{KiraEtAl2021}
Kira, O., Chang, C. Y.-Y., Gu, L., Wen, J., Hong, Z., and Sun, Y. (2021).
\newblock Partitioning net ecosystem exchange ({{NEE}}) of
  {{CO}}{\textsubscript{2}} using solar-induced chlorophyll fluorescence
  ({{SIF}}).
\newblock {\em Geophysical Research Letters}, 48(4):e2020GL091247.

\bibitem[Li et~al., 2018]{LiEtAl2018}
Li, X., Xiao, J., He, B., Altaf~Arain, M., Beringer, J., Desai, A.~R., Emmel,
  C., Hollinger, D.~Y., Krasnova, A., Mammarella, I., Noe, S.~M., Ortiz, P.~S.,
  {Rey-Sanchez}, A.~C., Rocha, A.~V., and Varlagin, A. (2018).
\newblock Solar-induced chlorophyll fluorescence is strongly correlated with
  terrestrial photosynthesis for a wide variety of biomes: First global
  analysis based on {{OCO-2}} and flux tower observations.
\newblock {\em Global Change Biology}, 24(9):3990--4008.

\bibitem[Liu et~al., 2017]{LiuEtAl2017}
Liu, J., Bowman, K.~W., Schimel, D.~S., Parazoo, N.~C., Jiang, Z., Lee, M.,
  Bloom, A.~A., Wunch, D., Frankenberg, C., Sun, Y., O'Dell, C.~W., Gurney,
  K.~R., Menemenlis, D., Gierach, M., Crisp, D., and Eldering, A. (2017).
\newblock Contrasting carbon cycle responses of the tropical continents to the
  2015--2016 {{El Ni{\~n}o}}.
\newblock {\em Science}, 358(6360):eaam5690.

\bibitem[MacBean et~al., 2018]{MacBeanEtAl2018}
MacBean, N., Maignan, F., Bacour, C., Lewis, P., Peylin, P., Guanter, L.,
  K{\"o}hler, P., {G{\'o}mez-Dans}, J., and Disney, M. (2018).
\newblock Strong constraint on modelled global carbon uptake using
  solar-induced chlorophyll fluorescence data.
\newblock {\em Scientific Reports}, 8(1):1973.

\bibitem[Magney et~al., 2020]{MagneyEtAl2020}
Magney, T.~S., Barnes, M.~L., and Yang, X. (2020).
\newblock On the covariation of chlorophyll fluorescence and photosynthesis
  across scales.
\newblock {\em Geophysical Research Letters}, 47(23):e2020GL091098.

\bibitem[Neal, 2003]{Neal2003}
Neal, R.~M. (2003).
\newblock Slice sampling.
\newblock {\em The Annals of Statistics}, 31(3):705--767.

\bibitem[Nelson et~al., 2024]{NelsonEtAl2024}
Nelson, J.~A., Walther, S., Gans, F., Kraft, B., Weber, U., Novick, K.,
  Buchmann, N., Migliavacca, M., Wohlfahrt, G., {\v S}igut, L., Ibrom, A.,
  Papale, D., G{\"o}ckede, M., Duveiller, G., Knohl, A., H{\"o}rtnagl, L.,
  Scott, R.~L., Zhang, W., Hamdi, Z.~M., Reichstein, M., \ldots\ Jung, M.
  (2024).
\newblock X-{{BASE}}: The first terrestrial carbon and water flux products from
  an extended data-driven scaling framework, {{FLUXCOM-X}}.
\newblock {\em Biogeosciences}, 21(22):5079--5115.

\bibitem[Norton et~al., 2019]{NortonEtAl2019}
Norton, A.~J., Rayner, P.~J., Koffi, E.~N., Scholze, M., Silver, J.~D., and
  Wang, Y.-P. (2019).
\newblock Estimating global gross primary productivity using chlorophyll
  fluorescence and a data assimilation system with the {{BETHY-SCOPE}} model.
\newblock {\em Biogeosciences}, 16(15):3069--3093.

\bibitem[{OCO-2 Science Team} et~al., 2020]{OCO-2ScienceTeamEtAl2020}
{OCO-2 Science Team}, Gunson, M., and Eldering, A. (2020).
\newblock {{OCO-2 Level}} 2 bias-corrected solar-induced fluorescence and other
  select fields from the {{IMAP-DOAS}} algorithm aggregated as daily files,
  {{Retrospective}} processing {{V10r}}.
\newblock {Goddard Earth Sciences Data and Information Services Center (GES
  DISC), Greenbelt, MD}.
\newblock \url{https://doi.org/10.5067/XO2LBBNPO010}.

\bibitem[O'Dell et~al., 2018]{ODellEtAl2018}
O'Dell, C.~W., Eldering, A., Wennberg, P.~O., Crisp, D., Gunson, M.~R., Fisher,
  B., Frankenberg, C., Kiel, M., Lindqvist, H., Mandrake, L., Merrelli, A.,
  Natraj, V., Nelson, R.~R., Osterman, G.~B., Payne, V.~H., Taylor, T.~E.,
  Wunch, D., Drouin, B.~J., Oyafuso, F., Chang, A., \ldots\ Velazco, V.~A.
  (2018).
\newblock Improved retrievals of carbon dioxide from {{Orbiting Carbon
  Observatory-2}} with the version 8 {{ACOS}} algorithm.
\newblock {\em Atmospheric Measurement Techniques}, 11(12):6539--6576.

\bibitem[Pakman and Paninski, 2014]{PakmanPaninski2014}
Pakman, A., and Paninski, L. (2014).
\newblock Exact {{Hamiltonian Monte Carlo}} for truncated multivariate
  {{Gaussians}}.
\newblock {\em Journal of Computational and Graphical Statistics},
  23(2):518--542.

\bibitem[Parazoo et~al., 2014]{ParazooEtAl2014}
Parazoo, N.~C., Bowman, K., Fisher, J.~B., Frankenberg, C., Jones, D. B.~A.,
  Cescatti, A., {P{\'e}rez-Priego}, {\'O}., Wohlfahrt, G., and Montagnani, L.
  (2014).
\newblock Terrestrial gross primary production inferred from satellite
  fluorescence and vegetation models.
\newblock {\em Global Change Biology}, 20(10):3103--3121.

\bibitem[Parazoo et~al., 2019]{ParazooEtAl2019}
Parazoo, N.~C., Frankenberg, C., K{\"o}hler, P., Joiner, J., Yoshida, Y.,
  Magney, T., Sun, Y., and Yadav, V. (2019).
\newblock Towards a harmonized long-term spaceborne record of far-red
  solar-induced fluorescence.
\newblock {\em Journal of Geophysical Research: Biogeosciences},
  124(8):2518--2539.

\bibitem[Peiro et~al., 2022]{PeiroEtAl2022}
Peiro, H., Crowell, S., Schuh, A., Baker, D.~F., O'Dell, C., Jacobson, A.~R.,
  Chevallier, F., Liu, J., Eldering, A., Crisp, D., Deng, F., Weir, B., Basu,
  S., Johnson, M.~S., Philip, S., and Baker, I. (2022).
\newblock Four years of global carbon cycle observed from the {{Orbiting Carbon
  Observatory}} 2 ({{OCO-2}}) version 9 and in situ data and comparison to
  {{OCO-2}} version 7.
\newblock {\em Atmospheric Chemistry and Physics}, 22(2):1097--1130.

\bibitem[Pierrat et~al., 2022]{PierratEtAl2022a}
Pierrat, Z., Magney, T., Parazoo, N.~C., Grossmann, K., Bowling, D.~R., Seibt,
  U., Johnson, B., Helgason, W., Barr, A., Bortnik, J., Norton, A., Maguire,
  A., Frankenberg, C., and Stutz, J. (2022).
\newblock Diurnal and seasonal dynamics of solar-induced chlorophyll
  fluorescence, vegetation indices, and gross primary productivity in the
  boreal forest.
\newblock {\em Journal of Geophysical Research: Biogeosciences},
  127(2):e2021JG006588.

\bibitem[{Porcar-Castell} et~al., 2014]{Porcar-CastellEtAl2014}
{Porcar-Castell}, A., Tyystj{\"a}rvi, E., Atherton, J., van~der Tol, C.,
  Flexas, J., Pf{\"u}ndel, E.~E., Moreno, J., Frankenberg, C., and Berry, J.~A.
  (2014).
\newblock Linking chlorophyll {\emph{a}} fluorescence to photosynthesis for
  remote sensing applications: Mechanisms and challenges.
\newblock {\em Journal of Experimental Botany}, 65(15):4065--4095.

\bibitem[Reichstein et~al., 2005]{ReichsteinEtAl2005}
Reichstein, M., Falge, E., Baldocchi, D., Papale, D., Aubinet, M., Berbigier,
  P., Bernhofer, C., Buchmann, N., Gilmanov, T., Granier, A., Gr{\"u}nwald, T.,
  Havr{\'a}nkov{\'a}, K., Ilvesniemi, H., Janous, D., Knohl, A., Laurila, T.,
  Lohila, A., Loustau, D., Matteucci, G., Meyers, T., \ldots\ Valentini, R.
  (2005).
\newblock On the separation of net ecosystem exchange into assimilation and
  ecosystem respiration: Review and improved algorithm.
\newblock {\em Global Change Biology}, 11(9):1424--1439.

\bibitem[Schimel et~al., 2015]{SchimelEtAl2015}
Schimel, D., Pavlick, R., Fisher, J.~B., Asner, G.~P., Saatchi, S., Townsend,
  P., Miller, C., Frankenberg, C., Hibbard, K., and Cox, P. (2015).
\newblock Observing terrestrial ecosystems and the carbon cycle from space.
\newblock {\em Global Change Biology}, 21(5):1762--1776.

\bibitem[Schuh et~al., 2019]{SchuhEtAl2019}
Schuh, A.~E., Jacobson, A.~R., Basu, S., Weir, B., Baker, D., Bowman, K.,
  Chevallier, F., Crowell, S., Davis, K.~J., Deng, F., Denning, S., Feng, L.,
  Jones, D., Liu, J., and Palmer, P.~I. (2019).
\newblock Quantifying the impact of atmospheric transport uncertainty on
  {{CO}}{\textsubscript{2}} surface flux estimates.
\newblock {\em Global Biogeochemical Cycles}, 33(4):484--500.

\bibitem[Shiga et~al., 2018]{ShigaEtAl2018}
Shiga, Y.~P., Tadi{\'c}, J.~M., Qiu, X., Yadav, V., Andrews, A.~E., Berry,
  J.~A., and Michalak, A.~M. (2018).
\newblock Atmospheric {{CO}}{\textsubscript{2}} observations reveal strong
  correlation between regional net biospheric carbon uptake and solar-induced
  chlorophyll fluorescence.
\newblock {\em Geophysical Research Letters}, 45(2):1122--1132.

\bibitem[Sitch et~al., 2015]{SitchEtAl2015}
Sitch, S., Friedlingstein, P., Gruber, N., Jones, S.~D., {Murray-Tortarolo},
  G., Ahlstr{\"o}m, A., Doney, S.~C., Graven, H., Heinze, C., Huntingford, C.,
  Levis, S., Levy, P.~E., Lomas, M., Poulter, B., Viovy, N., Zaehle, S., Zeng,
  N., Arneth, A., Bonan, G., Bopp, L., \ldots\ Myneni, R. (2015).
\newblock Recent trends and drivers of regional sources and sinks of carbon
  dioxide.
\newblock {\em Biogeosciences}, 12(3):653--679.

\bibitem[Sun et~al., 2018]{SunEtAl2018}
Sun, Y., Frankenberg, C., Jung, M., Joiner, J., Guanter, L., K{\"o}hler, P.,
  and Magney, T. (2018).
\newblock Overview of solar-induced chlorophyll fluorescence ({{SIF}}) from the
  {{Orbiting Carbon Observatory-2}}: Retrieval, cross-mission comparison, and
  global monitoring for {{GPP}}.
\newblock {\em Remote Sensing of Environment}, 209:808--823.

\bibitem[Sun et~al., 2017]{SunEtAl2017}
Sun, Y., Frankenberg, C., Wood, J.~D., Schimel, D.~S., Jung, M., Guanter, L.,
  Drewry, D.~T., Verma, M., {Porcar-Castell}, A., Griffis, T.~J., Gu, L.,
  Magney, T.~S., K{\"o}hler, P., Evans, B., and Yuen, K. (2017).
\newblock {{OCO-2}} advances photosynthesis observation from space via
  solar-induced chlorophyll fluorescence.
\newblock {\em Science}, 358(6360):eaam5747.

\bibitem[Sun et~al., 2023a]{SunEtAl2023a}
Sun, Y., Gu, L., Wen, J., van~der Tol, C., Porcar-Castell, A., Joiner, J.,
  Chang, C.~Y., Magney, T., Wang, L., Hu, L., Rascher, U., Zarco-Tejada, P.,
  Barrett, C.~B., Lai, J., Han, J., and Luo, Z. (2023a).
\newblock From remotely sensed solar-induced chlorophyll fluorescence to
  ecosystem structure, function, and service: {{Part I}}---{{Harnessing}}
  theory.
\newblock {\em Global Change Biology}, 29(11):2926--2952.

\bibitem[Sun et~al., 2023b]{SunEtAl2023}
Sun, Y., Wen, J., Gu, L., Joiner, J., Chang, C.~Y., van~der Tol, C.,
  Porcar-Castell, A., Magney, T., Wang, L., Hu, L., Rascher, U., Zarco-Tejada,
  P., Barrett, C.~B., Lai, J., Han, J., and Luo, Z. (2023b).
\newblock From remotely-sensed solar-induced chlorophyll fluorescence to
  ecosystem structure, function, and service: {{Part II}}---{{Harnessing}}
  data.
\newblock {\em Global Change Biology}, 29(11):2893--2925.

\bibitem[Tarantola, 2005]{Tarantola2005a}
Tarantola, A. (2005).
\newblock {\em Inverse Problem Theory and Methods for Model Parameter
  Estimation}.
\newblock {Society for Industrial and Applied Mathematics}, Philadelphia, PA.

\bibitem[Tramontana et~al., 2016]{TramontanaEtAl2016}
Tramontana, G., Jung, M., Schwalm, C.~R., Ichii, K., {Camps-Valls}, G.,
  R{\'a}duly, B., Reichstein, M., Arain, M.~A., Cescatti, A., Kiely, G.,
  Merbold, L., {Serrano-Ortiz}, P., Sickert, S., Wolf, S., and Papale, D.
  (2016).
\newblock Predicting carbon dioxide and energy fluxes across global {{FLUXNET}}
  sites with regression algorithms.
\newblock {\em Biogeosciences}, 13(14):4291--4313.

\bibitem[{Zammit-Mangion} et~al., 2022]{Zammit-MangionEtAl2022a}
{Zammit-Mangion}, A., Bertolacci, M., Fisher, J., Stavert, A., Rigby, M., Cao,
  Y., and Cressie, N. (2022).
\newblock {{WOMBAT}} v1.0: A fully {{Bayesian}} global flux-inversion
  framework.
\newblock {\em Geoscientific Model Development}, 15(1):45--73.

\bibitem[Zhang et~al., 2023]{ZhangEtAl2023a}
Zhang, M., Berry, J.~A., Shiga, Y.~P., Doughty, R.~B., Madani, N., Li, X.,
  Xiao, J., Wen, J., Sun, Y., and Miller, S.~M. (2023).
\newblock Solar-induced fluorescence helps constrain global patterns in net
  biosphere exchange, as estimated using atmospheric {{CO}}{\textsubscript{2}}
  observations.
\newblock {\em Journal of Geophysical Research: Biogeosciences},
  128(12):e2023JG007703.

\bibitem[Zhang et~al., 2018a]{ZhangEtAl2018}
Zhang, Y., Joiner, J., Alemohammad, S.~H., Zhou, S., and Gentine, P. (2018a).
\newblock A global spatially contiguous solar-induced fluorescence ({{CSIF}})
  dataset using neural networks.
\newblock {\em Biogeosciences}, 15(19):5779--5800.

\bibitem[Zhang et~al., 2018b]{ZhangEtAl2018a}
Zhang, Y., Xiao, X., Zhang, Y., Wolf, S., Zhou, S., Joiner, J., Guanter, L.,
  Verma, M., Sun, Y., Yang, X., {Paul-Limoges}, E., Gough, C.~M., Wohlfahrt,
  G., Gioli, B., van~der Tol, C., Yann, N., Lund, M., and De~Grandcourt, A.
  (2018b).
\newblock On the relationship between sub-daily instantaneous and daily total
  gross primary production: {{Implications}} for interpreting satellite-based
  {{SIF}} retrievals.
\newblock {\em Remote Sensing of Environment}, 205:276--289.

\bibitem[Zhu et~al., 2016]{ZhuEtAl2016}
Zhu, Z., Piao, S., Myneni, R.~B., Huang, M., Zeng, Z., Canadell, J.~G., Ciais,
  P., Sitch, S., Friedlingstein, P., Arneth, A., Cao, C., Cheng, L., Kato, E.,
  Koven, C., Li, Y., Lian, X., Liu, Y., Liu, R., Mao, J., Pan, Y., \ldots\
  Zeng, N. (2016).
\newblock Greening of the {{Earth}} and its drivers.
\newblock {\em Nature Climate Change}, 6(8):791--795.

\end{thebibliography}


\begin{thebibliography}{}

\bibitem[Bey et~al., 2001]{BeyEtAl2001}
Bey, I., Jacob, D.~J., Yantosca, R.~M., Logan, J.~A., Field, B.~D., Fiore,
  A.~M., Li, Q., Liu, H.~Y., Mickley, L.~J., and Schultz, M.~G. (2001).
\newblock Global modeling of tropospheric chemistry with assimilated
  meteorology: {{Model}} description and evaluation.
\newblock {\em Journal of Geophysical Research: Atmospheres},
  106(D19):23073--23095.

\bibitem[Enting, 2002]{Enting2002a}
Enting, I.~G. (2002).
\newblock {\em Inverse Problems in Atmospheric Constituent Transport}.
\newblock Cambridge University Press, Cambridge, UK.

\bibitem[Gelman et~al., 2013]{GelmanEtAl2013}
Gelman, A., Carlin, J.~B., Stern, H.~S., Dunson, D.~B., Vehtari, A., and Rubin,
  D.~B. (2013).
\newblock {\em Bayesian Data Analysis}, 3rd edition.
\newblock {Chapman and Hall/CRC}, New York, NY.

\bibitem[Hanke et~al., 2016]{HankeEtAl2016}
Hanke, M., Redler, R., Holfeld, T., and Yastremsky, M. (2016).
\newblock {{YAC}} 1.2.0: New aspects for coupling software in {{Earth}} system
  modelling.
\newblock {\em Geoscientific Model Development}, 9(8):2755--2769.

\bibitem[Koster et~al., 2015]{KosterEtAl2015}
Koster, R.~D., Bosilovich, M.~G., Akella, S., Lawrence, C., Cullather, R.,
  Draper, C., Gelaro, R., Kovach, R., Liu, Q., Molod, A., Norris, P., Wargan,
  K., Chao, W., Reichle, R., Takacs, L., Todling, R., Vikhliaev, Y., Bloom, S.,
  Collow, A., Partyka, G., \ldots\ Suarez, M. (2015).
\newblock {{MERRA-2}}: Initial evaluation of the climate.
\newblock Technical Report TM-2015-104606, NASA Goddard Space Flight Center,
  Greenbelt, MD.
\newblock \url{https://ntrs.nasa.gov/citations/20160005045}.

\bibitem[Landsch{\"u}tzer et~al., 2016]{LandschutzerEtAl2016}
Landsch{\"u}tzer, P., Gruber, N., and Bakker, D. C.~E. (2016).
\newblock Decadal variations and trends of the global ocean carbon sink.
\newblock {\em Global Biogeochemical Cycles}, 30(10):1396--1417.

\bibitem[Masarie et~al., 2014]{MasarieEtAl2014}
Masarie, K.~A., Peters, W., Jacobson, A.~R., and Tans, P.~P. (2014).
\newblock {{ObsPack}}: A framework for the preparation, delivery, and
  attribution of atmospheric greenhouse gas measurements.
\newblock {\em Earth System Science Data}, 6(2):375--384.

\bibitem[Nara et~al., 2017]{NaraEtAl2017}
Nara, H., Tanimoto, H., Tohjima, Y., Mukai, H., Nojiri, Y., and Machida, T.
  (2017).
\newblock Emission factors of {{CO}}{\textsubscript{2}}, {{CO}} and
  {{CH}}{\textsubscript{4}} from {{Sumatran}} peatland fires in 2013 based on
  shipboard measurements.
\newblock {\em Tellus B: Chemical and Physical Meteorology}, 69(1):1399047.

\bibitem[Nassar et~al., 2013]{NassarEtAl2013}
Nassar, R., Napier-Linton, L., Gurney, K.~R., Andres, R.~J., Oda, T., Vogel,
  F.~R., and Deng, F. (2013).
\newblock Improving the temporal and spatial distribution of
  {{CO}}{\textsubscript{2}} emissions from global fossil fuel emission data
  sets.
\newblock {\em Journal of Geophysical Research: Atmospheres}, 118(2):917--933.

\bibitem[Oda and Maksyutov, 2011]{OdaMaksyutov2011}
Oda, T., and Maksyutov, S. (2011).
\newblock A very high-resolution (1 km{\texttimes}1 km) global fossil fuel
  {{CO}}{\textsubscript{2}} emission inventory derived using a point source
  database and satellite observations of nighttime lights.
\newblock {\em Atmospheric Chemistry and Physics}, 11(2):543--556.

\bibitem[Oda et~al., 2018]{OdaEtAl2018}
Oda, T., Maksyutov, S., and Andres, R.~J. (2018).
\newblock The {{Open-source Data Inventory}} for {{Anthropogenic
  CO}}{\textsubscript{2}}, version 2016 ({{ODIAC2016}}): A global monthly
  fossil fuel {{CO}}{\textsubscript{2}} gridded emissions data product for
  tracer transport simulations and surface flux inversions.
\newblock {\em Earth System Science Data}, 10(1):87--107.

\bibitem[Schuldt et~al., 2021a]{SchuldtEtAl2021}
Schuldt, K.~N., Jacobson, A.~R., Aalto, T., Andrews, A., Lindauer, M.,
  Bergamaschi, P., Chen, H., Colomb, A., Conil, S., Cristofanelli, P.,
  Delmotte, M., Dlugokencky, E., Emmenegger, L., Fischer, M.~L., Hatakka, J.,
  Heliasz, M., Hermanssen, O., Holst, J., Jaffe, D., Karion, A., \ldots\
  De~Wekker, S. (2021a).
\newblock Multi-laboratory compilation of atmospheric carbon dioxide data for
  the years 2020-2021; obspack\_co2\_1\_{{NRT}}\_v6.1.1\_2021-05-17.
\newblock NOAA Global Monitoring Laboratory.
\newblock \url{https://doi.org/10.25925/20210517}.

\bibitem[Schuldt et~al., 2021b]{SchuldtEtAl2021a}
Schuldt, K.~N., Mund, J., Luijkx, I.~T., Aalto, T., Abshire, J.~B., Aikin, K.,
  Andrews, A., Aoki, S., Apadula, F., Baier, B., Bakwin, P., Bartyzel, J.,
  Bentz, G., Bergamaschi, P., Beyersdorf, A., Biermann, T., Biraud, S.~C.,
  Boenisch, H., Bowling, D., Brailsford, G., \ldots\ Zimnoch, M. (2021b).
\newblock Multi-laboratory compilation of atmospheric carbon dioxide data for
  the period 1957-2019; obspack\_co2\_1\_{{GLOBALVIEWplus}}\_v6.1\_2021-03-01.
\newblock NOAA Global Monitoring Laboratory.
\newblock \url{https://doi.org/10.25925/20201204}.

\bibitem[Tohjima et~al., 2005]{TohjimaEtAl2005}
Tohjima, Y., Mukai, H., Machida, T., Nojiri, Y., and Gloor, M. (2005).
\newblock First measurements of the latitudinal atmospheric
  {{O}}{\textsubscript{2}} and {{CO}}{\textsubscript{2}} distributions across
  the western {{Pacific}}.
\newblock {\em Geophysical Research Letters}, 32(17):2005GL023311.

\bibitem[Tukey, 1977]{Tukey1977a}
Tukey, J.~W. (1977).
\newblock {\em Exploratory Data Analysis}.
\newblock Addison-Wesley, Reading, MA.

\bibitem[van~der Werf et~al., 2017]{VanDerWerfEtAl2017}
van~der Werf, G.~R., Randerson, J.~T., Giglio, L., Van~Leeuwen, T.~T., Chen,
  Y., Rogers, B.~M., Mu, M., Van~Marle, M. J.~E., Morton, D.~C., Collatz,
  G.~J., Yokelson, R.~J., and Kasibhatla, P.~S. (2017).
\newblock Global fire emissions estimates during 1997--2016.
\newblock {\em Earth System Science Data}, 9(2):697--720.

\bibitem[Yantosca, 2019]{Yantosca2019}
Yantosca, B. (2019).
\newblock geoschem/geos-chem: {{GEOS-Chem}} 12.3.2.
\newblock Zenodo.
\newblock \url{https://doi.org/10.5281/zenodo.2658178}.

\bibitem[Yevich and Logan, 2003]{YevichLogan2003}
Yevich, R., and Logan, J.~A. (2003).
\newblock An assessment of biofuel use and burning of agricultural waste in the
  developing world.
\newblock {\em Global Biogeochemical Cycles}, 17(4):2002GB001952.

\end{thebibliography}

\clearpage
\renewcommand{\thesection}{S.\arabic{section}}
\renewcommand{\theequation}{S.\arabic{equation}}
\renewcommand{\thetable}{S.\arabic{table}}
\renewcommand{\thefigure}{S.\arabic{figure}}
\setcounter{page}{1}
\setcounter{section}{0}
\setcounter{equation}{0}
\setcounter{table}{0}
\setcounter{figure}{0}

\title{Supplementary Material for ``\titlevar''}

\maketitle
\vspace{2ex}

This document provides further information about the WOMBAT~v2.S framework, including auxiliary details of the hierarchical model (Section~\ref{supp:model-details}), the SIF--GPP sensitivity term (Section~\ref{supp:sif-gpp-sensitivity}), the implementation (Section~\ref{supp:application-details}), and the OSSE (Section~\ref{supp:alpha-setup}).
Additional figures and tables that support the results are given in Section~\ref{supp:additional-figures}.

\section{Remaining details of the hierarchical model}
\label{supp:model-details}

Section~\ref{sec:model} in the main text describes the updates from WOMBAT~v2.0 \citep{BertolacciEtAl2024} that are needed to accommodate the inclusion of SIF in WOMBAT~v2.S.
Model details that are not specific to SIF are included here.
Section~\ref{supp:basis-function-model} describes the basis-function representation for the component fluxes, and Section~\ref{supp:alpha-covariance} gives the covariance structure of the basis-function coefficients.
Sections \ref{supp:mole-frac-process-model} and \ref{supp:mole-frac-data-model} give full mathematical details of the CO\textsubscript{2} mole-fraction process model and the data model, respectively.
Parameters of the data model are specified in Section~\ref{supp:parameter-model}.

\subsection{Basis-function representation of component fluxes}
\label{supp:basis-function-model}

For computational tractability, WOMBAT uses a basis-function representation of the individual component fluxes.
In WOMBAT~v2.0 and v2.S, this reduced-rank construction partitions Earth's surface, $\mathbb{S}^2$, into $R$ disjoint regions denoted by $\{D_r \subset \mathbb{S}^2 : r = 1, \ldots, R\}$, and partitions the study time period, $\mathcal{T}$, into $Q$ disjoint sequential time periods denoted by $\{E_q \subset \mathcal{T} : q = 1, \ldots, Q\}$.
Define the $R$-dimensional vector of spatial indicators,
\begin{equation*}
  \wvec_{\textup{S}}(\svec) \equiv (\mathbbm{1}(\svec \in D_1), \ldots, \mathbbm{1}(\svec \in D_R))', \quad \svec \in \mathbb{S}^2,
\end{equation*}
on the spatial partition, and the $QR$-dimensional vector of spatiotemporal indicators,
\begin{equation*}
  \wvec_{\textup{ST}}(\svec, t) \equiv (\mathbbm{1}(\svec \in D_1 \cap t \in E_1), \mathbbm{1}(\svec \in D_1 \cap t \in E_2), \ldots, \mathbbm{1}(\svec \in D_R \cap t \in E_Q))', \quad \svec \in \mathbb{S}^2,\ t \in \mathcal{T}, 
\end{equation*}
on the spatiotemporal partition, where $\mathbbm{1}(\cdot)$ is the indicator function.
For component $c \in \mathcal{C}$, we model each of the unknown spatial and spatiotemporal processes in \eqref{eq:time-decomposition} as follows:
\begin{equation}
\label{eq:coefficient-scaling}
  \begin{aligned}
    \beta_{c,j}(\svec) &= (1 + \wvec_{\textup{S}}(\svec)'\alphavec_{c,j}) \beta_{c,j}^0(\svec), \quad j = 0, 1, \\
    \beta_{c,j,k}(\svec) &= (1 + \wvec_{\textup{S}}(\svec)'\alphavec_{c,j,k}) \beta_{c,j,k}^0(\svec), \quad j = 2, \ldots, 5,\ k = 1, \ldots, K_c, \\
    \epsilon_c(\svec, t) & = (1 + \wvec_{\textup{ST}}(\svec, t)'\alphavec_{c,6}) \epsilon_c^0(\svec, t),
  \end{aligned}
\end{equation}
for $\svec \in \mathbb{S}^2$ and $t \in \mathcal{T}$, where $\alphavec_{c,j} \equiv (\alpha_{c,j,1}, \ldots, \alpha_{c,j,R})'$ for $j = 0, 1$; $\alphavec_{c,j,k} \equiv (\alpha_{c,j,k,1}, \ldots, \alpha_{c,j,k,R})'$ for $j = 2, \ldots, 5$ and $k = 1, \ldots, K_c$; and $\alphavec_{c,6} \equiv (\alpha_{c,6,1,1}, \alpha_{c,6,1,2}, \ldots, \alpha_{c,6,R,Q})'$.
Consequently, the random and unknown $\alphavec$-vectors spatially adjust the known bottom-up estimates of the linear and seasonal coefficient fields, $\beta_{c,\cdot}^0(\cdot)$ and $\beta_{c,\cdot,\cdot}^0(\cdot)$, respectively, and spatiotemporally adjust the bottom-up estimate of the residual term, $\epsilon_c^0(\cdot\,, \cdot)$.
For instance, $\beta_{c,0}(\svec) = (1 + \alpha_{c,0,r})\beta_{c,0}^0(\svec)$ for all $\svec \in D_r$.

Following \citet{BertolacciEtAl2024}, we obtain the fields $\beta_{c,\cdot}^0(\cdot)$, $\beta_{c,\cdot,\cdot}^0(\cdot)$, and $\epsilon_c^0(\cdot\,, \cdot)$ by applying the decomposition in \eqref{eq:time-decomposition} to the bottom-up estimates of each component flux (see Section~\ref{supp:bottom-up} for details).
We construct basis functions by applying the partitioning method in \eqref{eq:coefficient-scaling} to each of the terms in \eqref{eq:time-decomposition}, and collect them in the basis-vector $\phivec_c(\cdot\,, \cdot)$ of dimension $(2R + 4K_cR + QR)$.
For example, the first three elements of $\phivec_c(\svec, t)$ are $\wvec_{\textup{S}}(\svec)'\beta_{c,0}^0(\svec)$, $\wvec_{\textup{S}}(\svec)'\beta_{c,1}^0(\svec)t$, and $\wvec_{\textup{S}}(\svec)'\beta_{c,2,1}^0(\svec)\cos(2\pi t / 365.25)$.

\subsection{Covariance of the basis-function coefficients}
\label{supp:alpha-covariance}

We model the flux basis-function coefficients $\alphavec$ in \eqref{eq:net-flux} using a constrained Gaussian distribution with covariance matrix $\Sigmavec_{\alpha}$ ($\textup{ConstrGau}(\zerovec, \Sigmavec_{\alpha}, F_{\alpha})$; see Section~\ref{sec:parameter-model}).
As illustrated in \eqref{eq:coefficient-scaling}, basis functions correspond to the linear, seasonal, and residual terms of the time-series decomposition, and hence the correlation structure of each term is represented through respective submatrices of $\Sigmavec_{\alpha}$.
We defer to Section~\ref{supp:reparameterization} our specification of the submatrices of $\Sigmavec_{\alpha}$ that correspond to the linear terms $\{\cov(\alphavec_{c,j}) : c \in \mathcal{C}, j = 0, 1\}$ since they are modified from those of WOMBAT~v2.0.
Submatrices corresponding to the seasonal and residual terms are unchanged from WOMBAT~v2.0, but we describe them below for completeness.

The submatrices of $\Sigmavec_{\alpha}$ that correspond to seasonal terms are specified through
\begin{equation}
  \label{eq:seasonal-covariance}
  \begin{aligned}
    \var(\alpha_{c,j,k,r}) &= 1/\tau^{\beta}_c, \\
    \corr(\alpha_{c,j,k,r}, \alpha_{c',j,k,r}) &= \rho^{\beta}_{c,c'},
  \end{aligned}
\end{equation}
for $c, c' \in \mathcal{C}$, $j = 2, \ldots, 5$, $k = 1, \ldots, K_c$, $r = 1, \ldots, R$, where $\tau^{\beta}_c > 0$, $\rho^{\beta}_{c,c'} \in [-1, 1]$, and any correlations not specified are equal to zero.
The precision parameter $\tau^{\beta}_c$ moderates how far the seasonal terms can vary from their bottom-up estimates, and the correlation parameter $\rho^{\beta}_{c,c'}$ governs cross-dependence between seasonal cycles of flux components $c$ and $c'$.
For instance, we expect anticorrelation between adjustments to the seasonal cycle of GPP and adjustments to the seasonal cycle of respiration.
This would correspond to a positive value of $\rho^{\beta}_{\textup{gpp}, \textup{resp}}$ since these components have opposite orientation.

The submatrices of $\Sigmavec_{\alpha}$ corresponding to residual terms have a similar structure, but with additional temporal correlations:
\begin{equation}
  \label{eq:residual-covariance}
  \begin{aligned}
    \var(\alpha_{c,6,r,q}) &= 1/\tau^{\epsilon}_c, \quad c \in \mathcal{C};\; r = 1, \ldots, R;\; q = 1, \ldots, Q; \\
    \corr(\alpha_{c,6,r,q}, \alpha_{c',6,r,q'}) &= \rho^{\epsilon}_{c,c'}(\kappa^{\epsilon}_{\textup{bio}})^{|q-q'|}, \quad c, c' \in \{\textup{gpp}, \textup{resp}\};\; r = 1, \ldots, R;\; q, q' = 1, \ldots, Q; \\
    \corr(\alpha_{\textup{ocean},6,r,q}, \alpha_{\textup{ocean},6,r,q'}) &= (\kappa^{\epsilon}_{\textup{ocean}})^{|q-q'|}, \quad r = 1, \ldots, R;\; q, q' = 1, \ldots, Q; \\
  \end{aligned}
\end{equation}
where $\tau^{\epsilon}_c > 0$, $\kappa^{\epsilon}_c \in [0, 1]$, $\rho^{\epsilon}_{c,c'} \in [-1, 1]$, $\rho^{\epsilon}_{c,c} = 1$, and unspecified correlations are equal to zero.
The precision parameter $\tau^{\epsilon}_c$ and the correlation parameter $\rho^{\epsilon}_{c,c'}$ function similarly to their seasonal counterparts in \eqref{eq:seasonal-covariance} above, and the parameter $\kappa^{\epsilon}_c$ models component-specific temporal correlation.
We assume that a single parameter can capture temporal dependencies in the residual terms for GPP and respiration components, and thus we set $\kappa^{\epsilon}_{\textup{gpp}} = \kappa^{\epsilon}_{\textup{resp}} \equiv \kappa^{\epsilon}_{\textup{bio}}$.
Through the interaction with $\rho^{\epsilon}_{c,c'}$, the GPP and respiration residual terms can be cross-correlated in time, but they are assumed to be independent of the ocean residual terms.

The covariance parameters in \eqref{eq:seasonal-covariance} and \eqref{eq:residual-covariance} are given the following independent hyperprior distributions that we retain from WOMBAT~v2.0:
\begin{equation*}
  \begin{gathered}
    \tau^{\beta}_c \sim \textup{Gamma}(\nu^{\beta}_{\tau},\, \omega^{\beta}_{\tau,c}),
    \qquad
    \rho^{\beta}_{c,c'} \sim \textup{Beta}(a^{\beta}_{\rho},\, b^{\beta}_{\rho}), \\
    \tau^{\epsilon}_c \sim \textup{Gamma}(\nu^{\epsilon}_{\tau},\, \omega^{\epsilon}_{\tau,c}),
    \qquad
    \rho^{\epsilon}_{c,c'} \sim \textup{Beta}(a^{\epsilon}_{\rho},\, b^{\epsilon}_{\rho}), \\
    \kappa^{\epsilon}_c \sim \textup{Beta}(a^{\epsilon}_{\kappa},\, b^{\epsilon}_{\kappa}),
  \end{gathered}
\end{equation*}
for $c, c' \in \mathcal{C}$, where Gamma$(\nu,\, \omega)$ is the gamma distribution with shape hyperparameter $\nu$ and rate hyperparameter $\omega$, and Beta$(a,\, b)$ is the beta distribution with shape hyperparameters $a$ and $b$.
For some covariance parameters, it is occasionally appropriate to assume a known value; for example, we assume that $\rho^{\beta}_{c,\textup{ocean}}$ and $\rho^{\epsilon}_{c, \textup{ocean}}$ are zero when $c$ corresponds to GPP or respiration (see Section~\ref{supp:basis-function-setup}).
We give choices for the fixed hyperparameters in Section~\ref{supp:hyperparameters} below.

\subsection{Mole-fraction process model}
\label{supp:mole-frac-process-model}

Let $Y_{\textup{co}2}(\svec, h, t)$ denote the CO\textsubscript{2} mole-fraction field at location $\svec \in \mathbb{S}^2$, geopotential height $h \geq 0$, and time $t \in \mathcal{T}$.
The mole-fraction field at time $t$ depends on all fluxes that occur over the period $\mathcal{T}_t \equiv [t_0, t]$, given by $X(\cdot\,, \mathcal{T}_t)$, and on the ``initial condition'' of the entire mole-fraction field, given by $Y_{\textup{co}2}(\cdot\,, \cdot\,, t_0)$.
This dependence is expressed as
\begin{equation}
\label{eq:mole-frac-transport}
  Y_{\textup{co}2}(\svec, h, t) = \mathcal{H}_{\textup{co}2}(Y_{\textup{co}2}(\cdot\,, \cdot\,, t_0),\, X(\cdot\,, \mathcal{T}_t);\; \svec, h, t), \quad \svec \in \mathbb{S}^2,\ h \geq 0,\ t \in \mathcal{T},
\end{equation}
where the operator $\mathcal{H}_{\textup{co}2}$ represents the true chemical transport of CO\textsubscript{2} through the atmosphere.
Since CO\textsubscript{2} is a long-lived trace-gas, $\mathcal{H}_{\textup{co}2}$ is nearly linear \citepsupp[e.g.,][Chapter~2]{Enting2002a}, meaning that the impact of the initial condition, given by $\mathcal{H}_{\textup{co}2}(Y_{\textup{co}2}(\cdot\,, \cdot\,, t_0), 0;\; \svec, h, t)$, is additive and can be separated from the impact of the fluxes over the time period $\mathcal{T}_t$, given by $\mathcal{H}_{\textup{co}2}(0, X(\cdot\,, \mathcal{T}_t);\; \svec, h, t)$.
In practice, the transport operator is not perfectly known and, as in previous versions of WOMBAT, we obtain an approximate transport operator, $\hat{\mathcal{H}}_{\textup{co}2}$, from the transport model GEOS-Chem (see Section~\ref{supp:transport-model}).
In the resulting approximation, the term $\hat{\mathcal{H}}_{\textup{co}2}(0, X(\cdot\,, \mathcal{T}_t);\; \svec, h, t)$ functions similarly to $\hat{\mathcal{H}}^{(1)}_{\textup{sif}}(X_{\textup{gpp}}^0(\svec, t);\; \svec, t)$ in that both terms describe the sensitivity of an observable process to the latent flux process.
However, where $\hat{\mathcal{H}}^{(1)}_{\textup{sif}}(X_{\textup{gpp}}^0(\svec, t);\; \svec, t)$ is relatively local and instantaneous, $\hat{\mathcal{H}}_{\textup{co}2}(0, X(\cdot\,, \mathcal{T}_t);\; \svec, h, t)$ depends on the entire flux field at every time in the period $\mathcal{T}_t$.

Like the basis-function representation of the flux process in \eqref{eq:net-flux}, the mole-fraction process model also has a basis-function representation.
Specifically,
\begin{equation}
\label{eq:mole-frac-process}
  Y_{\textup{co}2}(\svec, h, t) =
  Y_{\textup{co}2}^0(\svec, h, t) +
  \hat{\psivec}_{\textup{co}2}(\svec, h, t)'\alphavec +
  \xi_{Y_{\textup{co}2}}(\svec, h, t),
  \quad \svec \in \mathbb{S}^2,\ h \geq 0,\ t \in \mathcal{T},
\end{equation}
where $Y_{\textup{co}2}^0(\svec, h, t) \equiv \hat{\mathcal{H}}_{\textup{co}2}(\hat{Y}_{\textup{co}2}(\cdot\,, \cdot\,, t_0), 0;\; \svec, h, t) + \hat{\mathcal{H}}_{\textup{co}2}(0, X^0(\cdot\,, \mathcal{T}_t);\; \svec, h, t)$ for $\svec \in \mathbb{S}^2$, $h \geq 0$, $t \in \mathcal{T}$, with $\hat{Y}_{\textup{co}2}(\cdot\,, \cdot\,, t_0)$ an estimate of the initial condition.
The vector $\hat{\psivec}_{\textup{co}2}(\cdot\,, \cdot\,, \cdot)$ is a vector of response functions and, due to linearity in $\hat{\mathcal{H}}_{\textup{co}2}$, its $l$th element directly corresponds to the $l$th element of the basis-function vector $\phivec(\cdot\,, \cdot)$ through $\hat{\psi}_{\textup{co}2, l}(\svec, h, t) = \hat{\mathcal{H}}_{\textup{co}2}(0, \phi_l(\cdot\,, \mathcal{T}_t);\; \svec, h, t)$ for $\svec \in \mathbb{S}^2$, $h \geq 0$, $t \in \mathcal{T}$.
The term $\xi_{Y_{\textup{co}2}}(\cdot\,, \cdot\,, \cdot)$ is a residual mole-fraction process that accommodates spatiotemporally correlated errors that arise from the use of an approximate initial condition and an imperfect transport operator.

The model in \eqref{eq:mole-frac-process} links the essentially unobservable flux process to the potentially observable CO\textsubscript{2} mole-fraction process, which is defined on a four-dimensional space that can be observed incompletely by satellites and other instruments.
These observations are only available at a limited set of locations, heights, and times, and they are subject to measurement biases and errors.
The WOMBAT framework addresses these issues through a mole-fraction data model.

\subsection{Mole-fraction data model}
\label{supp:mole-frac-data-model}

Let $\{Z_{\textup{co}2, i} : i = 1, \ldots, N_{\textup{co}2}\}$ denote the available collection of CO\textsubscript{2} mole-fraction observations.
Those used in flux inversions derive from point-referenced in~situ and flask measurements, and column-averaged satellite retrievals.
Each point-referenced observation is made at a specific location $\svec_i$ typically near Earth's surface ($h_i$ small), involves some degree of time averaging, and is not subject to substantial bias or correlated measurement error.
Point-referenced observations are therefore modeled as $Z_{\textup{co}2, i} = \mathcal{A}_i(Y_{\textup{co}2}(\svec_i, h_i, \cdot)) + \epsilon_{\textup{co}2, i}$, where $\mathcal{A}_i$ represents averaging over a known, typically short, time period, and $\epsilon_{\textup{co}2, i}$ is a mean-zero uncorrelated measurement error that has different properties depending on the in~situ or flask measuring instrument.
In contrast, each column-averaged observation is made at a specific location $\svec_i$ and time $t_i$, involves a weighted average over a vertical column of the atmosphere, and is typically affected by both bias and correlated measurement error.
Thus, column-averaged observations are modeled as $Z_{\textup{co}2, i} = \mathcal{A}_i(Y_{\textup{co}2}(\svec_i, \cdot\,, t_i)) + b_{\textup{co}2, i} + \xi_{Z_{\textup{co}2, i}} + \epsilon_{\textup{co}2, i}$, where $\mathcal{A}_i$ represents averaging over the vertical dimension, $b_{\textup{co}2, i}$ is a bias term, $\xi_{Z_{\textup{co}2, i}}$ is a mean-zero spatiotemporally correlated error term, and $\epsilon_{\textup{co}2, i}$ is a mean-zero uncorrelated measurement-error term.

Substituting the mole-fraction process model from \eqref{eq:mole-frac-process} into the point-referenced and column-averaged data models above, the general data model for CO\textsubscript{2} mole-fraction observations is
\begin{equation}
\label{eq:mole-frac-data-model}
  Z_{\textup{co}2, i} = Z_{\textup{co}2, i}^0 + \hat{\psivec}_{\textup{co}2, i}'\alphavec + b_{\textup{co}2, i} + \xi_{\textup{co}2, i} + \epsilon_{\textup{co}2, i}, \quad i = 1, \ldots, N_{\textup{co}2},
\end{equation}
where $Z_{\textup{co}2, i}^0 \equiv \mathcal{A}_i(Y_{\textup{co}2}^0(\svec_i, \cdot\,, \cdot))$; $\hat{\psivec}_{\textup{co}2, i} \equiv \mathcal{A}_i(\hat{\psivec}_{\textup{co}2}(\svec_i, \cdot\,, \cdot))$, with the operator applied elementwise; $b_{\textup{co}2, i} = 0$ for point-referenced observations; and $\xi_{\textup{co}2, i} \equiv \mathcal{A}_i(\xi_{Y_{\textup{co}2}}(\svec_i, h_i, \cdot))$ if $Z_{\textup{co}2, i}$ is point-referenced, or $\xi_{\textup{co}2, i} \equiv \mathcal{A}_i(\xi_{Y_{\textup{co}2}}(\svec_i, \cdot\,, t_i)) + \xi_{Z_{\textup{co}2, i}}$ if $Z_{\textup{co}2, i}$ is column-averaged.
We note that the mole-fraction data model in \eqref{eq:mole-frac-data-model} shares the form of the SIF data model in \eqref{eq:sif-data-model}.

\subsection{Parameters of the grouped data model}
\label{supp:parameter-model}

As discussed in Section~\ref{supp:mole-frac-data-model} above, point-referenced measurements and column-averaged retrievals represent two sources of CO\textsubscript{2} mole-fraction observations.
The SIF retrievals described in Section~\ref{sec:data} represent another observation source.
As in previous versions of the framework, WOMBAT~v2.S splits observations derived from these sources into groups with similar bias and error properties (see Section~\ref{supp:error-properties} below) using the grouped data model given by \eqref{eq:overall-data-vector}.
The bias vector for each observation group $g \in \mathcal{G}$ takes the form of a linear model, $\bvec_g = \mathbf{A}_g\pivec_g$, where $\mathbf{A}_g$ is a design matrix and $\pivec_g$ is a vector of unknown coefficients, modeled as $\pivec_g \sim \textup{Gau}(\zerovec,\, \sigma_{\pi}^2\mathbf{I})$ \citep[see][Section~2.4.2]{Zammit-MangionEtAl2022a}.
Correlated and uncorrelated errors in group $g$ are represented by the vectors $\xivec_g$ and $\epsilonvec_g$, respectively.
Their distributional properties are governed by the parameters $\gamma^Z_g$, $\rho^Z_g$, and $\ell^Z_g$, as described in Section~\ref{sec:parameter-model}.
Prior distributions on these parameters are retained from WOMBAT~v2.0:
for the overall variance scaling, we use $\gamma^Z_g \sim \textup{Gamma}(\nu_{\gamma},\, \omega_{\gamma})$;
for the relative-contribution proportion, we use $\rho^Z_g \sim \textup{Uniform}(0, 1)$;
and for the temporal length scales, we use $\ell^Z_g \sim \textup{Gamma}(\nu_{\ell},\, \omega_{\ell,g})$, with $g \in \mathcal{G}$.
Values of the fixed hyperparameters are chosen to make these priors reasonably noninformative, as explained in Section~\ref{supp:hyperparameters}.

\section{SIF--GPP sensitivity}
\label{supp:sif-gpp-sensitivity}

To model the SIF--GPP relationship using \eqref{eq:sif-process-model}, we require an estimate $\hat{\eta}_1(\cdot\,, \cdot)$ of $\eta_1(\cdot\,, \cdot)$, the sensitivity of SIF to GPP.
The SiB4 biosphere model described in Section~\ref{supp:bottom-up} produces bottom-up estimates of GPP, $X_{\textup{gpp}}^0(\cdot\,, \cdot)$, and SIF, $Y_{\textup{sif}}^0(\cdot\,, \cdot)$, which we obtain at a \qtyproduct{1 x 1}{\degree}, hourly resolution.
To align with the spatial resolution of the transport model ($\hat{\mathcal{H}}_{\textup{co}2}$; see Section~\ref{supp:transport-model}), we aggregate these estimates to \qtyproduct{2 x 2.5}{\degree} (latitude--longitude) using a conservative remapping algorithm \citepsupp{HankeEtAl2016}.
From the 24 daily values in each \qtyproduct{2 x 2.5}{\degree} grid cell, we extract estimates from the three-hour interval centered on 13:30 local time.
This window captures the majority of nonzero GPP and SIF values in SiB4 and coincides with the OCO-2 overpass time.
This process yields three (hourly) values of GPP and SIF per day for each grid cell.
The resulting data set comprises pairs $(X_{\textup{gpp}}^0(\svec, t), Y_{\textup{sif}}^0(\svec, t))$, where $\svec$ is the centroid of a \qtyproduct{2 x 2.5}{\degree} grid cell $A \subset \mathbb{S}^2$, and $t$ is the midpoint of a one-hour interval $T \subset \mathcal{T}$.
Through analysis of these pairs, we estimate $\eta_1(\cdot\,, \cdot)$ for each calendar month and \qtyproduct{2 x 2.5}{\degree} grid cell.
This means that a given grid cell uses the same estimate $\hat{\eta}_1(\cdot\,, \cdot)$ for all instances of a particular month; for example, the same value applies to every July.
The analysis proceeds as follows.

We begin by organizing the available pairs of bottom-up estimates for each grid cell into groups by calendar month across the six-year study period.
As discussed in Section~\ref{sec:sif-process-model}, $\hat{\mathcal{H}}_{\textup{sif}}$ represents the bottom-up SiB4 relationship between SIF and GPP, expressed as $Y_{\textup{sif}}^0(\svec, t) = \hat{\mathcal{H}}_{\textup{sif}}\left(X_{\textup{gpp}}^0(\svec, t);\; \svec, t\right) + v_{\textup{sif}}^0(\svec, t)$.
We approximate $\hat{\mathcal{H}}_{\textup{sif}}$ using a linear model based on the within-month variability of the bottom-up estimates in a given grid cell:
\begin{equation}
\label{eq:sif-gpp-regression}
 Y_{\textup{sif}}^0(\svec, t) = a(\svec, M(t))X_{\textup{gpp}}^0(\svec, t) + c(\svec, M(t)) + e(\svec, M(t)),
\end{equation}
where $M(t)$ denotes the calendar month containing time $t$.
The slope $a(\cdot\,, \cdot)$ represents the approximate sensitivity of SIF to GPP in SiB4.
The intercept $c(\cdot\,, \cdot)$ absorbs both the mean of $v_{\textup{sif}}^0(\cdot\,, \cdot)$ and the mean of any error in the linear approximation of SiB4, which may each be nonzero.
The term $e(\cdot\,, \cdot)$ captures the SIF--GPP model error, which we assume to be Gaussian distributed with mean zero.

For each grid cell and calendar month, we estimate $a(\cdot\,, \cdot)$ and $c(\cdot\,, \cdot)$ by ordinary least squares using the available pairs of bottom-up GPP and SIF estimates.
We then interpret the estimated slope $\hat{a}(\cdot\,, \cdot)$ as the approximate SIF--GPP sensitivity from \eqref{eq:sif-process-model}, setting $\hat{\eta}_1(\svec, t) = \hat{a}(\svec, M(t))$.
The mean-squared error of each fit is retained as a measure of SIF--GPP model-error variance, which is incorporated into the data model as described in Section~\ref{sec:parameter-model}.

To complete the model in \eqref{eq:sif-process-model}, we use $\hat{a}(\cdot\,, \cdot)$ to compute the sensitivity vector $\hat{\psivec}_{\textup{sif}}(\cdot, \cdot)$, which also depends on the vector of GPP basis functions $\phivec_{\textup{gpp}}(\cdot\,, \cdot)$.
For this calculation, we spatially aggregate $\phivec_{\textup{gpp}}(\cdot\,, \cdot)$ from the \qtyproduct{1 x 1}{\degree} SiB4 resolution to the \qtyproduct{2 x 2.5}{\degree} transport resolution using a conservative remapping algorithm.

Equation \eqref{eq:sif-gpp-regression} is a useful approximation for the bottom-up SIF--GPP relationship, but there are circumstances where either the assumption of linearity fails or where the relationship is too weak.
We judge this for a given grid cell and calendar month using four criteria described below, all of which must be satisfied.
If they are not, we do not assimilate any SIF observations for that combination of grid cell and calendar month.

For a given grid cell and calendar month, the four criteria we impose are as follows.
First, to ensure the presence of a SIF--GPP relationship, we require that the bottom-up estimates contain at least 30 SIF values greater than \qty{0.1}{W.m^{-2}.\micro\meter^{-1}.sr^{-1}}.
Second, the bottom-up SIF--GPP relationship is occasionally nonlinear, so we perform an analysis-of-variance test to compare the linear model to a cubic polynomial model.
To justify the linear approximation, we must fail to reject the null hypothesis that the linear model is a better fit than the cubic polynomial model.
Third, even when the previous two criteria are met, nonlinear or excessively weak relationships can still exist, and thus we require a correlation of at least 0.5 between the bottom-up GPP and SIF values.
While high correlation alone does not guarantee linearity, this threshold helps exclude severe forms of nonlinearity such as exponential relationships.
Fourth, we require that the intercept estimate $\hat{c}(\cdot\,, \cdot)$ is greater than \qty{-0.6}{W.m^{-2}.\micro\meter^{-1}.sr^{-1}}.
Although negative SIF values are nonphysical, a negative intercept may be needed to capture both the mean of $v_{\textup{sif}}^0(\cdot\,, \cdot)$, which influences SIF independently of GPP, and any linear approximation error.
However, we have seen empirically that bottom-up SIF--GPP relationships with an intercept below this threshold are consistently nonlinear.

\begin{figure}[t!]
    \centering
    \includegraphics{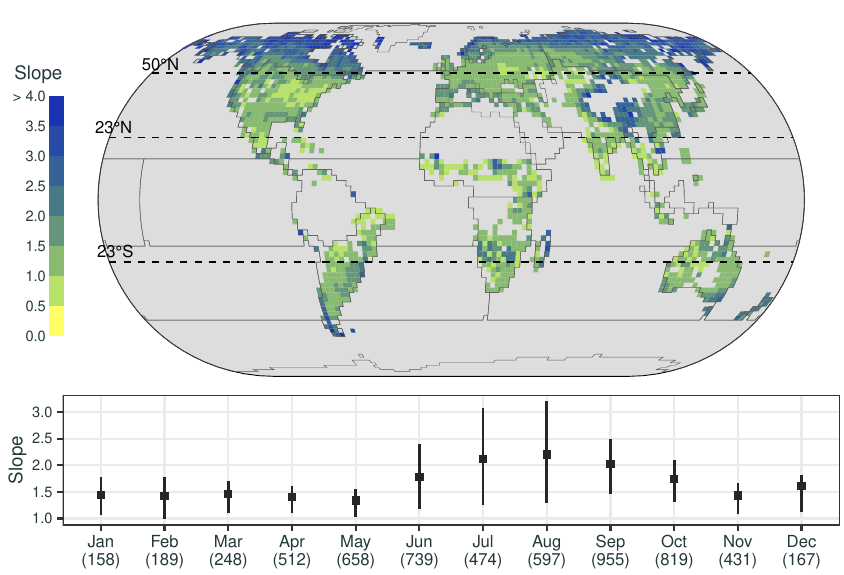}
    \caption{
        (Top panel) Map of estimated SIF--GPP sensitivities (the slope term in \eqref{eq:sif-gpp-regression}); each \qtyproduct{2 x 2.5}{\degree} grid cell shows the average over the 12 calendar-month slope estimates in that cell.
        The solid lines correspond to spatial-partition boundaries, and gray areas are missing values that correspond to ocean-only grid cells or where the four criteria for a valid linear model are not satisfied for any calendar month.
        (Bottom panel) The average and IQR for the set of global slope estimates in each calendar month; the total number of retained estimates in each set is given in parentheses below the month label.
        The slope units are (\unit{MW.m^{-2}.\micro\meter^{-1}.sr^{-1}}) / (\unit{kgCO_2.m^{-2}.s^{-1}}).
    }
    \label{fig:sif-gpp-slope-distribution}
\end{figure}

Figure~\ref{fig:sif-gpp-slope-distribution} shows summaries of the slope estimate $\hat{a}(\cdot\,, \cdot)$, including the seasonal average at each location and the global average for each month.
There is variability across the globe and throughout the season, with the largest sensitivities occurring spatially in high latitudes and temporally during the boreal summer growing season.
The spatial coverage is sparse in some regions, particularly in the tropics and at high latitudes, where the bottom-up SIF--GPP relationship tends to be nonlinear.
Maps of the exact slope and intercept terms, $\hat{a}(\cdot\,, \cdot)$ and $\hat{c}(\cdot\,, \cdot)$, are available for each calendar month in Section~\ref{supp:additional-figures}.
There, it is easier to see that coverage is sparse for winter months when GPP and SIF values are mostly zero.

\begin{figure}[t!]
    \centering
    \includegraphics{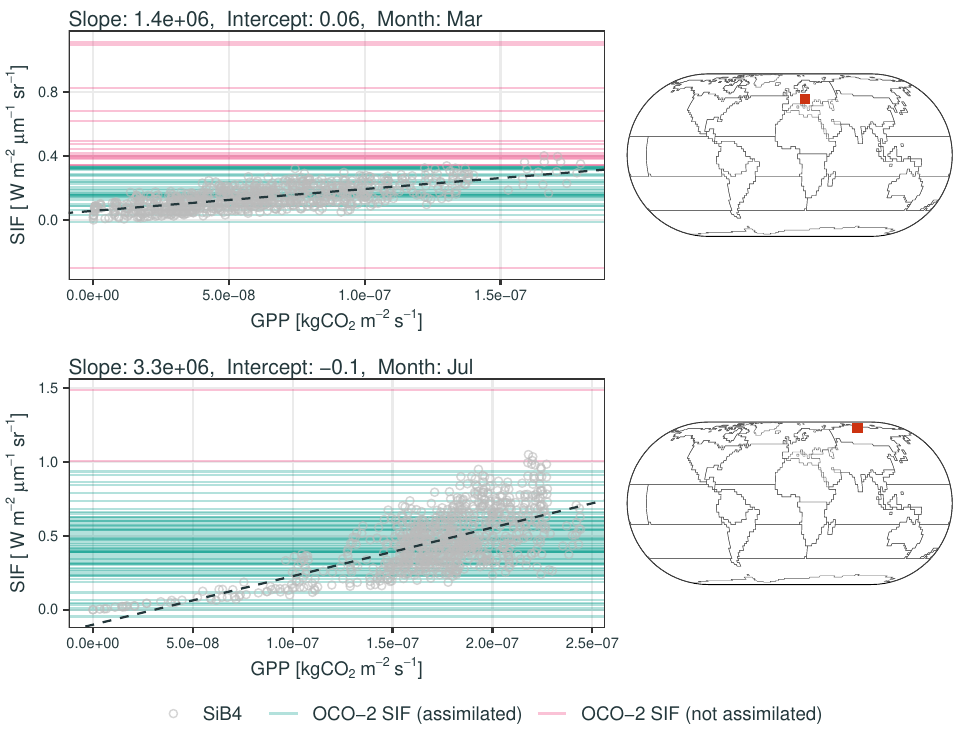}
    \caption{
        Bottom-up SIF--GPP relationship and corresponding OCO-2 SIF observations for two combinations of grid cell and calendar month.
        Points represent hourly bottom-up (SiB4) estimates of GPP and SIF, with the dashed line showing the fitted relationship.
        Solid lines show OCO-2 SIF observations, with assimilated observations in green and outlier observations in pink.
        The grid-cell location is indicated by the square on the adjacent map.
    }
    \label{fig:sif-gpp-model-examples}
\end{figure}

The four criteria above pertain to our linear approximation of the SIF--GPP relationship in SiB4.
Additionally, we observe that some OCO-2 SIF observations deviate significantly from SiB4 SIF values, falling outside the valid range for the approximate SIF--GPP relationship.
To identify these outliers, we employ Tukey's fences \citepsupp{Tukey1977a} for each grid cell and calendar month using the corresponding SiB4 SIF estimates.
Specifically, we only assimilate OCO-2 SIF observations that fall within $[Q_1 - 1.5 \times \textup{IQR}, Q_3 + 1.5 \times \textup{IQR}]$, where $Q_1$ and $Q_3$ represent the first and third quartiles of the SiB4 SIF estimates, and IQR is the interquartile range between them.
Figure~\ref{fig:sif-gpp-model-examples} illustrates this for two combinations of grid cell and calendar month.
In preliminary investigations, we found that the identified outliers do not cluster in space or time, indicating that this classification scheme is generally robust to different vegetation types and seasons.

\section{Remaining details of the inversion setup}
\label{supp:application-details}

This section provides additional information on how WOMBAT~v2.S was configured for inferring natural carbon fluxes using SIF and CO\textsubscript{2} mole-fraction observations.

\subsection{Transport model}
\label{supp:transport-model}

The mole-fraction process model in Section~\ref{supp:mole-frac-process-model} depends on the transport of CO\textsubscript{2} throughout the atmosphere.
As an approximation of transport, we use the GEOS-Chem global 3-D chemical transport model, version 12.3.2 \citepsupp{BeyEtAl2001, Yantosca2019}, retaining the configuration and the initial-condition estimate from WOMBAT~v2.0 (see Section~\ref{supp:mole-frac-process-model}).
The model is driven by MERRA-2 meteorological fields \citepsupp{KosterEtAl2015} and is run offline with a transport time step of 10 minutes and a flux time step of 20 minutes.
For efficiency, we use GEOS-Chem at an aggregated spatial resolution of \qty{2}{\degree} latitude by \qty{2.5}{\degree} longitude (with half-size polar cells of \qtyproduct{1 x 2.5}{\degree}) and 47 vertical levels of geopotential height.

\subsection{Bottom-up estimates}
\label{supp:bottom-up}

We rely on bottom-up estimates to inform the individual component fluxes in \eqref{eq:flux-components} and to inform the SIF process in \eqref{eq:sif-process-model}.
For GPP and respiration components, we obtain bottom-up estimates from the Simple Biosphere Model Version 4 (SiB4), which also provides bottom-up estimates of SIF, as described below.
We obtain bottom-up estimates of ocean-air exchange from \citetsupp{LandschutzerEtAl2016}, also described below.
Other fluxes are assumed known and fixed to their bottom-up estimates.
These include fossil-fuel emissions \citepsupp{OdaMaksyutov2011,OdaEtAl2018,NassarEtAl2013}, which are prescribed by the v10 MIP, as well as biomass burning emissions \citepsupp[GFED4.1s;][]{VanDerWerfEtAl2017} and biofuel fluxes \citepsupp{YevichLogan2003}, which are retained from WOMBAT~v2.0.

SiB4 is a mechanistic land-surface model that simulates the biophysical and biogeochemical processes governing the exchange of energy, water, and carbon between the terrestrial biosphere and the atmosphere \citep{HaynesEtAl2019, HaynesEtAl2019a}.
Using detailed representations of meteorology, plant type, soil, and land cover, SiB4 simulates several features of the terrestrial biosphere, including the terrestrial carbon cycle and SIF.
The simulations we use were made offline and cover the period from January 1, 2000 to December 31, 2020, with a temporal resolution of 10 minutes on a \qtyproduct{0.5 x 0.5}{\degree} latitude-longitude grid.
In each grid cell, SiB4 simulates values for 15 plant functional types.
To form our bottom-up estimates of SIF, GPP, and respiration, we first aggregate these simulations to a \qtyproduct{1 x 1}{\degree} latitude-longitude grid and an hourly time resolution, then sum over the 15 plant functional types.

The bottom-up estimates of ocean-air exchange given by \citetsupp{LandschutzerEtAl2016} are derived from observations of CO\textsubscript{2} partial-pressure difference at the ocean-air boundary.
Ocean CO\textsubscript{2} flux is proportional to this difference, and to sea-ice coverage and wind speed, which are included as inputs to the model.
These estimates cover the period from January 1982 to December 2019, and they are available on a \qtyproduct{1 x 1}{\degree} latitude-longitude grid at a monthly time resolution.

Together, SiB4 and \citetsupp{LandschutzerEtAl2016} provide the bottom-up estimates of SIF and all natural component fluxes.
Each set of estimates terminates before the end of our inversion time period, so we extend them to cover the full period.
The bottom-up SIF estimates from SiB4 are available through December 2020, and we extend them by repeating the values from January to March 2020 for January to March 2021.
For the component fluxes, we derive bottom-up estimates of $\beta_{c,\cdot}^0(\cdot)$, $\beta_{c,\cdot,\cdot}^0(\cdot)$, and $\epsilon_c^0(\cdot\,, \cdot)$ by applying the time-series decomposition in \eqref{eq:time-decomposition} to the available bottom-up estimates for each component.
We then extend the estimates to cover the missing period by reusing the residual term, $\epsilon_c^0(\cdot\,, \cdot)$, from the corresponding times in the most recent available year.
The details of this procedure can be found in \citet[Section~3.2]{BertolacciEtAl2024}, which we follow by setting the number of harmonics to $K_c = 3$ for GPP and respiration fluxes, and $K_c = 2$ for ocean flux.
Using these estimates, we populate the elements of the corresponding basis-vector $\phivec_c(\cdot\,, \cdot)$ as described in Section~\ref{supp:basis-function-model}.

\subsection{Basis functions and their coefficients}
\label{supp:basis-function-setup}

\begin{figure}[b!]
  \centering
  \includegraphics{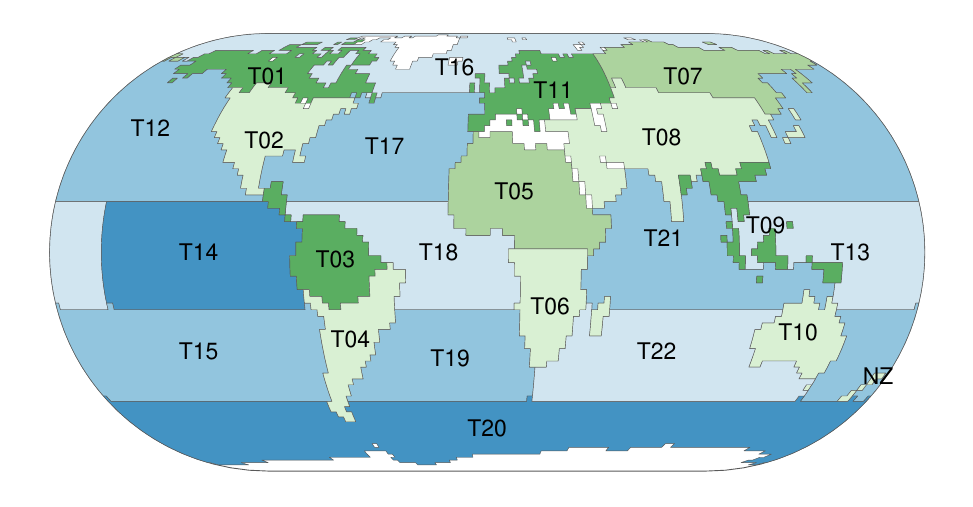}
  \caption{
    Map of the 22 TransCom3 regions (T01--T22) and the New Zealand region (NZ) used to create basis functions, adapted from \protect\citet{BertolacciEtAl2024}.
    Full names of these regions are given in Table~\ref{tab:transcom-regions}.
    White parts of the map correspond to areas assumed to have zero CO\textsubscript{2} surface flux.
  }
  \label{fig:transcom-regions}
\end{figure}

As in WOMBAT~v2.0, we specify the discrete spatiotemporal partition for the flux basis functions described in Section~\ref{sec:flux-process-model} by dividing Earth's surface into $R = 23$ regions and $Q = 79$ monthly time periods (September 2014 to March 2021, inclusive).
The spatial regions include the 22 TransCom3 regions \citep{GurneyEtAl2002}, plus a region containing the land area of New Zealand.
We show the spatial extents of these regions in Figure~\ref{fig:transcom-regions} and give their full names in Table~\ref{tab:transcom-regions}.
Half the TransCom3 regions are predominantly land regions, and half are predominantly ocean regions, but some region boundaries split coastal areas that contain both land and ocean.

\begin{table}[t!]
  \centering
  \begin{tabular}{lll||lll}
    \toprule
    Code & Name & Type & Code & Name & Type \\
    \midrule
    T01 & North American Boreal & Land & T12 & North Pacific Temperate & Ocean \\
    T02 & North American Temperate & Land & T13 & West Pacific Tropical & Ocean \\
    T03 & Tropical South America & Land & T14 & East Pacific Tropical & Ocean \\
    T04 & South American Temperate & Land & T15 & South Pacific Temperate & Ocean \\
    T05 & Northern Africa & Land & T16 & Northern Ocean & Ocean \\
    T06 & Southern Africa & Land & T17 & North Atlantic Temperate & Ocean \\
    T07 & Eurasia Boreal & Land & T18 & Atlantic Tropical & Ocean \\
    T08 & Eurasia Temperate & Land & T19 & South Atlantic Temperate & Ocean \\
    T09 & Tropical Asia & Land & T20 & Southern Ocean & Ocean \\
    T10 & Australia & Land & T21 & Indian Tropical & Ocean \\
    T11 & Europe & Land & T22 & South Indian Temperate & Ocean \\
    NZ & New Zealand & Land & & & \\
    \bottomrule
  \end{tabular}
  \caption{
    The code, name, and type of the 22 TransCom3 regions and the New Zealand region used to create basis functions.
    A map showing these regions is given in Figure~\ref{fig:transcom-regions}.
  }
  \label{tab:transcom-regions}
\end{table}

Land and ocean fluxes are commonly assumed to be uncorrelated in flux inversions \citep[e.g.,][]{BasuEtAl2013}.
We encode this assumption through our model for the basis-function coefficients, $\alphavec$, by setting the correlation between coefficients for the land basis functions and coefficients for the ocean basis functions to zero.
Specifically, $\rho^{\beta}_{\textup{gpp},\textup{ocean}} = \rho^{\beta}_{\textup{resp},\textup{ocean}} = \rho^{\epsilon}_{\textup{gpp},\textup{ocean}} = \rho^{\epsilon}_{\textup{resp},\textup{ocean}} = 0$ in \eqref{eq:seasonal-covariance} and \eqref{eq:residual-covariance}.

When attributing the land fluxes to GPP and respiration components, WOMBAT~v2.0 fixes the respiration linear term (RLT) to its bottom-up estimate.
As discussed in Section~\ref{sec:parameter-model}, the inclusion of SIF in WOMBAT~v2.S allows us to relax this assumption and infer the RLT in the inversion.
As an exception, in region T03 (Tropical South America) we keep the RLT fixed to its bottom-up estimate because nonlinearity in the bottom-up SIF--GPP relationship (see Figure~\ref{fig:sif-gpp-slope-distribution}) prevents us from assimilating SIF observations across most of this region (see Figure~\ref{fig:observation-count} in the main text).

As in WOMBAT~v2.0, we fix a number of ocean-related quantities to their bottom-up estimates.
This includes the linear and seasonal terms for GPP and respiration in the 11 ocean regions and the New Zealand region, as well as the linear and seasonal terms for the ocean fluxes in all regions.
Discussion of these choices and their implementation is given in Section~\ref{supp:ocean-prior} below.
After fixing these quantities, we make inference on the remaining 5,757 basis-function coefficients.
Importantly, residual terms are left to vary for all components, meaning that the GPP, respiration, and ocean fluxes are not fixed at any location or time.

\subsection{Physical constraints on GPP and respiration fluxes}
\label{supp:sign-constraints}

The constrained multivariate Gaussian prior on the basis-function coefficients, $\alphavec$, is partially specified through a constraint set, $F_{\alpha}$, which approximately enforces physical properties of GPP and respiration fluxes (see Section~\ref{sec:parameter-model}).
There are two types of constraints imposed by $F_{\alpha}$, and each corresponds to a different physical property.

The first type of constraint is a natural sign constraint specifying that GPP is always a sink, namely $X_{\textup{gpp}}(\cdot\,, \cdot) \leq 0$, and that respiration is always a source, namely $X_{\textup{resp}}(\cdot\,, \cdot) \geq 0$.
These sign constraints were applied in WOMBAT~v2.0 by aggregating $X_{\textup{gpp}}(\cdot\,, \cdot)$ and $X_{\textup{resp}}(\cdot\,, \cdot)$ onto a space-time grid with a spatial resolution matching the \qtyproduct{2 x 2.5}{\degree} transport-resolution grid cells and a monthly temporal resolution spanning the 79-month study period.
For WOMBAT~v2.S, we coarsen the spatial aggregation to the 23 basis-function regions (see Figure~\ref{fig:transcom-regions}).
This relaxes the constraint space and, as explained in Section~\ref{sec:parameter-model}, allows the GPP and respiration priors to remain close to their bottom-up estimates when the RLT is not fixed.
It is reasonable to constrain $\alphavec$ through these regional aggregates because $\alphavec$ adjusts flux estimates directly at the regional resolution.

Let $\xvec_{\textup{gpp}}$ and $\xvec_{\textup{resp}}$ be $QR$-dimensional vectors representing GPP and respiration values, respectively, aggregated to match the regional and monthly basis-function resolution.
For $c \in \{\textup{gpp}, \textup{resp}\}$, let $\xvec_c^0$ be the same aggregation applied to the bottom-up estimate.
Using the basis-function representation in \eqref{eq:basis-decomposition}, we note that
\begin{equation}
  \label{eq:flux-aggregation}
  \xvec_c = \xvec_c^0 + \Phivec_c\alphavec_c,
\end{equation}
where $\Phivec_c$ is a matrix whose rows are computed by applying the aggregation operation to the basis functions $\phivec_c(\cdot\,, \cdot)$.
Sign constraints for GPP and respiration components are enforced through the aggregated quantities as
\begin{equation}
  \label{eq:sign-constraints}
  \begin{aligned}
    \xvec_{\textup{gpp}}^0 + \Phivec_{\textup{gpp}}\alphavec_{\textup{gpp}} &\leq \zerovec, \\
    \xvec_{\textup{resp}}^0 + \Phivec_{\textup{resp}}\alphavec_{\textup{resp}} &\geq \zerovec,
  \end{aligned}
\end{equation}
where the inequalities are applied elementwise.
Though necessary, these constraints are not sufficient to ensure that GPP and respiration are physically plausible at fine resolutions.
That is, the inequalities can be violated at spatial or temporal scales finer than the aggregated basis-function resolution.
However, most results we report in this study are themselves spatial or temporal aggregates of the aggregation described above, and thus the constraints in \eqref{eq:sign-constraints} are sufficient for our reported GPP and respiration fluxes to satisfy the sign constraints.

The second type of constraint concerns the diurnal cycles of the GPP and respiration fluxes, which have a reliable phase driven by local solar time.
As discussed in Section~\ref{sec:flux-process-model}, the diurnal cycle appears in the residual term $\epsilon_c(\cdot\,, \cdot)$.
If left unconstrained, the parameterization of the corresponding basis-function coefficients, $\alphavec_{c,6}$ in \eqref{eq:coefficient-scaling}, allows the diurnal cycle to be flipped within a given month and region, implying a nonphysical 12-hour phase shift in the cycle.
To avoid this flip, we follow WOMBAT~v2.0 and impose the elementwise constraint that $\alphavec_{c,6} \geq -\onevec$ for $c \in \{\textup{gpp}, \textup{resp}\}$.

Since the two types of constraints are both linear, they can be jointly represented in matrix-vector form as $\dvec + \Phivec\alphavec \geq \zerovec$.
There are a total of 7,268 constraints in WOMBAT~v2.S (23 regions $\times$ 79 months $\times$ 2 constrained flux components $\times$ 2 constraint types), which is about \qty{1}{\percent} of the number of constraints used in WOMBAT~v2.0.

\subsection{Reparameterization for linear basis-function coefficients}
\label{supp:reparameterization}

Relaxing the fixed respiration linear term (RLT) assumption from WOMBAT~v2.0 revealed an issue with the priors on the intercepts, $\beta_{c,0}(\cdot)$, and trends, $\beta_{c,1}(\cdot)$, in the time-series decomposition given by \eqref{eq:time-decomposition} for the GPP and respiration components.
The intercept and trend determine the linear term of the time-series decomposition which, from \eqref{eq:time-decomposition} and \eqref{eq:coefficient-scaling}, is expressed as
\begin{equation}
  \label{eq:linear-term}
  (1 + \alpha_{c,0,r})\beta_{c,0}^0(\svec) + (1 +\alpha_{c,1,r})\beta_{c,1}^0(\svec) t, \quad c \in \mathcal{C},\ \svec \in D_r,\ t \in \mathcal{T},
\end{equation}
where $D_r \subset \mathbb{S}^2$ is a basis-function region (see Figure~\ref{fig:transcom-regions}).
In WOMBAT~v2.0, the RLT is fixed to its bottom-up estimate by setting $\alpha_{\textup{resp},j,r} = 0$ for $j = 0, 1$ across all regions $r = 1, \ldots, 23$.
This also forces the GPP linear term to remain close to its bottom-up estimate due to the NEE relationship in \eqref{eq:nee-definition}, which is constrained by CO\textsubscript{2} mole-fraction observations.

When we relax the fixed RLT assumption in WOMBAT~v2.S, an issue arises due to the temporal reference point of the bottom-up estimates $\beta_{c,0}^0(\cdot)$ and $\beta_{c,1}^0(\cdot)$.
These estimates derive from a 20-year period that includes the six-year inversion period at its end (see Section~\ref{supp:bottom-up}), with time $t$ given in days since January 1, 2000.
This reference point, being 17 years before the middle of the inversion period, complicates the specification of interpretable prior variances for both intercepts and trends.

The prior on $\beta_{c,j}(\svec)  = (1 + \alpha_{c,j,r})\beta_{c,j}^0(\svec)$ is determined through the prior on $\alpha_{c,j,r}$.
In WOMBAT~v2.0, the prior variances of $\alpha_{c,0,r}$ and $\alpha_{c,1,r}$ are too small, which restricts the effective range of the linear term during the inversion period compared to what would be expected from a line pivoting around the middle of the inversion period.
The resulting prior distribution on the linear term forms a cone that widens as time progresses across the inversion period, rather than expanding symmetrically around the midpoint.

To address this, we reparameterize the elements of $\alphavec$ associated with the GPP and respiration linear terms.
This reparameterization introduces an analogous vector $\alphavec^*$, defined in a target space where the corresponding linear terms pivot around the middle day of the inversion period, $t_m  \in \mathcal{T}$, making their scales more interpretable.
We map the prior on $\alphavec^*$ to an updated prior on $\alphavec$ through a transformation matrix $\Pvec$, as $\alphavec = \Pvec\alphavec^*$.
The target-space prior on $\alphavec^*$ is specified to yield appropriate prior variances when mapped to the original space.
In what follows, we detail the construction of $\Pvec$ and then the prior for $\alphavec^*$.

We reparameterize only the basis-function coefficients associated with GPP and respiration linear terms in land regions, specifically the subset $\{\alpha_{c,j,r} : c = \textup{gpp}, \textup{resp};\; j = 0, 1;\; r = 1, \ldots, 11\}$.
Because $\alpha_{c,j,r}$ is defined at the basis-function resolution, it adjusts $\beta_{c,j}^0(\svec)$ by the same amount at every location $\svec \in D_r$.
Therefore, we specify the reparameterization in terms of regional aggregates of $\beta_{c,j}^0(\cdot)$, which are computed as
\begin{equation}
  \label{eq:coefficient-aggregation}
  B_{c,j,r}^0 \equiv \sum_{\svec \in D_r} w(\svec)\beta_{c,j}^0(\svec),
  \quad c \in \{\textup{gpp}, \textup{resp}\},\ j = 0, 1,\ r = 1, \ldots, 11,
\end{equation}
where $D_r \subset \mathbb{S}^2$ is a terrestrial basis-function region, and $w(\svec)$ is a weighting for the area of the \qtyproduct{2 x 2.5}{\degree} grid cell centered at $\svec$.

At the regional scale, the linear term in \eqref{eq:linear-term} can be expressed as
\begin{equation}
  \label{eq:linear-term-aggregated}
  B_{c,0,r}^0 + \alpha_{c,0,r}B_{c,0,r}^0 + B_{c,1,r}^0 t + \alpha_{c,1,r}B_{c,1,r}^0 t, 
  \quad c \in \{\textup{gpp}, \textup{resp}\},\ r = 1, \ldots, 11.
\end{equation}
The reparameterization transforms the coefficients $\alpha_{c,0,r}$ and $\alpha_{c,1,r}$ from a target space in which the analogous coefficients $\alpha_{c,0,r}^*$ and $\alpha_{c,1,r}^*$ adjust a line that pivots around $t_m$.
That is, in the target space, the linear term is given by
\begin{equation}
  \label{eq:linear-term-target}
  B_{c,0,r}^0 + \alpha_{c,0,r}^* (B_{c,0,r}^0 + B_{c,1,r}^0 t_m) +
  B_{c,1,r}^0 t + \alpha_{c,1,r}^* B_{c,1,r}^0(t - t_m),
  \quad c \in \{\textup{gpp}, \textup{resp}\},\ r = 1, \ldots, 11,
\end{equation}
and thus $\alpha_{c,0,r}^*$ and $\alpha_{c,1,r}^*$ adjust the bottom-up estimate of the linear term through a line that pivots around the point $(t_m,\, B_{c,0,r}^0 + B_{c,1,r}^0 t_m)$.

We obtain the original linear term in \eqref{eq:linear-term-aggregated} from the target space in \eqref{eq:linear-term-target} through the following relationships.
For $c \in \{\textup{gpp}, \textup{resp}\}$ and $r = 1, \ldots, 11$,
\begin{equation}
  \label{eq:transformation-equations}
  \begin{aligned}
    \alpha_{c,0,r} &= \alpha_{c,0,r}^* \left(1 + B_{c,1,r}^0 t_m / B_{c,0,r}^0\right) - \alpha_{c,1,r}^* B_{c,1,r}^0 t_m / B_{c,0,r}^0, \\
    \alpha_{c,1,r} &= \alpha_{c,1,r}^*,
  \end{aligned}
\end{equation}
which can be represented in matrix-vector form as
\begin{equation}
  \label{eq:transformation-regional}
  \begin{pmatrix}
    \alpha_{c,0,r} \\
    \alpha_{c,1,r}
  \end{pmatrix} =
  \begin{pmatrix}
    1 + B_{c,1,r}^0 t_m/B_{c,0,r}^0 & -B_{c,1,r}^0 t_m/B_{c,0,r}^0 \\
    0 & 1
  \end{pmatrix}
  \begin{pmatrix}
    \alpha_{c,0,r}^* \\
    \alpha_{c,1,r}^*
  \end{pmatrix}.
\end{equation}
The $2 \times 2$ matrix in \eqref{eq:transformation-regional} defines the submatrices of $\Pvec$ that transform elements of $\alphavec$ corresponding to the linear terms of GPP and respiration in each land region.
For all other elements of $\alphavec$, the corresponding submatrices of $\Pvec$ are identity matrices.

In the target space, we model the basis-function coefficients as $\alphavec^* \sim \textup{Gau}(\zerovec,\, \Sigmavec_{\alpha}^*)$.
The submatrices of $\Sigmavec_{\alpha}^*$ corresponding to seasonal and residual terms remain identical to their counterparts in $\Sigmavec_{\alpha}$ (specified in Section~\ref{supp:alpha-covariance}).
The submatrices for linear terms have unit diagonal elements, except for elements corresponding to GPP and respiration trends.
For these trend-related elements, we found empirically that setting $\var(\alpha^*_{c,1,r}) = 10{,}000$ (no units), for $c \in \{\textup{gpp}, \textup{resp}\}$ and $r = 1, \ldots, 11$, provides appropriate flexibility when mapped to the original space.
We map this Gaussian prior from the target space to a constrained Gaussian prior in the original space through
\begin{equation}
  \label{eq:transformation}
  \alphavec \sim \textup{ConstrGau}(\zerovec,\, \Pvec\Sigmavec_{\alpha}^*\Pvec',\, F_{\alpha}),
\end{equation}
which affects only the elements of $\alphavec$ corresponding to the GPP and respiration linear terms; elements related to seasonal and residual terms of GPP and respiration retain their zero mean and original covariance structure.
The entire ocean flux decomposition is also unaffected.

\subsection{Additional information to improve ocean flux estimation}
\label{supp:ocean-prior}

In each of the 23 basis-function regions, we treat the linear and seasonal terms for ocean flux as known and fix them to their bottom-up estimates.
This choice is retained from WOMBAT~v2.0 since these terms cannot be reliably constrained by the observations, which are mainly available over land.
Similarly, in the 11 predominantly ocean regions and the New Zealand region, the linear and seasonal terms of natural land component fluxes (i.e., GPP and respiration) are fixed to their bottom-up estimates.
As explained in \citet{BertolacciEtAl2024}, these regions have small land areas, and hence their land fluxes are weakly identifiable.
Holding the linear and seasonal terms fixed to their bottom-up estimates in these regions, inference is made through the residual term.

The residual term for ocean flux is also difficult to constrain, owing to the dearth of ocean observations.
WOMBAT~v2.0 uses the parameterization of the CAMS inversion framework \citep[adapted for the OCO-2 MIP protocol in][]{CrowellEtAl2019} to specify $\Sigmavec_{\alpha, \textup{ocean}}^{\epsilon}$, a submatrix of $\Sigmavec_{\alpha}$ that helps inform the basis-function coefficients for ocean residual fluxes.
See Section~S.2.2 in the Supplementary Material of \citet{BertolacciEtAl2024} for details.
In WOMBAT~v2.S, we retain the v2.0 specification of $\Sigmavec_{\alpha, \textup{ocean}}^{\epsilon}$, but we inflate each element by a factor of 10.
This relaxes the ocean prior, which we found necessary for appropriate coverage of the true ocean residual term during initial simulation testing in the OSSE of Section~\ref{sec:osse}.

\subsection{\texorpdfstring{CO\textsubscript{2} mole-fraction observations}{CO2 mole-fraction observations}}
\label{supp:mole-frac-observations}

The v10 MIP prescribes CO\textsubscript{2} mole-fraction observations from the OCO-2 satellite and from in~situ and flask sources.
Onboard OCO-2 are three spectrometers that measure sunlight reflected from Earth's surface in three different spectral bands.
These measurements are input to an optimal estimation algorithm to retrieve CO\textsubscript{2} mole fraction at 20 vertical levels \citep{CrispEtAl2017,ElderingEtAl2017a}.
The retrieved levels are then column-averaged to produce XCO\textsubscript{2} (column-average CO\textsubscript{2}), which is later bias-corrected using external validation data \citep{ODellEtAl2018}.
The XCO\textsubscript{2} retrievals have the same spatial and temporal resolution as the SIF retrievals described in Section~\ref{sec:data}.
While retrievals are made over land and ocean, only land nadir and land glint retrievals are used in the v10 MIP; note that ocean fluxes can still be inferred from land retrievals due to atmospheric transport.
Further, the MIP protocol stipulates the use of ``good'' retrievals (quality flag 0) that are post-processed into 10-second averages along orbit tracks; see \citet{ByrneEtAl2023} for details of the post-processing procedure.
The resulting XCO\textsubscript{2} observations may not represent the average of the coarse 3-D grid-cells used in transport models, and thus they are accompanied by estimated transport-model error variances \citep[based on][]{SchuhEtAl2019}.
They are also accompanied by estimated observation-error variances and, following the convention of the v10 MIP, we add both types of error variance to construct an error budget for each observation.
The final 530,201 XCO\textsubscript{2} observations span the period from September 6, 2014 to March 31, 2021.

For in~situ and flask observations, the v10 MIP prescribes three data sets in the ObsPack format \citepsupp{MasarieEtAl2014}.
The primary data set is the CO\textsubscript{2} GLOBALVIEWplus v6.1 ObsPack \citepsupp{SchuldtEtAl2021a}, and additional data sets include the CO\textsubscript{2} NRT v6.1.1 ObsPack \citepsupp{SchuldtEtAl2021} and the CO\textsubscript{2} NIES Shipboard v3.0 ObsPack \citepsupp{TohjimaEtAl2005,NaraEtAl2017}.
In total, there are 1,053,457 in~situ and flask CO\textsubscript{2} mole-fraction observations designated for assimilation, with 15,569 from aircraft instruments, 218,256 from shipboard instruments, 407,479 from surface instruments, and 412,153 from tower instruments.
Since these observations have well-known error characteristics, their error budgets are based on estimated transport-model error variances only.
The designated observations typically represent sub-daily averages, and they provide coverage from September 1, 2014 to January 31, 2021.

The two types of CO\textsubscript{2} mole-fraction observations have different strengths and weaknesses.
In~situ and flask observations are made at a specific location and altitude, so their accompanying transport-model error variances tend to be larger than those for XCO\textsubscript{2} observations, which have a broader spatial footprint and represent an average over the atmospheric column.
Conversely, in~situ and flask observations are generally considered to be unbiased and to have minimal measurement error, whereas XCO\textsubscript{2} observations are more affected by these error sources.
As shown in Figure~\ref{fig:observation-count} in the main text, in~situ and flask observations are highly abundant in North America, Western Europe, and parts of the Pacific Ocean, however they are mostly absent in other regions.
The XCO\textsubscript{2} observations have global land coverage, although they are less abundant in predominantly cloudy regions such as parts of the tropics, and in high latitudes during winter.

\subsection{Observation groups and error properties}
\label{supp:error-properties}

As mentioned in Section~\ref{sec:data-model}, WOMBAT splits the available observations into groups, $g \in \mathcal{G}$, based on common biases and error properties.
In WOMBAT~v2.S, we retain the five groups from WOMBAT~v2.0 and introduce a sixth group for SIF observations.
The original CO\textsubscript{2} mole-fraction groups are (1) aircraft observations, (2) shipboard observations, (3) surface observations, (4) tower observations, and (5) OCO-2 land XCO\textsubscript{2} observations.
The new group is (6) OCO-2 land SIF observations.
Aircraft and shipboard observations are treated separately because they are from instruments that move at different speeds, while surface and tower observations are separated because they are from stationary instruments that make readings at different heights.
OCO-2 XCO\textsubscript{2} and SIF observations are different groups because they are retrieved by independent algorithms and because the OCO-2 instrument is very different from the in~situ and flask instruments.
Within a group $g$, all observations share the parameters $\pivec_g$, $\gamma^Z_g$, $\rho^Z_g$, and $\ell^Z_g$, and therefore they are assumed to have identical biases and error properties.

Biases are generally considered absent in the in~situ and flask observations prescribed by the v10 MIP, as discussed in Section~\ref{supp:mole-frac-observations} above.
The prescribed OCO-2 XCO\textsubscript{2} observations and the version 10r OCO-2 SIF observations that we use are both bias-corrected offline \citep[see][respectively]{ODellEtAl2018, DoughtyEtAl2022}.
In WOMBAT~v1.0, \citet{Zammit-MangionEtAl2022a} found that estimating additional biases in bias-corrected observations made little difference to flux estimates.
Hence, although the WOMBAT framework is capable of handling the raw, uncorrected OCO-2 observations and estimating the biases, we follow \citet{BertolacciEtAl2024} and omit the bias coefficients $\pivec_g$, setting the bias term to $b_{g,i} = 0$ for all observations $i = 1, \ldots, N_g$ in all groups $\{g \in \mathcal{G}\}$.

Recall from Section~\ref{sec:parameter-model} that accompanying each observation is an estimated error variance that we term the error budget.
We partition the error associated with an error budget into a correlated error term, $\xivec_g$, and an uncorrelated error term, $\epsilonvec_g$, using the parameters $\{\rho^Z_g : g \in \mathcal{G}\}$.
For the in~situ and flask observation groups, we assume that $\rho^Z_g = 1$, which sets the variance of the uncorrelated error term to zero.
This is because these observations have minimal measurement error, but they can still be affected by correlated errors due to misrepresentation in the coarse-resolution transport model.
For the OCO-2 XCO\textsubscript{2} and SIF observation groups, we treat $\rho^Z_g$ as unknown and nonnegative.

To specify the covariance of the correlated error terms $\{\xivec_g : g \in \mathcal{G}\}$, the observations from each group are further subdivided into separate time series according to instrument or location.
The observations from a single ship or a single aircraft constitute a time series, the observations from a single surface site or a single tower also constitute a time series, and the OCO-2 XCO\textsubscript{2} and SIF observations constitute a pair of time series.
Due to differences in their retrieval algorithms and the four criteria for SIF observations to be assimilated (see Section~\ref{supp:sif-gpp-sensitivity}), the XCO\textsubscript{2} and SIF observations are rarely available simultaneously.
We therefore assume that their observation errors are uncorrelated and treat them as separate time series.
Let $\xivec_{g,s}$, $s = 1, \ldots, S_g$, be the vector containing the elements of $\xivec_g$ that belong to time series $s$ in group $g$, where $S_g$ is the number of time series in group $g$.
WOMBAT models the within-series correlation for $\xivec_{g,s}$ using a stationary exponential function of temporal lag, with $e$-folding timescale $\ell^Z_g$.
The elements of different time series are assumed to be mutually independent, so all other correlations between elements of $\{\xivec_g : g \in \mathcal{G}\}$ are set to zero.

\subsection{Hyperparameters}
\label{supp:hyperparameters}

Hyperpriors on the covariance parameters for the basis-function coefficients (see Section~\ref{supp:alpha-covariance}) are retained from WOMBAT~v2.0.
For the parameters $\tau^{\beta}_c$ and $\tau^{\epsilon}_c$, we use hyperparameters $\nu^{\beta}_{\tau} = \nu^{\epsilon}_{\tau} = 0.35428$ and $\omega^{\beta}_{\tau,c} = \omega^{\epsilon}_{\tau,c} = 0.01534$, which are standard choices in the WOMBAT framework since they construct reasonably noninformative hyperpriors \citep{Zammit-MangionEtAl2022a}.
For $\rho^{\beta}_{c,c'}$, $\rho^{\epsilon}_{c,c'}$, and $\kappa^{\epsilon}_c$, we use hyperparameters $a^{\beta}_{\rho} = b^{\beta}_{\rho} = a^{\epsilon}_{\rho} = b^{\epsilon}_{\rho} = a^{\epsilon}_{\kappa} = b^{\epsilon}_{\kappa} = 1$, which yields uniform hyperpriors over $[0, 1]$.

Priors on data-model parameters are described in Section~\ref{supp:parameter-model}.
For all observation groups, $\{g \in \mathcal{G}\}$, we specify a reasonably noninformative prior on $\gamma^Z_g$ using parameters $\nu_{\gamma} = 1.62702$ and $\omega_{\gamma} = 2.17124$ \citep[see][for justification of these choices]{Zammit-MangionEtAl2022a}.
For the length-scale parameters $\{\ell^Z_g: g \in \mathcal{G}\}$, we let $\nu_{\ell} = 1$ and set $\omega_{\ell,g} = \qty{1}{day^{-1}}$ for in~situ and flask groups, and $\omega_{\ell,g} = \qty{1}{min^{-1}}$ for OCO-2 XCO\textsubscript{2} and SIF groups.

\subsection{Computation}
\label{supp:computation}

\begin{figure}[ht!]
  \centering
  \includegraphics{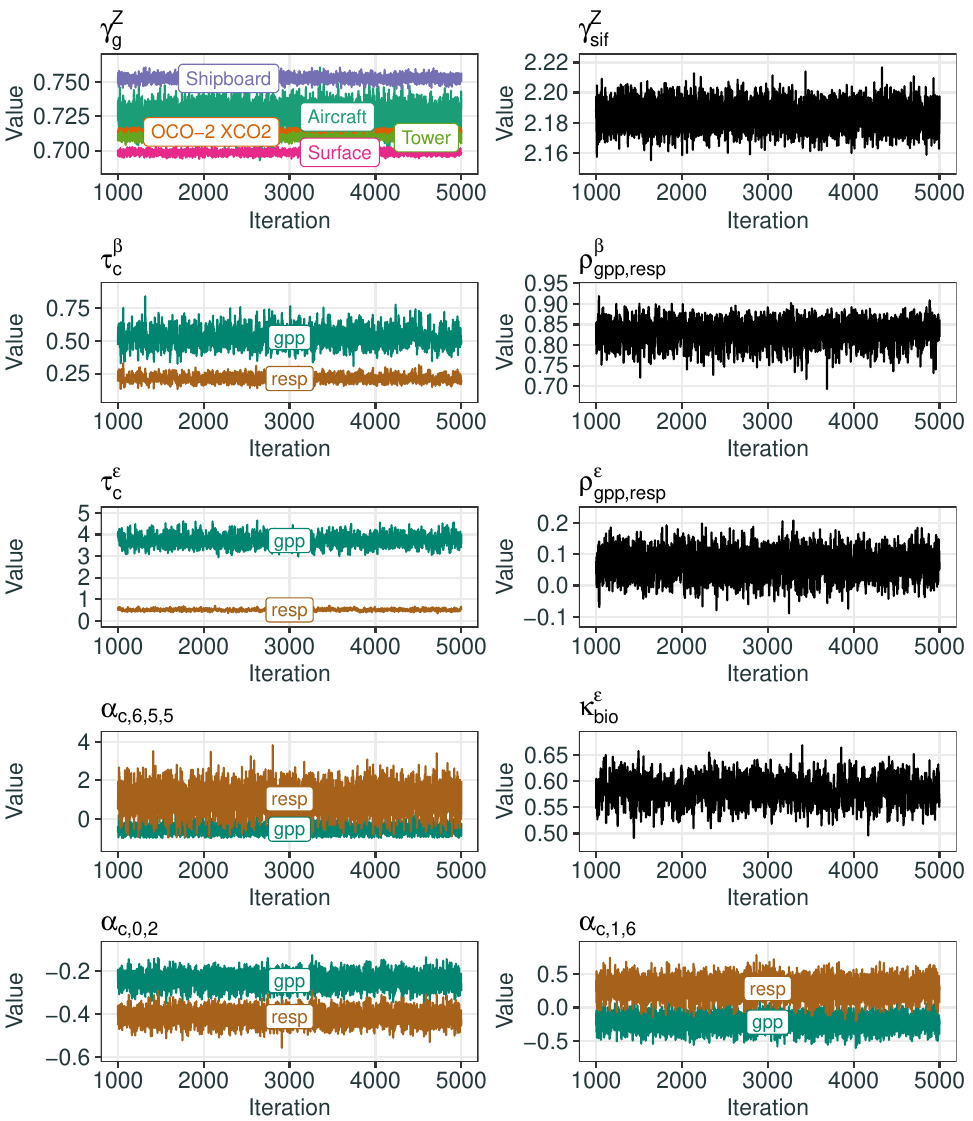}
  \caption{
    MCMC trace plots for the variance scaling factors $\{\gamma^Z_g : g \in \mathcal{G}\}$, for the covariance parameters of $\alphavec$, and for six elements of $\alphavec$, following the 1,000-iteration warm-up.
    For reference, the bottom-left panel shows the intercept coefficient for GPP and respiration components in region T02, and the bottom-right panel shows the trend coefficient for GPP and respiration in region T06 (see Table~\ref{tab:transcom-regions}).
  }
  \label{fig:traceplots}
\end{figure}

Obtaining the basis functions was the most computationally demanding step and took several weeks on the Gadi supercomputer at the Australian National Computational Infrastructure.
Fortunately, the basis functions are shared between the different inversions that we perform in this study, so they only need to be computed once.
All other inversion steps were performed on a high-performance computing machine with 56 computing cores.
The first-stage inversion to obtain estimates of $\{\rho^Z_g\}$ and $\{\ell^Z_g\}$ took roughly two hours and, once computed, these estimates were fixed for the subsequent inversions.
We performed each OSSE inversion in Section~\ref{sec:osse} by running the MCMC sampler for 1,000 iterations, 200 of which were discarded as warm-up iterations.
For the real-data inversion in Section~\ref{sec:results}, we ran the sampler for 5,000 iterations and discarded the first 1,000.
Each OSSE inversion took roughly four hours, and the real-data inversion took roughly 43 hours.
In the real-data inversion, the effective sample size \citepsupp{GelmanEtAl2013} is greater than 600 for all parameters and greater than 2,000 for \qty{95}{\percent} of the parameters.
Figure~\ref{fig:traceplots} shows trace plots for a subset of the parameters from the real-data inversion, including: the data-model parameter $\{\gamma^Z_g\}$; all basis-function covariance parameters described in Section~\ref{supp:alpha-covariance}; and six basis-function coefficients.

\section{OSSE fluxes and observations}
\label{supp:alpha-setup}

Throughout the OSSE described in Section~\ref{sec:osse}, we consider four true-flux cases, with each case defined by the choice of flux basis-function coefficients, $\alphavec$.
These cases are intended to represent plausible flux scenarios, allowing us to test the performance of WOMBAT~v2.S under various inversion setups.
The first true-flux case sets $\alphavec = \zerovec$ and the second true-flux case sets $\alphavec = \tilde{\alphavec}^{(\textup{v}2)}$, the posterior mean obtained from the WOMBAT~v2.0 inversion in \citet{BertolacciEtAl2024}.
In both these initial cases, the true respiration linear term (RLT) is equal to its bottom-up estimate, with respiration coefficients $\alpha_{\textup{resp},j,r}$ set to zero for both intercept ($j = 0$) and trend ($j = 1$) terms in all regions ($r = 1, \ldots, 23$).
The third and fourth true-flux cases explore scenarios where the true RLT intentionally deviates from its bottom-up estimate, which we achieve by modifying the WOMBAT~v2.0 posterior mean.
The modification involves adjusting a subset of the coefficients in $\tilde{\alphavec}^{(\textup{v}2)}$ to shift the RLT from its bottom-up estimate by a small fraction, but without changing the true NEE linear term.

The third and fourth true-flux cases are constructed as follows.
For $j = 0, 1$ and $r = 1, \ldots, 11$, let $B_{\textup{gpp},j,r}^0$ be the area-weighted sum of $\beta_{\textup{gpp},j}^0(\cdot)$ over every \qtyproduct{2 x 2.5}{\degree} grid cell in region $r$, and let $B_{\textup{resp},j,r}^0$ be defined similarly; see also \eqref{eq:coefficient-aggregation}.
The NEE linear term consists of an intercept term ($j = 0$) and a trend term ($j = 1$), and each is the sum of the corresponding GPP and respiration terms.
At the regionally aggregated resolution, the WOMBAT~v2.0 posterior mean of each term is given by
\begin{equation}
  \label{eq:nee-balance}
  \begin{aligned}
    \tilde{B}_{\textup{nee},j,r}^{(\textup{v}2)}
    &= (1 +\tilde{\alpha}_{\textup{gpp},j,r}^{(\textup{v}2)})B_{\textup{gpp},j,r}^0
      + B_{\textup{resp},j,r}^0 \\
    &= (1 + \dot{\alpha}_{\textup{gpp},j,r})B_{\textup{gpp},j,r}^0
    + (1 + \dot{\alpha}_{\textup{resp},j,r})B_{\textup{resp},j,r}^0\,,
  \end{aligned}
\end{equation}
for $j = 0, 1$, and $r = 1, \ldots, 11$.
To ensure plausible NEE values in each region, we select both $\dot{\alpha}_{\textup{gpp},j,r}$ and $\dot{\alpha}_{\textup{resp},j,r}$ to retain the NEE linear term from the WOMBAT~v2.0 posterior mean.
We first specify $\dot{\alpha}_{\textup{gpp},j,r}$ by adding a small fraction to $\tilde{\alpha}_{\textup{gpp},j,r}^{(\textup{v}2)}$, which proportionally shifts $B_{\textup{gpp},j,r}^0$.
Then, using \eqref{eq:nee-balance}, we determine the corresponding value of $\dot{\alpha}_{\textup{resp},j,r}$ that preserves $\tilde{B}_{\textup{nee},j,r}^{(\textup{v}2)}$ within each region.
Specifically, we define
\begin{equation}
  \label{eq:rlt-shift}
  \dot{\alpha}_{\textup{gpp},j,r} \equiv \tilde{\alpha}_{\textup{gpp},j,r}^{(\textup{v}2)} + \delta
  \quad \text{and} \quad
  \dot{\alpha}_{\textup{resp},j,r} \equiv -\delta B_{\textup{gpp},j,r}^0 / B_{\textup{resp},j,r}^0,
\end{equation}
for $j = 0, 1$, $r = 1, \ldots, 11$, and where $\delta \in \mathbb{R}$ is a fractional shift.

The third and fourth true-flux cases represent \qty{10}{\percent} shifts in $B_{\textup{gpp},j,r}^0$, implemented using $\delta = 0.1$ and $\delta = -0.1$, respectively.
As a result, the nonzero values $\{\dot{\alpha}_{\textup{resp},j,r} : j = 0, 1;\, r = 1, \ldots, 11\}$ perturb the RLT from its bottom-up estimate across all terrestrial basis-function regions, with one exception.
In Tropical South America ($r = 3$), we set $\delta = 0$ to maintain the RLT at its bottom-up estimate; the RLT is not inferred for this region in the inversion due to our inability to assimilate sufficient SIF observations there (see Section~\ref{supp:basis-function-setup}).
Hence, $\dot{\alpha}_{\textup{resp},j,3}$ remains zero for both intercept and trend terms ($j = 0,1$) in Tropical South America only.
This treatment ensures that using a fixed RLT for this region in the inversion does not confound our analysis of intentional RLT perturbations in other regions.

\begin{figure}[t!]
  \centering
  \includegraphics{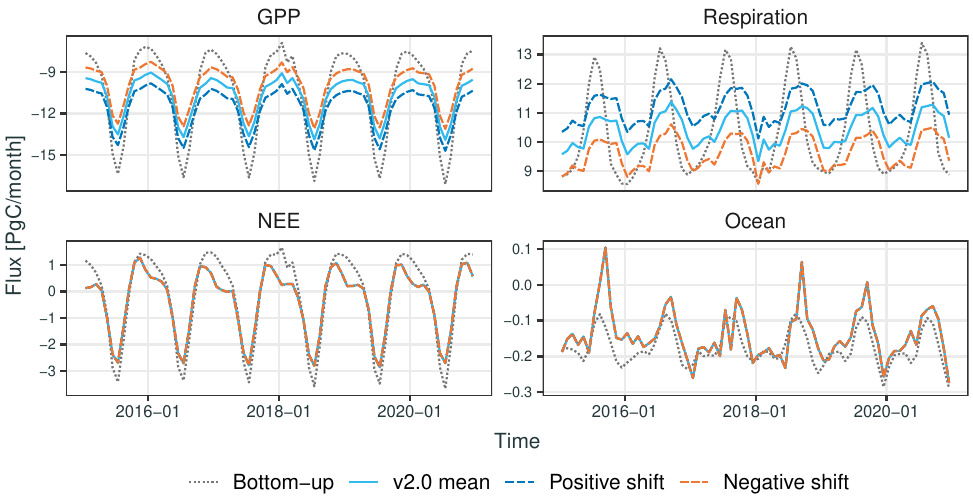}
  \caption{
    Monthly global totals of the GPP, respiration, NEE and ocean components of the four true-flux cases used in the OSSE of Section~\ref{sec:osse}.
    The GPP and respiration components vary between the ``v2.0 mean,'' ``Positive shift,'' and ``Negative shift'' cases, while their sum (NEE) and the ocean component remain unchanged, as designed in \eqref{eq:rlt-shift}.
  }
  \label{fig:osse-true-flux}
\end{figure}

Having selected a true basis-function coefficient vector, $\alphavec^{\textup{true}}$, from one of the four cases described above, the corresponding true-flux field is obtained by substituting $\alphavec^{\textup{true}}$ into the flux process model in \eqref{eq:net-flux}.
Figure~\ref{fig:osse-true-flux} plots monthly global totals of the true GPP, respiration, NEE, and ocean component fluxes for each of the four true-flux cases.
By applying \eqref{eq:sif-process-model} and \eqref{eq:mole-frac-process}, the true-flux field implies a true SIF field and a true CO\textsubscript{2} mole-fraction field, respectively.
We simulate groups of noisy observations of the true SIF and mole-fraction fields at the same locations and times as the actual observations (described in Section~\ref{sec:data}) using the data model given in \eqref{eq:overall-data-vector}.
As described in Section~\ref{supp:error-properties} above, we assume the observations are unbiased and set the true bias vector, $\bvec^{\textup{true}}_g$, to zero for each group.
Recall from Section~\ref{supp:parameter-model} that the observation error budget is scaled by the factor $\gamma^Z_g$, and its corresponding error term is then apportioned to correlated ($\xivec_g$) and uncorrelated ($\epsilonvec_g$) errors by $\rho^Z_g$.
The covariance matrix of the correlated errors is controlled by the temporal length-scale parameter $\ell^Z_g$.
We simulate the true error vectors, $\xivec^{\textup{true}}_g$ and $\epsilonvec^{\textup{true}}_g$, using estimates of $\gamma^Z_g$, $\rho^Z_g$, and $\ell^Z_g$ found from the first stage of the inversion (see Section~\ref{sec:inference}).
These estimates are given in Table~\ref{tab:hyperparameter-table} below.

\newpage
\section{Additional tables and figures}
\label{supp:additional-figures}

\begin{figure}[ht!]
  \centering
  \includegraphics{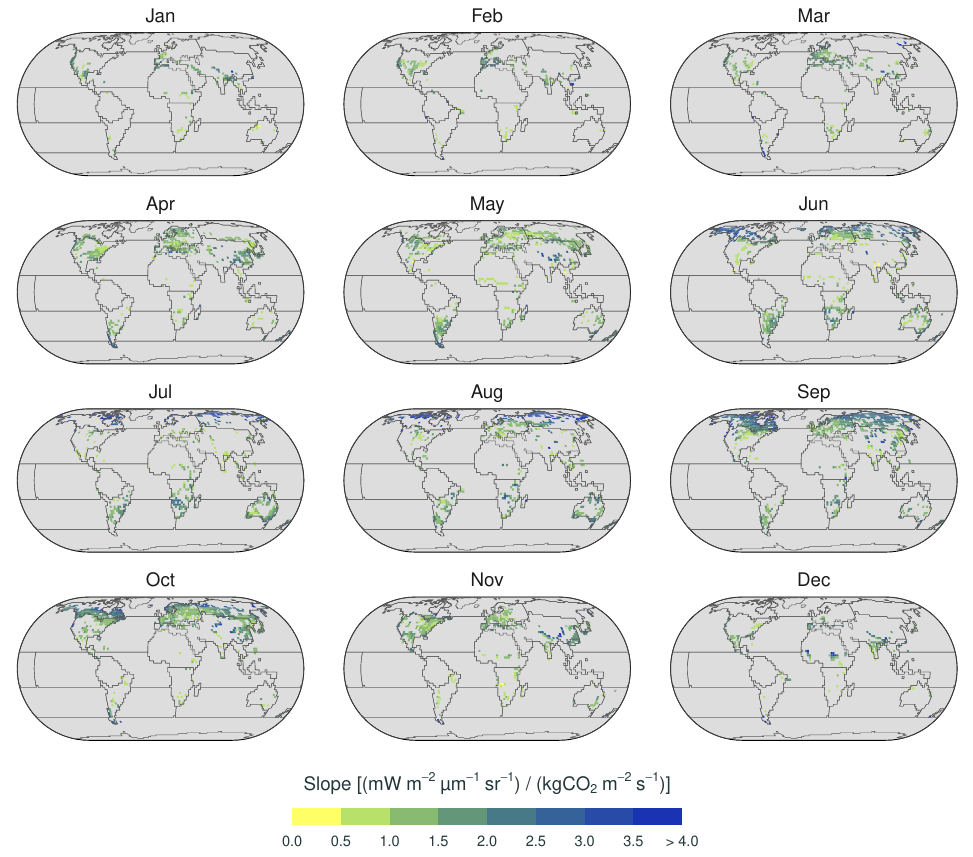}
  \caption{
    Maps of estimated SIF--GPP sensitivities (the slope term in \eqref{eq:sif-gpp-regression}) in each \qtyproduct{2 x 2.5}{\degree} grid cell and calendar month.
    Gray areas are missing values that correspond to ocean-only grid cells or where the four criteria for a valid linear model are not satisfied by the bottom-up SiB4 estimates of GPP and SIF.
  }
  \label{fig:sif-gpp-map-slope}
\end{figure}

\begin{figure}[ht!]
  \centering
  \includegraphics{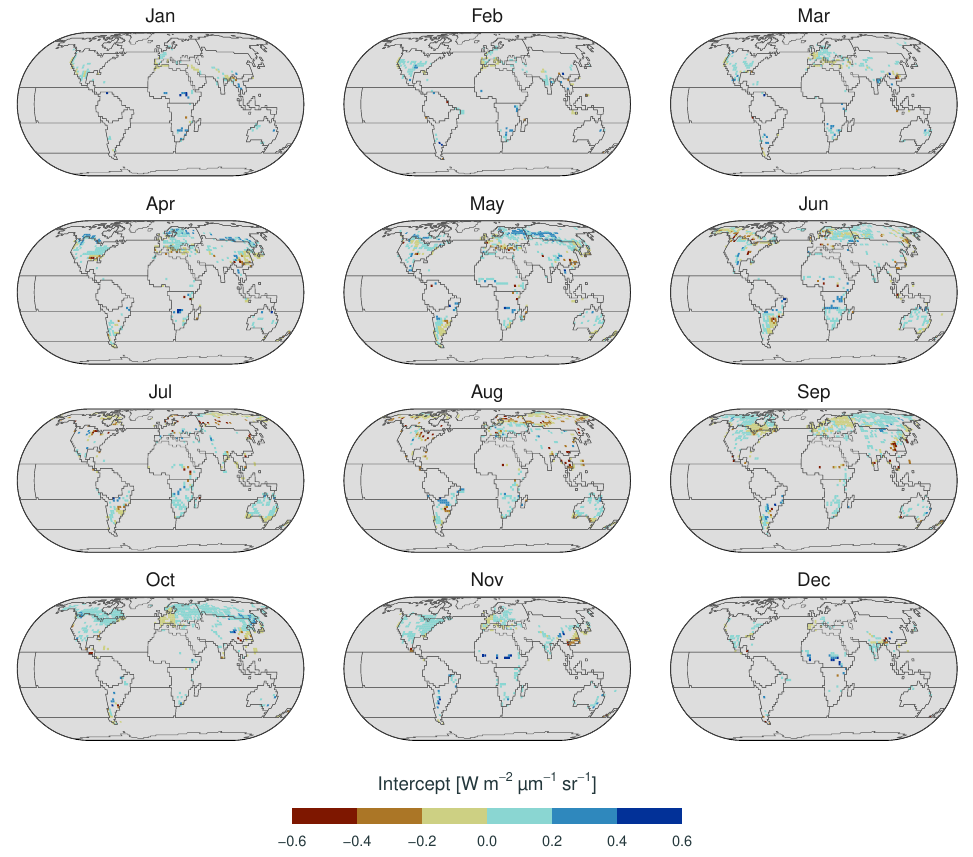}
  \caption{
    Analogous to Figure~\ref{fig:sif-gpp-map-slope}, but for the intercept term in \eqref{eq:sif-gpp-regression}.
  }
  \label{fig:sif-gpp-map-intercept}
\end{figure}

\begin{table}[hb!]
  \renewcommand{\arraystretch}{0.9}
  \centering
  \begin{tabular}{lccl}
\toprule
Observation group & $\hat{\gamma}^Z_g$ & $\hat{\rho}^Z_g$ & \multicolumn{1}{c}{$\hat{\ell}^Z_g$} \\
\midrule
Aircraft & 0.704 & 1.000 & 7.4 minutes \\
Shipboard & 0.704 & 1.000 & 4.8 hours \\
Surface & 0.704 & 1.000 & 15.9 hours \\
Tower & 0.704 & 1.000 & 14.9 hours \\
OCO-2 XCO\textsubscript{2} & 0.705 & 0.934 & 59.1 seconds \\
OCO-2 SIF & 2.032 & 0.931 & 63.2 seconds \\
\bottomrule
\end{tabular}

  \caption{
    Estimates of the error parameters $\gamma^Z_g$, $\rho^Z_g$, and $\ell^Z_g$ from the first stage of the WOMBAT~v2.S inversion with real data.
    Recall that $\hat{\rho}^Z_g = 1$ is fixed for the first four observation groups, and that $\gamma^Z_g$ is re-estimated in the second stage for all groups (see Figure~\ref{fig:traceplots}).
  }
  \label{tab:hyperparameter-table}
\end{table}

\begin{table}[ht!]
  \renewcommand{\arraystretch}{0.9}
  \renewrobustcmd{\bfseries}{\fontseries{b}\selectfont}
  \centering
  \begin{threeparttable}
    \begin{tabular}{lllCCCC}
\toprule
True Flux & \multicolumn{2}{c}{Inversion Setup} & \multicolumn{4}{c}{Mean CRPS [PgC/year]} \\
\cmidrule(lr){2-3} \cmidrule(l{10pt}){4-7}
& RLT\textsuperscript{a} & Includes SIF & GPP & Resp. & NEE & Ocean \\
\midrule
Bottom-up\textsuperscript{b} & Fixed & Yes & \textbf{0.02} & \textbf{0.02} & 0.01 & 0.02 \\
 &  & No & 0.03 & 0.03 & 0.02 & 0.02 \\
 & Inferred & Yes & 0.07 & 0.06 & 0.01 & 0.02 \\
 &  & No & 0.65 & 0.65 & 0.01 & 0.02 \\
\midrule
v2.0 mean\textsuperscript{b} & Fixed & Yes & \textbf{0.10} & \textbf{0.11} & 0.04 & 0.02 \\
 &  & No & 0.15 & 0.15 & 0.04 & 0.02 \\
 & Inferred & Yes & 0.18 & 0.18 & 0.04 & 0.02 \\
 &  & No & 0.70 & 0.70 & 0.04 & 0.02 \\
\midrule
Positive shift & Fixed & Yes & 0.37 & 0.36 & 0.04 & 0.02 \\
 &  & No & 0.35 & 0.35 & 0.04 & 0.02 \\
 & Inferred & Yes & \textbf{0.17} & \textbf{0.16} & 0.04 & 0.02 \\
 &  & No & 0.80 & 0.80 & 0.04 & 0.02 \\
\midrule
Negative shift & Fixed & Yes & 0.35 & 0.36 & 0.04 & 0.02 \\
 &  & No & 0.34 & 0.35 & 0.04 & 0.02 \\
 & Inferred & Yes & \textbf{0.16} & \textbf{0.16} & 0.04 & 0.02 \\
 &  & No & 0.79 & 0.79 & 0.04 & 0.02 \\
\bottomrule
\end{tabular}

    \begin{tablenotes}
      \item [a] {\scriptsize The respiration linear term (RLT) is either fixed to its bottom-up estimate or inferred for land regions.}
      \item [b] {\scriptsize The bottom-up RLT estimate is the true RLT in these cases.}
    \end{tablenotes}
  \end{threeparttable}
  \caption{
    Mean continuous ranked probability score (CRPS) when estimating monthly regional flux components in the OSSE of Section~\ref{sec:osse}.
    A lower score indicates better performance, and the lowest score is given in bold typeface for each true-flux case unless there is a tie.
    We obtain each value over the same regions and time periods as those used for constructing the flux basis functions (see Section~\ref{supp:basis-function-setup}).
  }
  \label{tab:osse-metrics-crps}
\end{table}

\begin{figure}[ht!]
  \centering
  \includegraphics{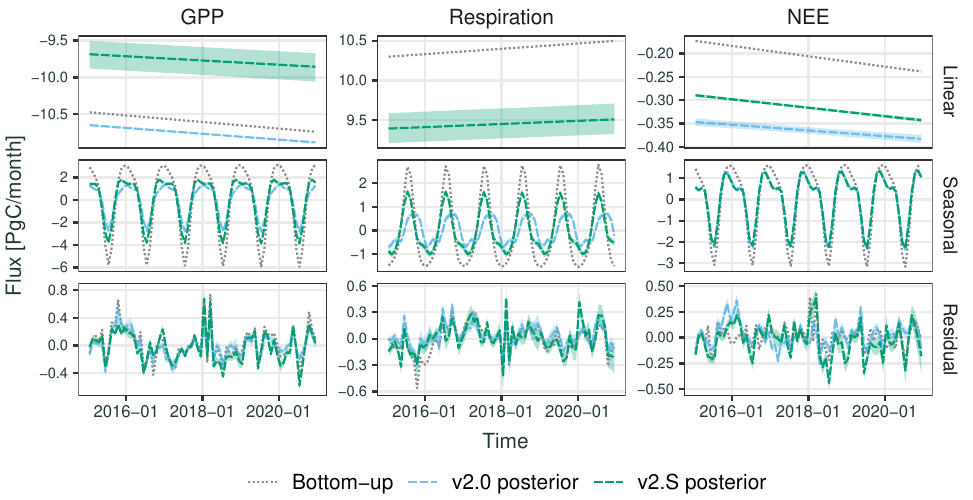}
  \caption{
    Decomposition of monthly global total GPP, respiration, and NEE posterior flux estimates into linear, seasonal, and residual terms.
    Posterior estimates from WOMBAT~v2.0 and v2.S are shown, but the respiration linear term (RLT) is inferred for land regions in v2.S only.
    Shaded areas show the region between the 2.5th and 97.5th posterior percentiles (i.e., \qty{95}{\percent} prediction intervals).
    Bottom-up estimates are included for reference.
  }
  \label{fig:flux-decomposition}
\end{figure}

\begin{figure}[ht!]
  \centering
  \includegraphics{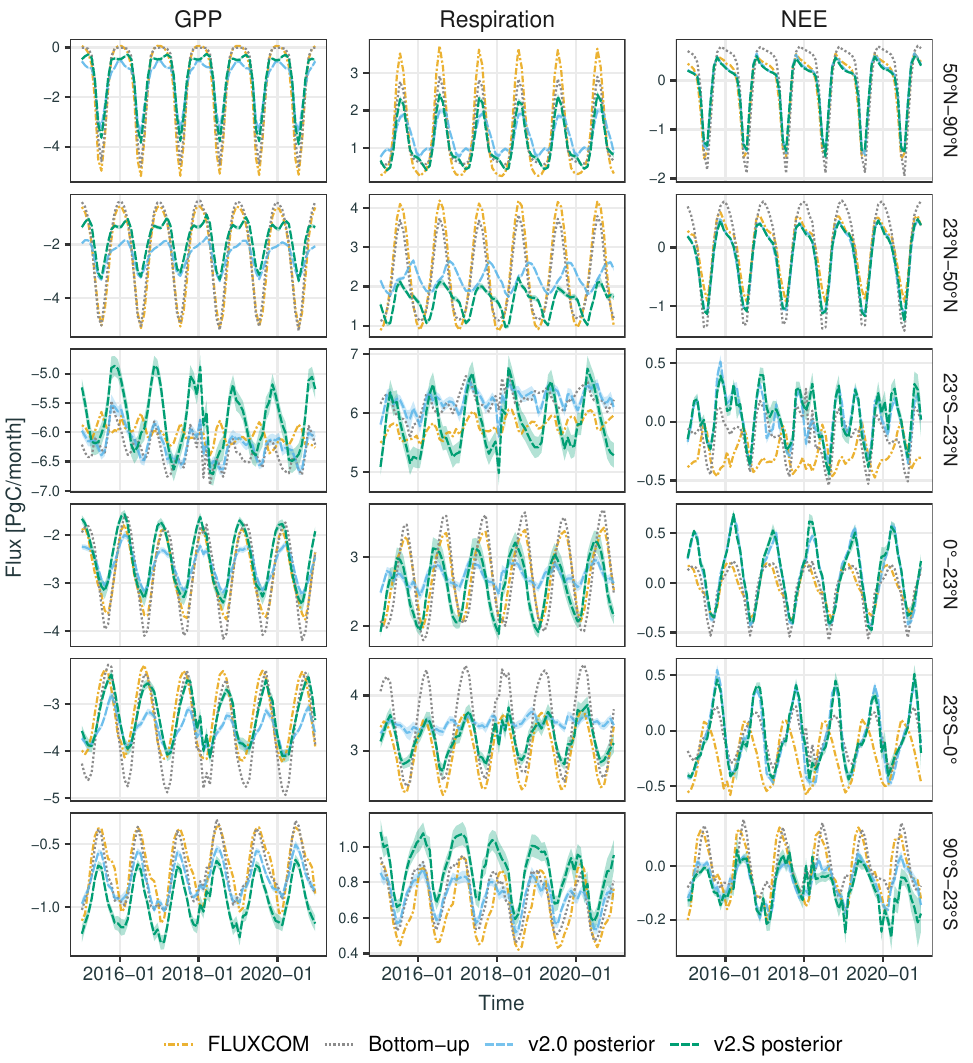}
  \caption{
    Monthly total GPP, respiration, and NEE component fluxes during the period from January 2015 to December 2020, divided into six latitude bands: the northern boreal (\ang{50}N--\ang{90}N), the northern temperate (\ang{23}N--\ang{50}N), the tropics (\ang{23}S--\ang{23}N), the northern tropics (\ang{0}--\ang{23}N), the southern tropics (\ang{23}S--\ang{0}), and the southern extratropics (\ang{90}S--\ang{23}S).
    Lines show the FLUXCOM and bottom-up (SiB4) estimates, and the WOMBAT~v2.0 and v2.S posterior means.
    Shaded areas around WOMBAT estimates depict the region between the 2.5th and 97.5th posterior percentiles.
  }
  \label{fig:flux-net-zonal}
\end{figure}

\begin{figure}[ht!]
  \centering
  \includegraphics{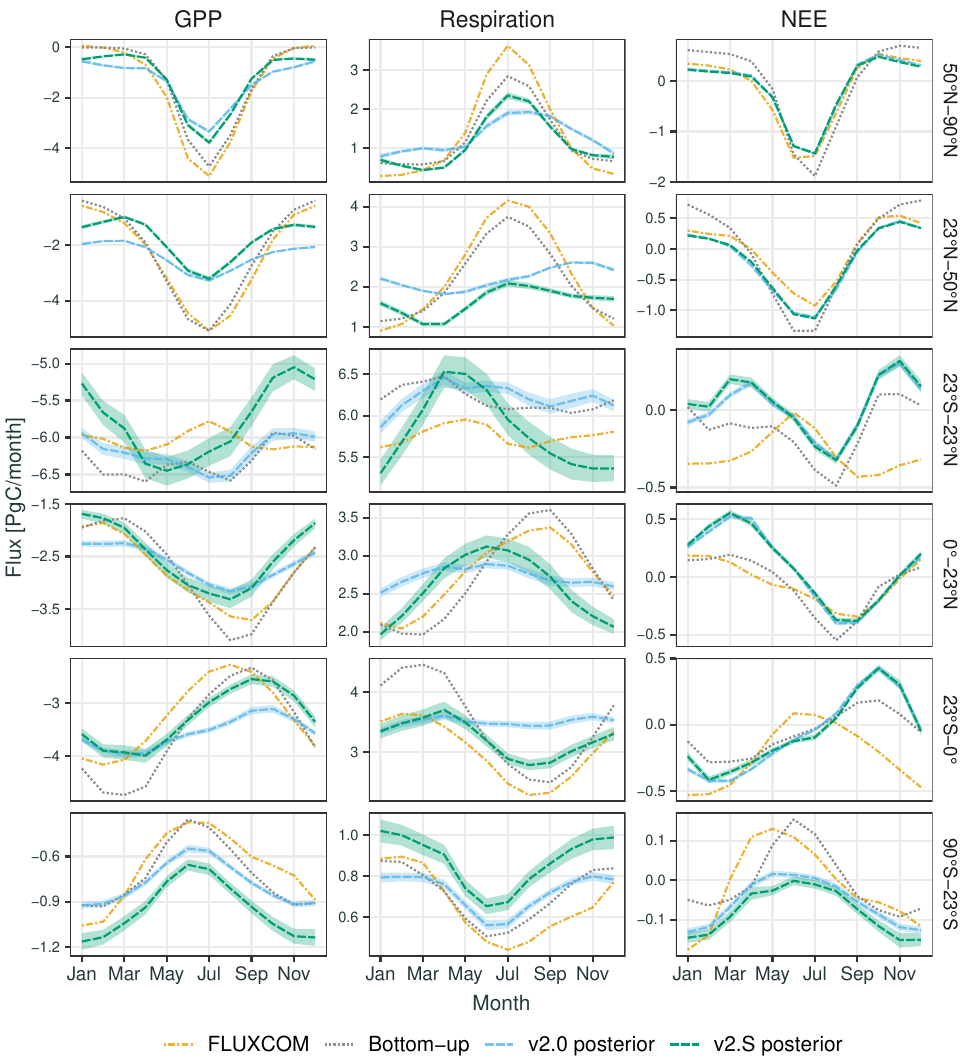}
  \caption{
    Analogous to Figure~\ref{fig:flux-net-zonal}, but for the seasonal profile of each component flux, where each month shows the average of the six corresponding estimates during the six-year period.
    Due to the smaller scale in the tropics band, the wider posterior prediction intervals in WOMBAT~v2.S are more evident there.
  }
  \label{fig:seasonal-cycle-zonal}
\end{figure}

\begin{figure}[ht!]
  \centering
  \includegraphics{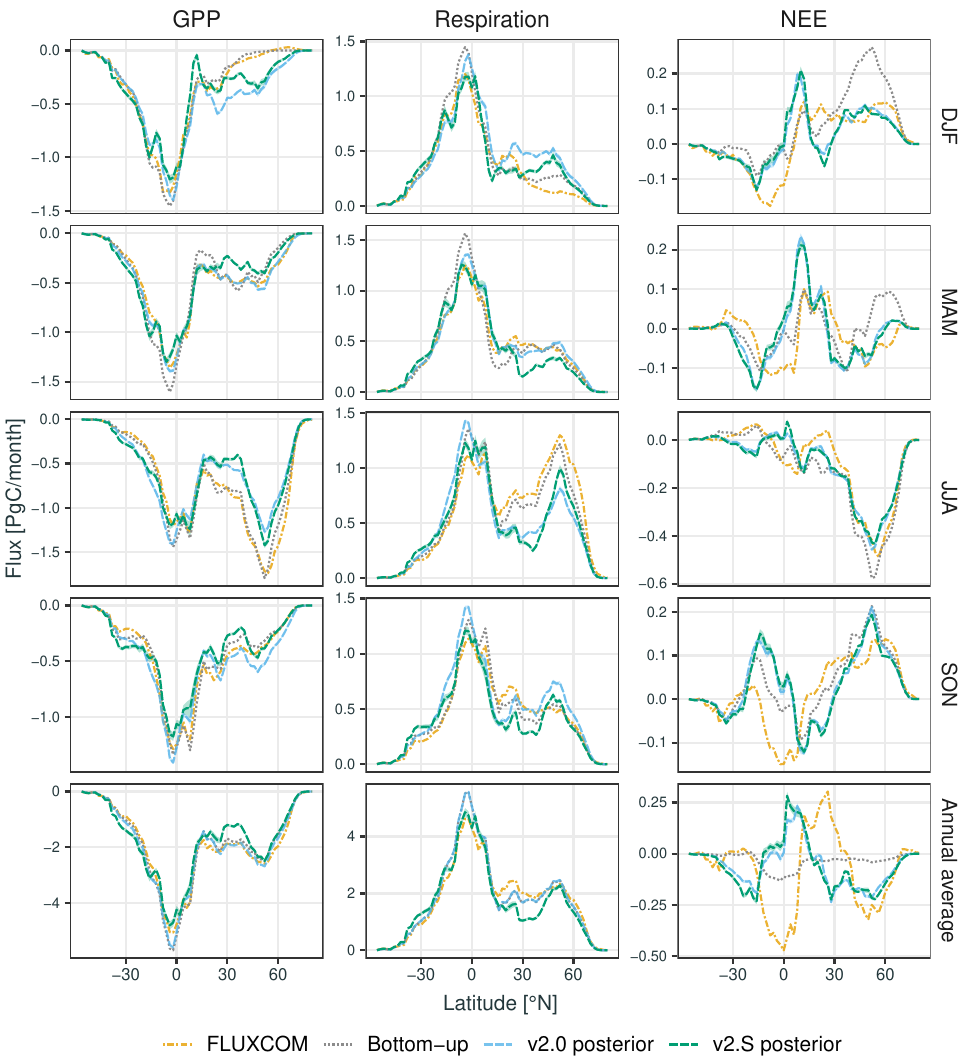}
  \caption{
    Averaged latitudinal profile of GPP, respiration, and NEE monthly flux estimates in boreal winter (DJF), spring (MAM), summer (JJA), autumn (SON), and annually.
    Averages are taken over seasons and \qty{2}{\degree} latitude bands for estimates during the period from January 2015 to December 2020.
    Lines show the FLUXCOM and bottom-up (SiB4) estimates, and the WOMBAT~v2.0 and v2.S posterior means.
    Shaded areas around the WOMBAT estimates depict the region between the 2.5th and 97.5th posterior percentiles (i.e., \qty{95}{\percent} prediction intervals).
  }
  \label{fig:seasonal-latitude-profile}
\end{figure}

\begin{figure}[ht!]
  \centering
  \includegraphics{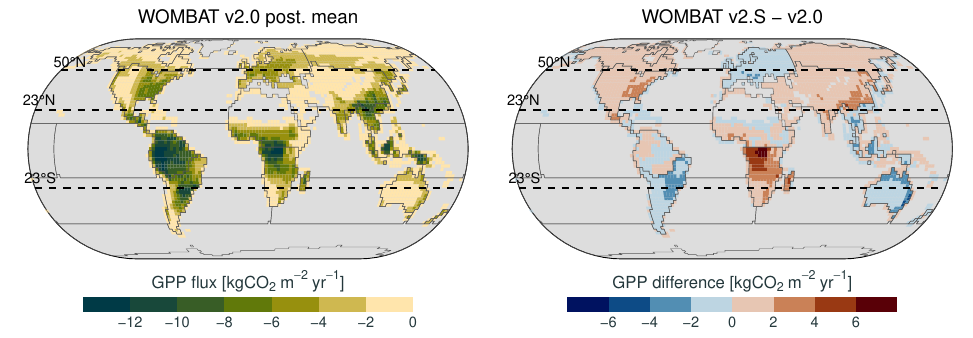}
  \caption{
    (Left) WOMBAT~v2.0 posterior mean of the average GPP flux over January 2015--December 2020.
    (Right) Difference between the WOMBAT~v2.S and v2.0 posterior means of the average GPP flux.
    Grid cells are \qtyproduct{2 x 2.5}{\degree} and those shown in gray have zero GPP flux.
    Solid gray lines delineate the 23 regions used to partition the spatial domain.
  }
  \label{fig:average-maps-gpp-v2}
\end{figure}

\begin{figure}[ht!]
  \centering
  \includegraphics{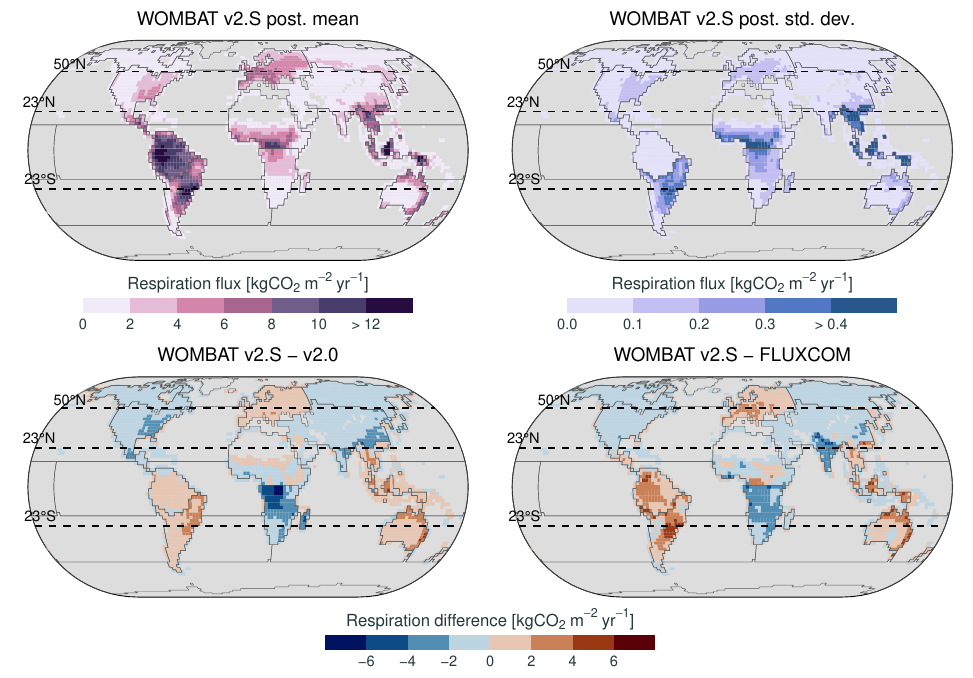}
  \caption{
    (Top row) WOMBAT~v2.S posterior mean and posterior standard deviation of the average respiration flux over January 2015--December 2020.
    (Bottom row) Difference between the WOMBAT~v2.S and v2.0 posterior mean of the average respiration flux, and difference between the WOMBAT~v2.S posterior mean and the FLUXCOM estimate of the average respiration flux.
    Grid cells are \qtyproduct{2 x 2.5}{\degree} and those shown in gray have zero respiration flux.
    Solid gray lines delineate the 23 regions used to partition the spatial domain.
  }
  \label{fig:average-maps-resp}
\end{figure}

\begin{figure}[ht!]
  \centering
  \includegraphics{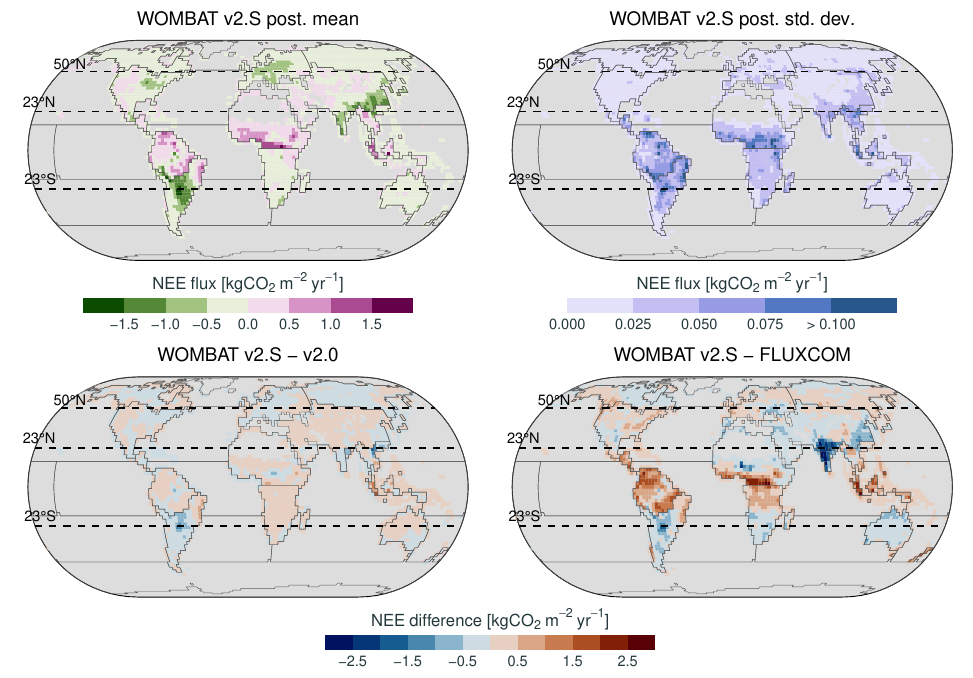}
  \caption{
    Analogous to Figure~\ref{fig:average-maps-resp}, but for NEE.
  }
  \label{fig:average-maps-nee}
\end{figure}

\clearpage
\bibliographystylesupp{apalike-maxbibnames20}
\bibliographysupp{references}

\end{document}